%% file: main.tex
\RequirePackage[2018-12-01]{latexrelease}

\documentclass{rQUF2e}

\usepackage{multirow}
\usepackage{adjustbox}
\usepackage{amsmath}
\usepackage{pdflscape}
\usepackage{rotating}
\usepackage{graphicx} 
\usepackage{xcolor}
\usepackage{booktabs}
\usepackage[normalem]{ulem}
\usepackage{hyperref}
\usepackage{cleveref}

\begin{document}


\title{Temporal Representation Learning for Stock Similarities and Its Applications in Investment Management}

\author{Yoontae Hwang$^{1}$,  
Stefan Zohren$^{2}$\thanks{Corresponding author. Email: stefan.zohren@eng.ox.ac.uk}, and
Yongjae Lee$^{1}$\thanks{Corresponding author. Email: yongjaelee@unist.ac.kr}\
\affil{$^{1}$Ulsan National Institute of Science and Technology, Ulsan, Republic of Korea\
$^{2}$University of Oxford, Oxford, UK}}

\received{June 17, 2024}

\maketitle

\begin{abstract}
In the era of rapid globalization and digitalization, accurate identification of similar stocks has become increasingly challenging due to the non-stationary nature of financial markets and the ambiguity in conventional regional and sector classifications. To address these challenges, we examine \texttt{SimStock}, a novel temporal self-supervised learning framework that combines techniques from self-supervised learning (SSL) and temporal domain generalization to learn robust and informative representations of financial time series data. The primary focus of our study is to understand the similarities between stocks from a broader perspective, considering the complex dynamics of the global financial landscape. We conduct extensive experiments on four real-world datasets with thousands of stocks and demonstrate the effectiveness of \texttt{SimStock} in finding similar stocks, outperforming existing methods. The practical utility of \texttt{SimStock} is showcased through its application to various investment strategies, such as pairs trading, index tracking, and portfolio optimization, where it leads to superior performance compared to conventional methods. Our findings empirically examine the potential of data-driven approach to enhance investment decision-making and risk management practices by leveraging the power of temporal self-supervised learning in the face of the ever-changing global financial landscape.

\end{abstract}


\begin{classcode} C45, C58, C67, G11, G17 \end{classcode}

\label{ch:ch1}

\input{content/chap1_intro.tex}

\label{ch:ch2}
\input{content/chap2_related.tex}

\label{ch:ch3}
\input{content/chap3_simstock.tex}

\label{ch:ch4}
\input{content/chap4_exp.tex}


\label{ch:ch5}
\input{content/chap5_con.tex}



\bibliographystyle{rQUF}
\bibliography{rQUFguide}

\appendix
\clearpage
\label{ch:appendix_A}
\input{Appendix/Appendix_A.tex}

\clearpage
\label{ch:appendix_B}
\input{Appendix/Appendix_B.tex}

\clearpage
\label{ch:appendix_C}
\input{Appendix/Appendix_C.tex}

\clearpage
\label{ch:appendix_D}
\input{Appendix/Appendix_D.tex}

\clearpage
\label{ch:appendix_E}
\input{Appendix/Appendix_E.tex}

\clearpage
\label{ch:appendix_F}
\input{Appendix/Appendix_F.tex}

\end{document}

%% file: content/chap1_intro.tex
\section{Introduction}
The identification of similar stocks is an essential task in finance, with far-reaching implications for portfolio diversification and risk management. Traditionally, the similarity between stocks has been determined based on factors such as industry sector, geographical location, and market capitalization. However, in the era of rapid globalization and digitization, these conventional approaches have become increasingly inadequate in capturing the complex dynamics of the global financial landscape. The rapid pace of globalization has led to the emergence of intricate relationships among companies across different sectors and regions. Multinational corporations now operate in multiple countries, blurring the lines between traditional sector classifications. Additionally, the rise of digital technologies has given birth to new industries and business models that transcend geographical boundaries. As a result, the similarities between stocks can no longer be accurately captured by relying solely on regional or sector-based classifications. Moreover, the non-stationary nature of financial markets poses significant challenges to the identification of similar stocks. The statistical properties of financial time series, including returns and correlations, are subject to change over time, a phenomenon known as concept drift or temporal shift \citep{lu2018learning, bai2022temporal}. Consequently, similarity measures based on historical data may become unreliable and fail to capture the evolving dynamics of the market. Indeed, \citep{wan2021sentiment} showed that news data can be useful for modelling evolving company relationships.

Recent advances in deep learning-based methods in finance domain have shown promise in predicting the desired parameters from the given data \citep{zhang2020deep}. However, financial time series data presents unique challenges. The parameters of interest, such as expected returns, are not only difficult to predict due to the presence of excessive noise, but also the labels that can help the model extract hidden patterns in the data are often not well-defined. Moreover, the complex interactions among various financial assets and the impact of external factors, such as macroeconomic variables and market sentiment, further complicate the estimation process.

One promising solution to tackle these challenges is to leverage representation learning techniques to extract meaningful embeddings from unlabeled financial time series data, taking into account for temporal distribution shifts. By learning robust and informative representations, we can capture the complex relationships between different assets and exploit the inherent structure in the data. Self-supervised learning (SSL) has emerged as a powerful paradigm for learning such representations from unlabeled data, with successful applications in various domains, including computer vision \citep{chen2020simple, he2020momentum} and natural language processing \citep{devlin2018bert, yang2019xlnet}. SSL enables the model to learn meaningful representations by solving pretext tasks that do not require explicit labels, making it particularly suitable for scenarios where labeled data is scarce or expensive to obtain. However, while these methods are useful for exploiting inductive biases in data used in any domain, their use for financial time series data has been less explored.

Time series data exhibits distinct characteristics, such as seasonality, trend, and interaction between stocks, which require careful consideration when designing SSL frameworks. Furthermore, the non-stationary nature of financial markets necessitates the development of techniques that can adapt to distribution shifts and generalize well to future time periods. Most existing SSL methods focus on learning invariant representations \citep{chen2020simple, he2020momentum}, assuming that the data distribution remains stationary. However, this assumption does not hold in the context of financial markets, where the underlying dynamics can change over time.

To address these challenges, we combine techniques from SSL and the field of temporal domain generalization. Our approach aims to learn general model representations that can adapt to temporal shifts over time, thereby enhancing the accuracy and robustness of financial parameter estimation using learned embeddings. By incorporating temporal domain generalization into the SSL framework, we enable the model to learn representations that are not only informative but also resilient to distribution shifts.

The main contributions of this paper are as follows:
\begin{itemize}
\item We examine a novel temporal Self-Supervised Learning (SSL) framework, \texttt{SimStock}, that combines SSL with temporal domain generalization to learn robust and informative representations of financial time series data. \texttt{SimStock} leverages the power of SSL to capture the complex relationships between different financial assets while accounting for temporal shifts in the data distribution;
\item We introduce a new corruption method for SSL of stock data, termed dimension corruption, which integrates temporal patterns into the corruption process. By corrupting the input data along different dimensions, such as time, asset, and feature, \texttt{SimStock} learns representations that are robust to noise and non-stationarity in the data;
\item We conduct extensive experiments on four real-world benchmarks with thousands of stocks to demonstrate the effectiveness of \texttt{SimStock} in finding similar stocks. Our results show that \texttt{SimStock} achieves state-of-the-art performance, outperforming existing methods in terms of accuracy and robustness. We showcase the practical utility of \texttt{SimStock} in simplifying the process of screening potential investment opportunities; 
\item We demonstrate the practical utility of \texttt{SimStock} in various financial applications, including pairs trading, index tracking, and portfolio optimization. We show that using the similar stocks identified by \texttt{SimStock} leads to superior performance compared to conventional methods in these applications, highlighting the potential of our approach to enhance investment strategies and risk management.
\end{itemize}

%% file: content/chap2_related.tex
\section{Related work}
This section explores relevant research in two key areas: Self-Supervised Learning (SSL) for time series data and the estimation of financial parameters from historical data. We discuss the challenges and limitations of applying SSL techniques to time series and review various approaches aimed at improving the robustness and accuracy of parameter estimation in finance.

\subsection{Self-supervised learning for time series data} Self-Supervised Learning (SSL) has emerged as a promising paradigm for learning robust and generalizable representations from unlabeled data, with successful applications in both computer vision and natural language processing (NLP) \citep{chen2020simple, jing2020self, zhai2019s4l, lee2019latent, qiu2020pre, ruder2018strong, song2020mpnet}. SSL methods aim to overcome the limitations of traditional supervised learning, which requires large amounts of manually labeled data and can produce models that are non-robust and sensitive to small variations in the inputs.

First and foremost, in the field of NLP, Self-Supervised Learning (SSL) methods, such as BERT \citep{devlin2018bert}, GPT \citep{radford2019language}, and their variants \citep{mann2020language, touvron2023llama, jiang2024mixtral}, have achieved state-of-the-art results earlier than in other fields by pre-training on large corpora of unlabeled text data to learn to predict masked words or generate realistic text sequences. One reason for the success of SSL in NLP is the abundance of unlabeled text data. Later, SSL techniques were proposed to learn useful visual features by solving pre-tasks on unlabeled images, such as predicting the relative positions of image patches or identifying distorted versions of an image, in order to alleviate the label requirements and enable powerful feature extraction in computer vision as well \citep{wei2022masked, fang2022corrupted}. This sequential adaptation is not only due to the abundance of unlabeled text data in NLP, but also because language data is discrete and structured (i.e., words), while image data is high-dimensional, continuous, and amorphous, as discussed by \citep{he2020momentum}, the success of SSL methodologies in NLP followed by success in computer vision is a natural result.

In the temporal domain, recent research on SSL methods has predominantly concentrated on video understanding \citep{jenni2020video} or action classification \citep{qian2021spatiotemporal}. Video-based data is particularly suitable for SSL approaches because of the high correlation between frames, despite its temporal nature. Also, the temporal coherence and continuity present in videos provide a rich source of information that can be exploited to learn meaningful representations without the need for extensive manual labeling. Therefore, methodologies used in computer vision, such as contrastive learning and predictive modeling, can be easily adapted to the video domain. For instance, recent research leverage SSL techniques to learn patch-level \citep{yun2022patch, caron2021emerging}  or region-level \citep{xiao2021region} representations from videos. These approaches aim to capture the inherent structure and relationships within video frames, enabling the learning of meaningful features without relying on explicit labels. By exploiting the temporal coherence and spatial consistency present in videos, these self-supervised methods can discover emergent properties and capture rich semantic information. 

However, limited research has been conducted regarding the application of such techniques to financial data reflecting its temporal characteristics. One of the main reasons for this is the challenge of generating different \textit{views} (both positive and negative), which play a key role in self-supervised representation, in non-stationary time-series data. In the case of financial price data, it is not straightforward to generate meaningful positive and negative views that capture the inherent temporal dependencies and preserve the essential characteristics of the original time series. Unlike images or text, where various augmentation techniques can be applied to create different \textit{views} (e.g., cropping, flipping, masking or word substitution) without significantly altering the core information, financial time series data is highly sensitive to temporal order and contextual relationships. Previous studies have applied various methods to generate views for time series data in order to solve this problem. The work in \citep{yue2022ts2vec} utilized timestamp masking and random cropping, while \citep{choi2023multi} employed scaling and permutation techniques. However, these approaches may not be suitable for time series data that require fine-grained representations.

Next, multivariate time series, which are common in most time series data, present a unique challenge for Self-Supervised Learning (SSL) methods. Although these time series contain multiple dimensions, the relevant and informative patterns are often confined to a limited number of dimensions. Consequently, directly applying SSL techniques that have been successful with other data types to extract meaningful information from time series data becomes difficult, as the sparse nature of useful information in multivariate time series hinders the effectiveness of these methods. 

Third, time series data possess distinct characteristics, including seasonality, trend, and frequency domain information \citep{wen2020fast, wu2021autoformer}. Considering how SSL works, in sequential data, the hidden patterns of the data are used based on the "current time" that the data is being learned. In other words, this assumes that these characteristics will apply the same in the future. However, the non-stationary nature of time series data poses a significant challenge for SSL methods. The underlying patterns and relationships in the data can change over time, leading to a phenomenon known as concept drift or temporal shift. This means that the representations learned by SSL methods based on historical data may not generalize well to future time periods. To address this issue, techniques from the field of Domain Generalization (DG) can be employed. DG aims to learn general model representations that can adapt to temporal shifts over time. To improve the generalization ability of the model when there is temporal change, methodologies such as DRAIN \citep{bai2022temporal} have been proposed, but they are only applicable to supervised learning. 

This work extends an earlier prototype of \texttt{SimStock} which was introduced in \citep{hwang2023simstock} and explored the possibility of using SSL for identifying similar stocks based on their temporal dynamics. Here we show that such an architecture indeed finds applicability and effectiveness across various financial tasks, such as pairs trading, index tracking, and portfolio optimization, by conducting rigorous comparative analyses of different models within this framework.

\subsection{Parameter estimation in non-stationary financial markets} In the field of investment management, estimating future returns and risk is a fundamental challenge due to the inherent uncertainty and non-stationary nature of financial markets. Traditionally, researchers and practitioners have relied on historical financial data to estimate various parameters, such as expected returns, volatility, and covariance matrices, which are crucial inputs for portfolio optimization, risk management, and asset pricing models.

One common approach is to use a rolling window of historical returns to estimate the expected returns and covariance matrix of a set of financial assets. For instance, in the classic mean-variance portfolio optimization (MVO) framework proposed by \citep{Markowitz1952}, the optimal portfolio weights are determined based on the estimated expected returns and covariance matrix using a sample of historical returns. See \citep{kim2021mean} for more explanations. Similarly, the Capital Asset Pricing Model (CAPM) \citep{sharpe1964capital, lintner1975valuation} and the Fama-French factor models \citep{fama1993common, fama2015five} rely on historical data to estimate the beta coefficients and factor premiums. 

However, using historical data for parameter estimation has several limitations. First, financial markets are known to exhibit non-stationarity, meaning that the statistical properties of the data, such as the mean and variance, can change over time \citep{cont2001empirical, lo2017adaptive}. This implies that the estimates based on historical data may not be representative of the future. Second, the sample size of historical data is often limited, leading to estimation errors and potential over-fitting \citep{kan2004optimal, kan2007optimal}. For instance, in the context of portfolio optimization, an inherent drawback of MVO is the high sensitivity of the optimal portfolio to estimation errors in the input parameters \citep{michaud1989markowitz}, particularly in the expected returns \citep{chopra2013effect} and correlations \citep{chung2022effects}. Small changes in the estimated expected returns can lead to significant shifts in the optimal portfolio weights, resulting in portfolios that may be suboptimal or unstable out-of-sample.

To mitigate these issues, researchers have proposed various techniques to improve the robustness and accuracy of parameter estimation. One popular approach is the shrinkage method, which combines the sample estimates with a structured estimator to reduce the estimation error \citep{ledoit2003improved, ledoit2004well, ledoit2004honeyorigin} Another technique is the use of robust estimators, such as the minimum covariance determinant (MCD) estimator \citep{rousseeuw1999fast} and the minimum volume ellipsoid (MVE) estimator \citep{van2009minimum}, which are less sensitive to outliers and heavy-tailed distributions. More recently, \citep{gerber2021gerber} proposed the Gerber statistic for estimating the covariance matrix between assets, a robust co-movement measure that extends Kendall's Tau by counting the proportion of simultaneous co-movements when their amplitudes exceed data-dependent thresholds, capturing meaningful co-movements while being insensitive to extreme values and noise.

In addition to these techniques, researchers have also explored various approaches to extrapolate past data and model future market dynamics. For example, \citep{barberis2000investing} proposed a Bayesian approach that combines the sample estimates with prior beliefs about the asset returns to improve the out-of-sample performance of portfolio optimization. \citep{rapach2010out} employed combination forecasts, which aggregate individual forecasts based on different predictors, to enhance the accuracy of out-of-sample stock return predictions. Also, \citep{welch2008comprehensive} found that the historical average excess returns of stocks over bonds, which is often used as an estimate of the equity premium, is sensitive to the choice of the sample period and the assumptions about survivorship bias. Moreover, they showed that this historical average is a poor predictor of future returns, as the equity premium exhibits substantial time-variation and mean-reversion.

Despite these advancements, the fundamental challenge of estimating future returns and risk from historical data remains. The assumption that future behavior will be similar to the past is often violated in practice, as financial markets are subject to regime shifts, structural breaks, and extreme events \citep{ang2002international, guidolin2007asset}. See \citep{lee2023overview} for more detailed review of machine learning for asset management.

%% file: content/chap3_simstock.tex
\section{\texttt{SimStock}}

\begin{figure}
\centering
\includegraphics[width=\textwidth]{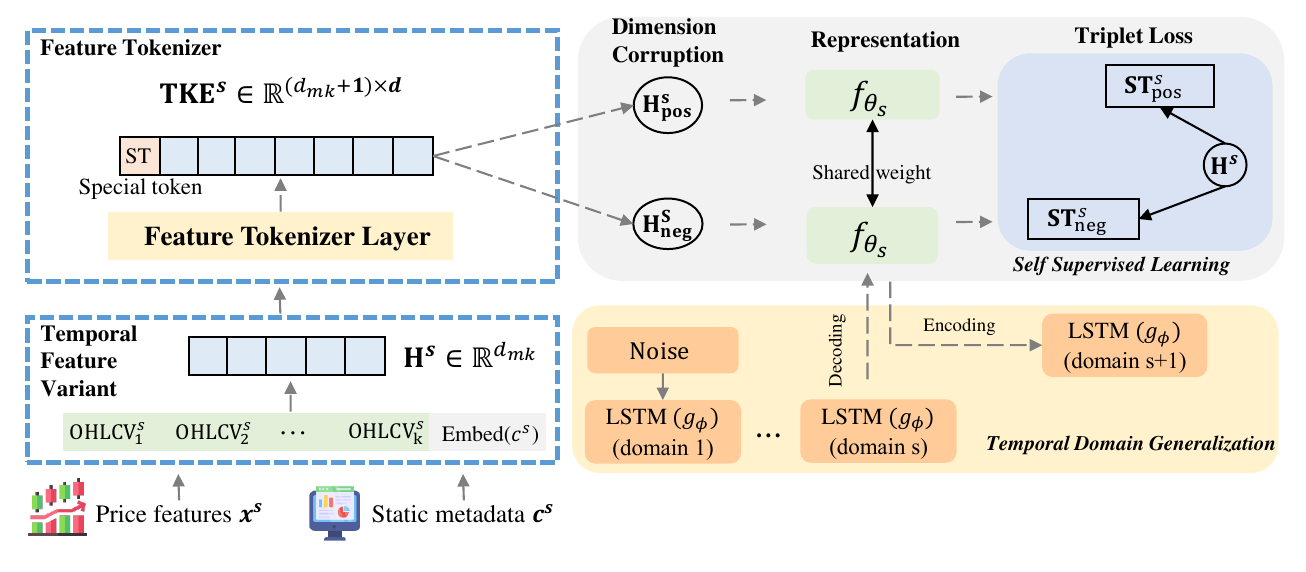}
\caption{\texttt{SimStock} combines self-supervised learning framework with temporal domain generalization for more robust and comprehensive stock representations. }
\label{chap3_model}
\end{figure}

In this section, we introduce \texttt{SimStock}\footnote{Its earlier prototype introduced in \citep{hwang2023simstock}.}, a novel SSL framework that incorporates temporal domain generalization to learn robust and comprehensive representations of financial time series data. The various components of the model are graphically illustrated in \cref{chap3_model}.

\subsection{Preliminaries}
We consider a self-supervised task where the stock data distribution evolves over time. In the training phase, we are given $\mathrm{\mathit{T}}$ observed source domains $ \mathcal{D}_{1:T} = \{ \mathrm{\mathcal{D}_1}, \mathrm{\mathcal{D}_2}, ..., \mathrm{\mathcal{D}_\textit{T}} \}$, which are sampled from distributions at $\mathrm{\mathit{T}}$ different time points $t_{1} \le t_{2} \le ... \le t_{T}$. 
Each source domain is denoted as $\mathrm{\mathcal{D}_s} = \{x_{i}^{s}, c_{i}^{s}\}_{i=1}^{N_{s}}$, for $s=1,2,..., \mathrm{\mathit{T}}$, where $x^{s}\in \mathbb{R}^{d_{m}}$ represent the $d_{m}$-dimensional temporal features, $c^{s} \in \mathbb{R}^{d_n} $ is $d_{n}$-dimensional static metadata, and $N_{s}$ is the sample size at timestamp $t_{s}$. We have omitted the sample index $i$ for simplicity. The model will only be tested on a target domain in the future, i.e., $\mathrm{\mathcal{D}_{\textit{T}+1}}$ where $t_{T+1} \geq t_{T}$. 

Our goal is to proactively capture the drift from temporal domains to find stock representations that are robust with respect to temporal distribution shifts. We presume that the representation model, denoted as $f_{\theta_{s}}$, is characterized by a deep neural network with function parameters $\theta_{s}$ at timestamp $t_{s}$. Consequently, we can get the representation embedding $z_{s}=f_{\theta_{s}}(x^{s},c^{s})$, where $z_{s}$ represent both temporal and static features of stock data. In the next section, we show that the representation model serves as a mapping function during training, which predicts the dynamics across the parameter $\theta_{1:T} = \{\theta_{1}, \theta_{2},..,\theta_{T} \}$ at each domain $D_{s}$ for $s=1,2,..., \mathrm{\mathit{T}}$. 

\subsection{Training Dynamics of Temporal Domain Generalization} \label{chap3_TDG}
We are motivated by DRAIN \citep{bai2022temporal}, which first proposed the concept of temporal domain generalization. In each temporal domain $\mathrm{\mathcal{D}}_{s}$, the representation network $f_{\theta_{s}}$ can be trained by maximizing the conditional probability $\mathbb{P}(\theta_{s}| \mathrm{\mathcal{D}}_{s})$. Here, $\theta_{s}$ represents the state of the model parameters at timestamp $t_s$. Given the dynamic nature of $\mathrm{\mathcal{D}}_{s}$, the conditional probability $\mathbb{P}(\theta_{s} | \mathrm{\mathcal{D}}_{s})$ will also change over time. The objective, in the context of temporal domain generalization, is to estimate $\theta_{\textit{T}+1}$ utilizing all the training data from $\mathrm{\mathcal{D}_{1:T}}$. From a probabilistic perspective, we can express this as:
\begin{equation}
    \mathbb{P}(\theta_{T+1} | \mathrm{\mathcal{D}}_{1:T}) = \int_{\Omega} \mathbb{P}(\theta_{T+1} | \theta_{1:T},\mathrm{\mathcal{D}}_{1:T}) \cdot \mathbb{P}(\theta_{1:T} | \mathrm{\mathcal{D}}_{1:T}) d\theta_{1:T},
    \label{eq1}
\end{equation}
where $\Omega$ denotes the space for model parameters $\theta_{1:T}$. In \cref{eq1}, the first term inside the integral $\mathbb{P}(\theta_{T+1} | \theta_{1:T},\mathrm{\mathcal{D}}_{1:T})$ represents the inference phase, which is the process of predicting the future state of the target representation network (i.e., $\theta_{T+1}$) given all historical states (i.e., $\theta_{1:T},\mathrm{\mathcal{D}}_{1:T}$). The second term $\mathbb{P}(\theta_{1:T} | \mathrm{\mathcal{D}}_{1:T})$ signifies the training phase, which involves leveraging all training data $D_{1:T}$ to ascertain the state of the model on each source domain. More specifically, the training phase can decomposed as follows: 

\begin{equation}
\begin{aligned}
\mathbb{P}\left(\bm{\theta}_{1:T} \mid \mathcal{D}_{1:T}\right) &= \prod_{s=1}^T \mathbb{P}\left(\bm{\theta}_s \mid \bm{\theta}_{1:s-1}, \mathcal{D}_{1:s}\right) \\
&= \mathbb{P}\left(\bm{\theta}_1 \mid \mathcal{D}_1\right) \prod_{s=2}^T \mathbb{P}\left(\bm{\theta}_s \mid \bm{\theta}_{1:s-1}, \mathcal{D}_{1:s}\right).
\end{aligned}
\end{equation}

This formula breaks down the training process into $T$ steps, where the first step corresponds to learning the model parameter on the first domain $\mathcal{D}_1$, and each subsequent step corresponds to learning the model parameter on the new domain $\mathcal{D}_s$ conditional on parameter statuses from the previous domains and training data, i.e., $\mathbb{P}\left(\bm{\theta}_s \mid \bm{\theta}_{1:s-1}, \mathcal{D}_{1:s}\right), \forall s \in \{2,\dots,T\}$.

Suppose that we are at time $t_s$. In order to effectively address the temporal drift present across the domain, the next parameters $\theta_{s+1}$ need to be updated on the current and previous domains $D_{1:s}$. The main problem is how to actually update $\theta_{s+1}$. In this regard, DRAIN introduces a sequential learning process using a LSTM \citep{hochreiter1997long} to describe the stochastic process of ${\theta_s}$. Within the LSTM, each unit $g_\phi$ defined by its parameters $\phi$ is used to generate $\theta_{s+1}$ while taking into consideration the preceding context $D_{1:s}$ and $\theta_{1:s}$. This process is illustrated with yellow boxes in \cref{chap3_model}.

\subsection{Temporal Representation Learning} \label{TRL}
Our ultimate goal is to learn a representation model, $f_{\theta_{s}}$, which captures the stock data distribution that evolves over time. To achieve this, we develop an SSL framework for temporal representation learning of stock data.

{\bfseries Temporal feature variant.} \label{feature_variant} Let $x^{s} \in \mathbb{R}^{d_m}$ be the price feature of a stock, where $d_m$ is the dimension of the feature. The time-varying patterns of stock prices are essential for identifying short- and long-term characteristics of stocks. To learn more rich representations, the price feature $x^{s}$ is processed by a temporal transformation module $\mu$. Specifically, the price feature $x^{s}$ is provided with $k$ variations, denoted as:

\begin{equation}
\mu(x^{s}) = \text{CONCAT}(\mu_{1}(x^{s}), \mu_{2}(x^{s}),...,\mu_{k}(x^{s})) \in \mathbb{R}^{d_{mk}},
\end{equation}
where $d_{mk} = d_m \times k$, and each $\mu_{i}: \mathbb{R}^{d_m} \to \mathbb{R}^{d_m}$ for $i \in \{1, 2, ..., k\}$ is a temporal transformation function from the collection $\mathit{U}$. In other words, $\mu_{1}, \mu_{2}, ..., \mu_{k} \in \mathit{U}$, where $\mathit{U}$ denotes the collection of temporal transformations. This module is used to create temporal features that incorporate various time intervals. For example, $\mu_{1}(x^{s})$ and $\mu_{2}(x^{s})$ would reflect temporal patterns within a day and a week. Various methods, such as the moving average \citep{etsformer, autoformer}, Fourier transform \citep{zhou2022fedformer}, moving average convergence/divergence (MACD) features for momentum \citep{lim2019enhancing} and mixtures of experts \citep{zhou2022fedformer}, can be utilized to create these temporal features. In this study, we use moving average, which is the most common choice. 

More specifically, for each temporal transformation $\mu_i$, we calculate the moving average of the price feature $x^s$ over a window of size $w_i$:
\begin{equation}
\mu_i(x^s) = \frac{1}{w_i} \sum_{j=0}^{w_i-1} x^s_{t-j}, \quad i=1,2,\ldots,k.
\end{equation}
where $w_i$ is the window size for the $i$-th transformation, and $x^s_{t-j}$ denotes the price feature at time $t-j$ in domain $s$. In our implementation, we use five different window sizes: $w_1=5$, $w_2=10$, $w_3=15$, $w_4=20$, and $w_5=25$. These window sizes correspond to weekly, bi-weekly, tri-weekly, four-weekly, and monthly moving averages, respectively.

{\bfseries Combined embedding with static metadata.} In our framework, static metadata $c^{s}$, which can include a variety of data such as firm description, 3-statement financial information, and more, is handled in the static embedding layer. However, for the purpose of this study, we have only included sector information in $c^{s}$. As a result, an embedding $\text{Embed}(c^{s}) \in \mathbb{R}^{d_{mk}}$ is obtained. 
Next, we create a combined embedding that incorporates both the temporal feature variant $\mu(x^{s})$ and the embedded static metadata $\text{Embed}(c^{s})$. The resulting combined embedding is denoted as follows:
\begin{equation}
\mathbf{H}^{s} = \mu(x^{s}) + \text{Embed}(c^{s}) \in \mathbb{R}^{d_{mk}}.
\end{equation}

{\bfseries Feature Tokenizer module.} We draw inspiration from the tokenizer approach \cite{gorishniy2021revisiting}, which transforms input features into token embeddings to obtain more meaningful representations. This method \cite{gorishniy2021revisiting} do not reflect the temporal aspect, but we utilize temporal feature variants in the Feature Tokenizer module to capture the time-varying patterns in the stock price data. The feature-wise token embeddings $\textbf{TKE}^{s}_{j}$ for a given feature index $j$ are computed as follows: 
\begin{equation}
    \textbf{TKE}^{s}_{j} = b^{s}_{j} + \mathbf{H}^{s}_{j}W_{j}^{s} 
\end{equation}
where $b^{s}_{j} \in \mathbb{R}^{d}$ is the $j$-th feature bias term and $W_{j}^{s} \in \mathbb{R}^{d}$ is the weight vector for the $j$-th feature. Consequently, the token embeddings $\textbf{TKE}^{s} \in \mathbb{R}^{d_{c} \times d}$  can be obtained by stacking all of the feature embeddings and adding a special [\texttt{ST}] token, which is known to possess the essence of information after training. This is represented as:
\begin{equation}
\textbf{TKE}^{s} = \texttt{\text{STACK}}([\texttt{\text{ST}}], \textbf{TKE}^{s}_{1}, ..., \textbf{TKE}^{s}_{d_{mk}})
\end{equation}
where $\mathbb{R}^{d_{c} \times d}$ = $\mathbb{R}^{(d_{mk}+1) \times d}$ denotes the dimension of the combined token embeddings $\textbf{TKE}^{s}$. Therefore, the feature tokenizer plays a crucial role in learning meaningful temporal representations by transforming combined embedding.

{\bfseries Dimension corruption.} \label{dimension} In self-supervised learning, the main objective is to learn an embedding space where positive pairs (or views) remain close to each other, while negative pairs (or views) are far apart. When generating views, mixup \citep{zhang2017mixup} or cutmix \citep{yun2019cutmix} methods are most commonly used. These methods are suitable for invariant augmentation of static data (e.g., images), however, these are not suitable for time-series data (e.g., stocks). We generate views for temporal variants on the same instance, unlike conventional SSL methods that use invariant augmentation by using different instances together. For time-series data, mixing different sequences would break the entire temporal structure. Therefore, we propose a dimension corruption method for the augmentation of temporal data.

First, we create positive and negative \textit{views}, $\mathbf{H}_{pos}^{s}$ and $\mathbf{H}_{neg}^{s}$, by randomly shuffling the dimensions within the token embeddings $\textbf{TKE}^{s}$. Here, we define two permutation matrices, $\mathbf{P}_{pos}^{s}$ and $\mathbf{P}_{neg}^{s}$, both of size $d \times d$. \footnote{A permutation matrix is a square 0-1 matrix that has exactly one entry of 1 in each row and each column and 0s elsewhere.}

\begin{equation}
\begin{aligned}
\mathbf{H}_{pos}^{s} &= \lambda \textbf{TKE}^{s} + (1 - \lambda) \textbf{TKE}^{s} \mathbf{P}_{pos}^{s}\\
\mathbf{H}_{neg}^{s} &= (1 - \lambda) \textbf{TKE}^{s} + \lambda \textbf{TKE}^{s} \mathbf{P}_{neg}^{s}
\end{aligned}
\label{eq8}
\end{equation}

In this case, the formula \eqref{eq8} generates positive and negative views for self-supervised learning. 

\begin{figure}
\centering
\includegraphics[width=0.6\textwidth]{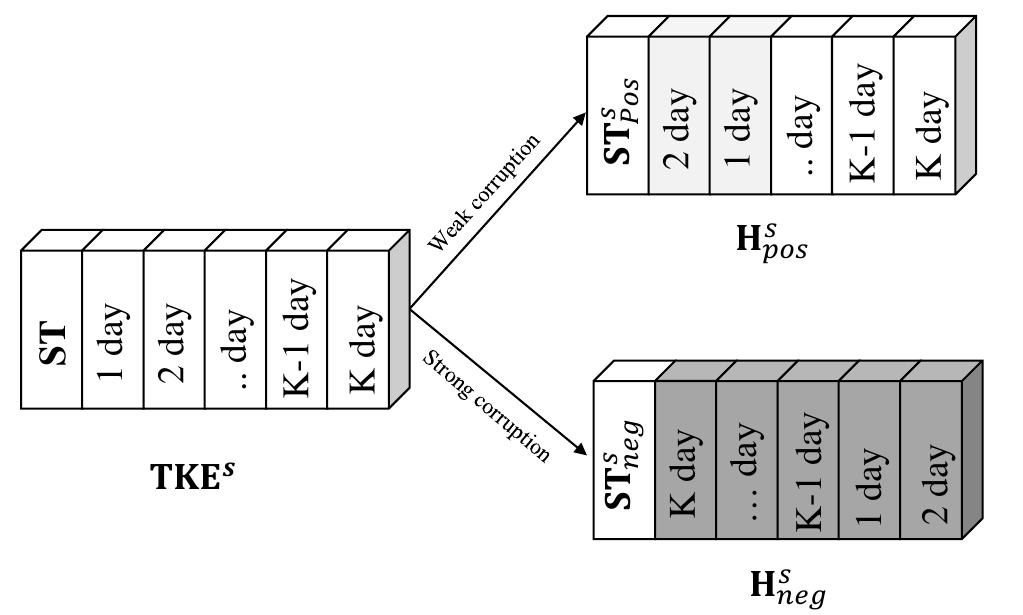}
\caption{Dimension corruption method for generating positive and negative views from token embeddings ($\mathbf{TKE}^{s}$). The positive view $\mathbf{H}_{pos}^{s}$ is created by applying a small perturbation in dimension order to the original TKE, while the negative view $\mathbf{H}_{neg}^{s}$ is generated with a larger perturbation in dimension order, preserving more of the original temporal structure in the positive view compared to the negative view.}
\label{chap3_tke}
\end{figure}

The degree of this perturbation in both views is determined by the mixing parameter $\lambda$. With $\lambda > 0.5$, the positive view $\mathbf{H}_{pos}^{s}$ has minor perturbations, maintaining much of the original token embedding. The negative view $\mathbf{H}_{neg}^{s}$ is more altered, with greater dimension shuffling, deviating more from the original. More specifically, as shown in \cref{chap3_tke}, the positive view corrupted less variation in temporal features, while the negative view corrupted more variation in temporal features. These views allow the model to consider the chronological order when obtaining embeddings. We set $\lambda = 0.7$ as the default value in this study. An ablation study on the impact of different values of $\lambda$ can be found in Appendix A.

{\bfseries Representation module.} The representation module $f_{\theta_s}$ aims to characterize the shift between different domains by refining the parameters $\theta_s$ through the process described in Section 3.2. In order to effectively reflect temporal patterns of corrupted token embeddings ($\mathbf{H}_{pos}^{s}$ and $\mathbf{H}_{neg}^{s}$), we use the self-attention mechanism \citep{vaswani2017attention}. The self-attention mechanism aggregates corrupted token embeddings with normalized importance as follows:
\begin{equation}
    \text{Attention}(Q,K,V) = \text{Softmax}\left(\frac{QK^T}{\sqrt{d}}\right)V.
\end{equation}
Here, $Q$ = $\textbf{H}_{*}^{s} W_{Q} \in \mathbb{R}^{d \times d_{k}}$, $K = \textbf{H}_{*}^{s} W_{K} \in \mathbb{R}^{d \times d_{k}}$ and $V = \textbf{H}_{*}^{s} W_{V} \in \mathbb{R}^{d \times d_{v}}$ represent queries, keys, and values, respectively. Note that $\textbf{H}_{*}^{s}$ represents token embeddings (either positive or negative), while $W_Q$, $W_K$, and $W_V$ are learnable matrices that share weights between positive and negative token embeddings. The output, which has a dimension of $d_v$, is then transformed back into an embedding of dimension $d$ through a fully connected layer. Finally, the outputs $\texttt{\text{ST}}_{neg}^{s}$ and $\texttt{\text{ST}}_{pos}^{s}$ are obtained.

{\bfseries Triplet loss.}  
For \texttt{SimStock}, we train it to minimize a triplet loss \citep{balntas2016learning}, which is a popular choice in SSL. The key idea behind triplet loss is the use of triplets, each of which consists of an anchor, and positive and negative views. Here, the anchor is the embeddings for the unperturbed combined embedding.

For the triplet $(\texttt{ST}_{pos}^{s}, \texttt{ST}_{neg}^{s}, \textbf{H}^{s})$, where $\texttt{ST}_{pos}^{s}$ is the positive view, $\texttt{ST}_{neg}^{s}$ is the negative view, and $\textbf{H}^{s}$ is the  combined embedding (anchor), the triplet loss is defined as follows:
\begin{equation}
L_{\text{triplet}} = \max(0, \text{sim}(\textbf{H}^{s}, \texttt{ST}_{pos}^{s}) - \text{sim}(\textbf{H}^{s}, \texttt{ST}_{neg}^{s}) + \alpha)
\end{equation}
In the above equation, $\text{sim}(\cdot, \cdot)$ denotes a similarity measure (e.g., cosine similarity or Euclidean distance), and $\alpha > 0$ is a margin that is introduced to separate positive pairs from negative pairs. The intuition behind this loss function is that we want to ensure that the anchor point gets closer to the positive sample than to the negative sample by at least the margin $\alpha$.

{\bfseries Inference phase.} In our framework, the inference phase is particularly important. Unlike most existing contrastive representation learning studies \citep{chen2020simple, grill2020bootstrap}, our model is specifically designed to be robust with respect to temporal distribution shifts. The inference phase consists of passing the target domain $D_{s+1}$ through the embedding module to obtain the combined embedding $\textbf{H}^{s+1}$ and further processed by the feature tokenizer module to obtain the token embeddings $\textbf{TKE}^{s+1}$. The stock representation is then obtained by feeding $\textbf{TKE}^{s+1}$ into the representation model $f_{\theta_{s+1}}$, which is updated with the optimal parameters $\theta_{s+1}$ generated by the TDG method described in Section \cref{chap3_TDG}

%% file: content/chap4_exp.tex
\section{Experiment}
Now we present experiment results to thoroughly demonstrate the performance of \texttt{SimStock} on real-world benchmark datasets. Our code and configurations will be publicly available on GitHub. 

\subsection{Implementation details} \label{section_4_1}
We present the details of tasks, datasets, baseline models, hyperparameter selection, evaluation metrics, and the experiment setting.

\subsubsection{Tasks} We conduct four main experiments to assess the effectiveness of \texttt{SimStock} in various financial applications:
\begin{itemize}
\item \textbf{Finding similar stocks:} We evaluate how well \texttt{SimStock} can identify stocks that exhibit similar price movements, compared to baseline methods. Given a query stock, we consider both same exchange (finding similar stocks within the same exchange) and different exchanges (finding similar stocks in a different exchange from the query stock) scenarios;
\item \textbf{Pairs trading:} We evaluate the effectiveness of pairs trading based on the similar stocks identified by \texttt{SimStock}. We form pairs of stocks that are found to be similar, and then execute a pairs trading strategy;
\item \textbf{Index tracking:} Given an ETF as the query, we find a basket of similar stocks using \texttt{SimStock} and assess the performance of using these stocks for index tracking. The goal is to see if the selected stocks can closely mimic the returns of the target ETF;
\item \textbf{Portfolio optimization:} We investigate whether \texttt{SimStock} embeddings can enhance portfolio optimization. Specifically, we replace the correlation matrix of stock returns using the \texttt{SimStock} embedding as a similarity measure for mean-variance portfolio optimization. We compare the performance of the resulting portfolios with those obtained using conventional methods for estimating.
\end{itemize}

\subsubsection{Datasets} We collect daily stock price features (Open, High, Low, Close, and Volume) and sector information from Yahoo Finance for stocks listed on five major exchanges: \textbf{NYSE} (New York Stock Exchange), \textbf{NASDAQ} (National Association of Securities Dealers Automated Quotations), \textbf{SSE} (Shanghai Stock Exchange), \textbf{SZSE} (Shenzhen Stock Exchange), and \textbf{TSE} (Tokyo Stock Exchange). In our study, we combine NYSE and NASDAQ stocks (4,231 in total) and refer to them as the \textbf{US} exchange, while treating SSE (1,407 stocks), SZSE (1,696 stocks), and TSE (3,882 stocks) separately. The data spans three periods: a training period from January 1, 2018, to December 31, 2021; a reference period from January 1, 2022, to December 31, 2022; and a test period from January 1, 2023, to December 31, 2023. This timeline allows us to verify whether stocks identified as similar in the reference period maintain their similarity in the subsequent one-year test period.

Our dataset is less susceptible to survivor bias because the source domain is organized as a sequence of domains \{$\mathcal{D}_1, \mathcal{D}_2, ..., \mathcal{D}_T$\} over time. Each source domain $\mathcal{D}_s$ consists of a distinct set of stocks that existed at time step $s$. This allows the model to reflect the realistic entry and exit of stocks at each time step. For example, $\mathcal{D}_{s-1}$ will contain stocks that existed at time $s-1$, but may not contain all stocks from $\mathcal{D}_s$ if some were delisted. Similarly, $\mathcal{D}_s$ can include newly added stocks that were not present in $\mathcal{D}_{s-1}$. Furthermore, the price data used for each stock is adjusted to account for corporate actions such as splits and dividends.

We generate normalized input features describing the trend of a stock on day t. The variables $z_\text{Open}$, $z_\text{High}$ and $z_\text{Low}$ represent the comparison values of the opening, highest, and lowest prices, respectively, relative to the closing price of the same day. Also, $z_\text{Close}$ and $z_\text{Volume}$ represent the comparative values of the closing prices and the volume values compared with day $t-1$, respectively. Refer to \cref{chap4_prepro} for the formulas used to calculate each feature. In addition, we calculated stock price features for 5, 10, 15, 20, 25, and 30-day intervals for the temporal feature variant as discussed in \cref{TRL}.

\begin{table}[!h]
\centering
    \input{Tables/chap4_prepro.tex}
    \caption{Normalized temporal price features.}
    \label{chap4_prepro}
\end{table}


\subsubsection {Baseline models}

For the task of finding similar stocks, index tracking and pairs trading task, we compare \texttt{SimStock} with the following baselines:

\begin{itemize}
\item \textbf{Corr1}: Calculates the correlation of stock returns using the past one-year returns (i.e., from January 1, 2022 to December 31, 2022).
\item \textbf{Corr2}: Calculates the correlation of stock returns using returns from the beginning of the test period (i.e., from January 1, 2018 to December 31, 2022).
\item \textbf{Peer}: Uses the list of similar stocks provided by Financial Modeling Prep.\footnote{https://site.financialmodelingprep.com/}
\item \textbf{TS2VEC} \citep{yue2022ts2vec}: A state-of-the-art method based on self-supervised learning for finding similar time-series data.
\end{itemize}

For the portfolio optimization task, we compare the performance of portfolios constructed using the \texttt{SimStock} similarity matrix with those constructed using the following conventional correlation matrices:

\begin{itemize}
\item \textbf{Historical covariance matrix(HC)} \citep{jobson1980estimation}: The sample correlation matrix of the most recent past stock returns.
\item \textbf{Shrinkage method(SM)} \citep{ledoit2004honeyorigin}: A shrinkage estimator of the covariance matrix proposed by \citep{ledoit2004honeyorigin}.
\item \textbf{Gerber statistic(GS)} \citep{gerber2021gerber}: A robust correlation measure that counts the proportion of simultaneous co-movements between assets when their amplitudes exceed data-dependent thresholds.
\end{itemize}

\subsection{Can \texttt{SimStock} find similar stocks?}  \label{section_4_2}

In this section, we explore two distinct scenarios for identifying similar stocks using our proposed model: the \textit{same exchange scenario} and the \textit{different exchanges scenario}. The same exchange scenario focuses on finding similar stocks within the same exchange given a query stock. This approach allows for the identification of stocks with comparable characteristics and behaviors within a specific market. On the other hand, the different exchanges scenario involves finding similar stocks within another exchange given a query stock. This scenario leverages the concept of transfer learning, where the trained weights of a model from one exchange are applied to stock data from a different exchange. Transfer learning is a machine learning technique that leverages the knowledge gained from a model trained on a source task to improve the performance of a model on a related target task \citep{pan2009survey}. In our study, we apply transfer learning by utilizing a model trained on one stock exchange (source domain) to identify similar stocks in another exchange (target domain). For example, models trained on the US exchange can be used to find similar stocks in the SSE, SZSE, or TSE exchanges. Notice that the query is not restricted to individual stocks. It can be either sector indices or ETFs.

\subsubsection{Evaluation metrics} To evaluate the performance of models in finding similar stocks compared to other models, we employ two widely used metrics: Correlation and Dynamic Time Warping (DTW). 
DTW is a measure for measuring the similarity between two temporal sequences, allowing for non-linear alignments and capturing the similarities in the overall patterns of stock price movements \citep{berndt1994using}. Consider two time-series sequences $X = \{x_{1}, ..., x_{m}\}$ and $Y = \{y_{1}, ..., y_{n}\}$. The DTW between $X$ and $Y$ is defined as: 
\begin{equation}
 \text{DTW}(X, Y) = \sqrt{\sum_{(i, j) \in \pi}\left\|x_i-y_j\right\|^2}.
\end{equation}
Here, an alignment path $\pi$ of length $K$ is a sequence of $K$ index pairs $(i, j)_{K}$, where $\max (m, n) \leq K \leq m+n-1$. Also, $||.||$ is the Euclidean distance. DTW uses global path constraints while comparing two time-series sequences $X$ and $Y$. That is, the pairs $i$ and $j$ are constrained so that $|i - j| \leq r$, where $r$ is a predefined radius, in the case of the Sakoe–Chiba band. 

To assess the effectiveness of our approach in capturing similarities and predicting future stock behaviors, we evaluate the model's performance using an out-of-sample period. The dataset used for inference is the year immediately preceding the out-of-sample period. By finding similar stocks in the previous year and comparing their performance to the target stock in the out-of-sample period, we measure the model's performance using Correlation and Dynamic Time Warping (DTW) measures. We consider the top 9, 7, 5, 3, and 1 similar stocks (namely, TOP@9, TOP@7, TOP@5, TOP@3, and TOP@1) for evaluation.

\subsubsection{Quantitative evaluation: \textit{same exchange scenario}} The diagonal plots in \cref{chap4_find_simstock_corr} and \cref{chap4_find_simstock_dtw} illustrate the performance (Correlation and DTW) of different models in same exchange scenario. The diagonal plots show that our proposed \texttt{SimStock} consistently outperforms the baselines (Corr1, Corr2, Peer, and TS2VEC) in finding similar stocks within the same exchange, across all four stock markets (US, SSE, SZSE, and TSE).  For example, in the US to US scenario, \textit{SimStock} achieves a correlation of around 0.8 for the top similar stock (TOP@1), substantially higher than the 0.6 correlation of the best performing TS2VEC. 

Similarly, for the SSE to SSE and TSE to TSE scenarios, \texttt{SimStock} consistently shows superior performance compared to the baselines, with correlations close to 0.8 for TOP@1, but \texttt{SimStock} and TS2VEC gave equivalent results in SZSE to SZSE. Note that the performances of all the baseline models were not much different. It is interesting that the peer stocks picked by investment platforms (Peer) were not quite close to the query stocks in terms of DTW. The DTW results also indicate that \texttt{SimStock} identifies more similar stocks, as evidenced by the lower DTW distances compared to the baselines. These findings demonstrate that our \texttt{SimStock} method effectively captures the underlying patterns and dynamics of stock price movements, enabling it to find highly similar stocks within the same exchange.

\subsubsection{Quantitative evaluation: \textit{different exchanges scenario}} 
\begin{figure*}[!htbp]
\centering
\includegraphics[width=\textwidth]{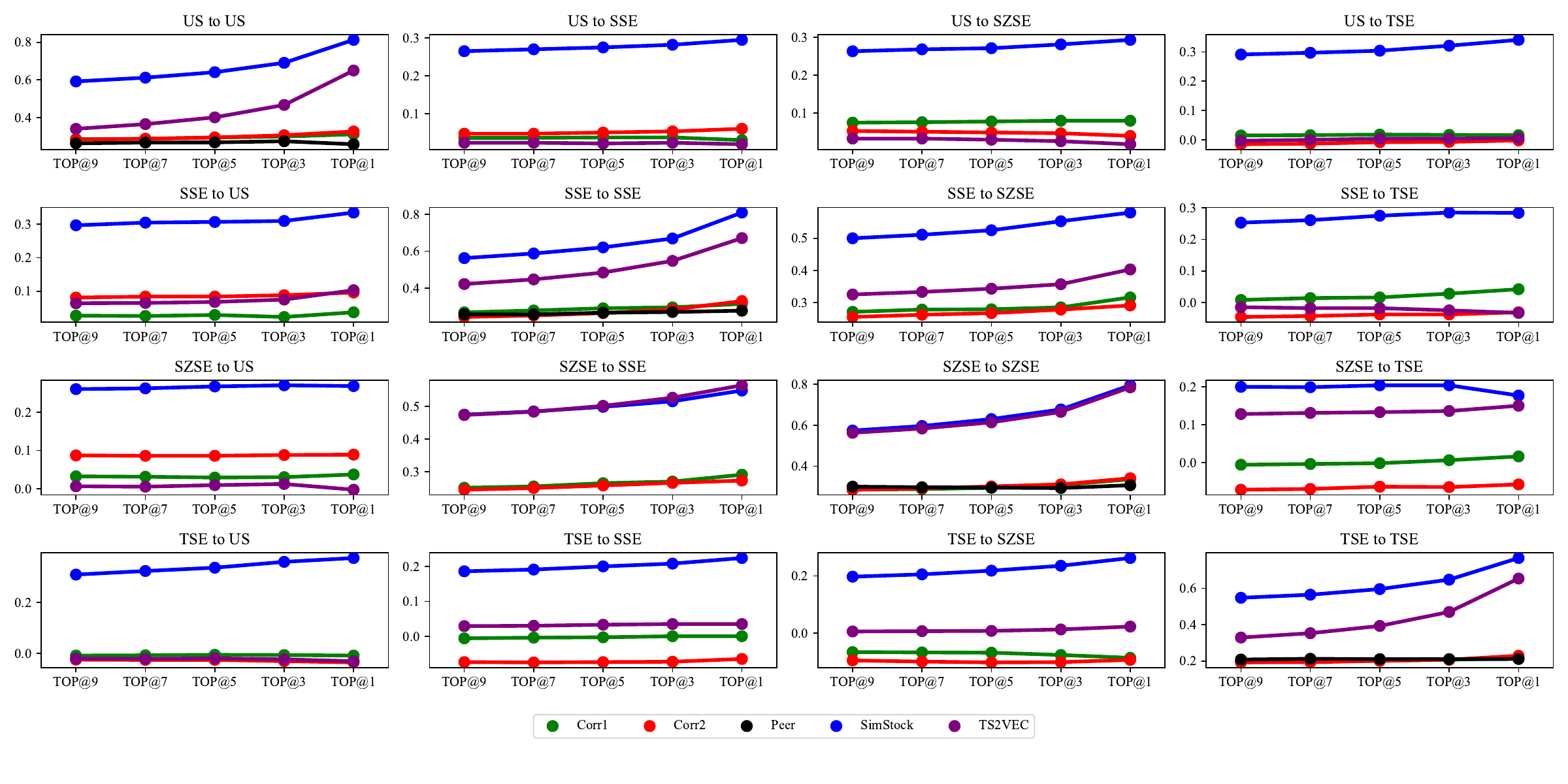}
\caption{Performance of models in same exchange (diagonal) and different exchanges (off-diagonal) scenarios for finding similar stocks. The performance is evaluated using TOP@$k$ Correlation metrics, where $k$ = 9, 7, 5, 3, and 1. Each data point represents the average correlation between the target stock and the top $k$ similar stocks identified by the respective model.}
\label{chap4_find_simstock_corr}
\end{figure*}

\begin{figure*}[!htbp]
\centering
\includegraphics[width=\textwidth]{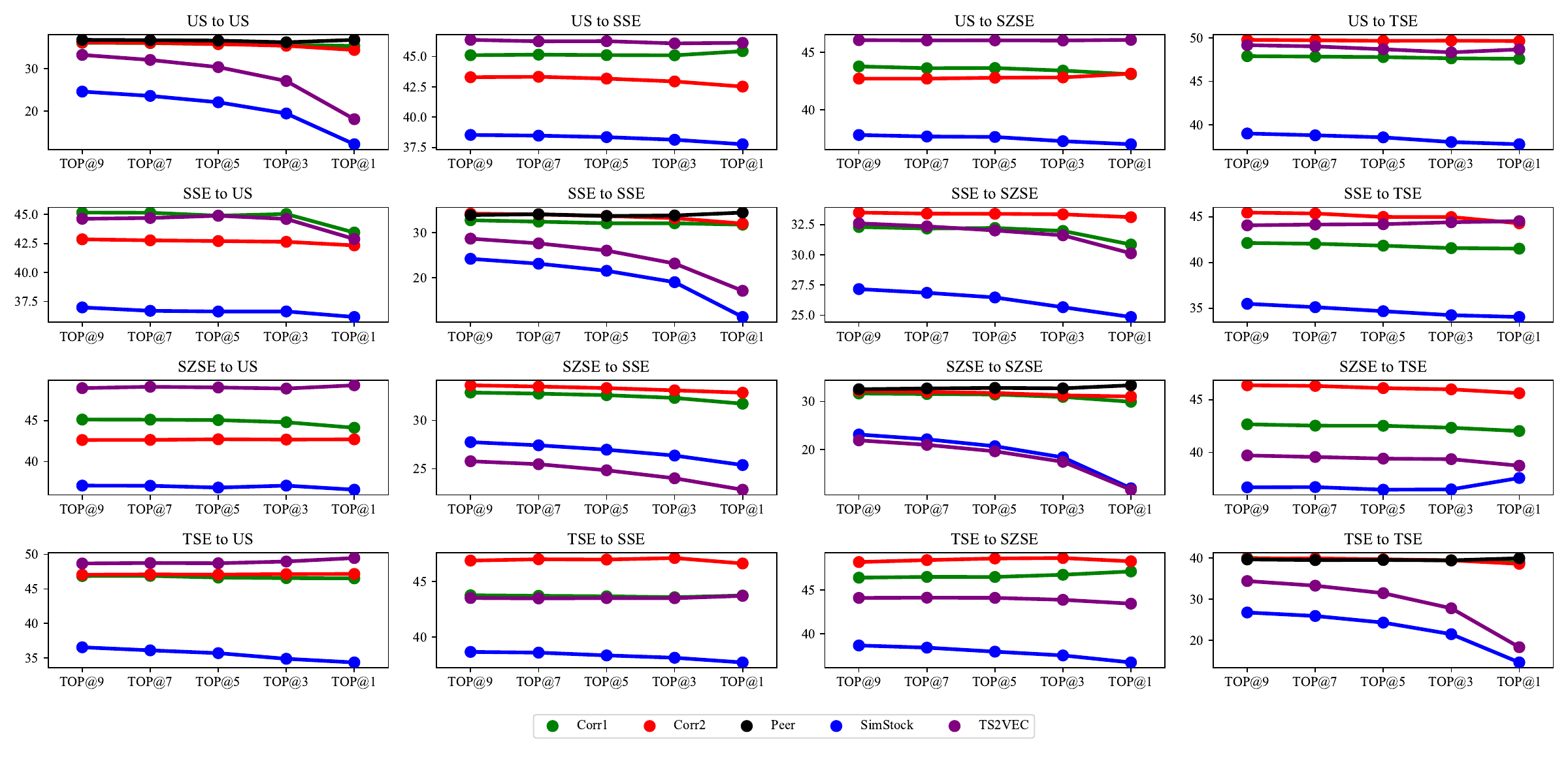}
\caption{Performance of models in same exchange (diagonal) and different exchanges (off-diagonal) scenarios for finding similar stocks. The performance is evaluated using TOP@$k$ DTW metrics, where $k$ = 9, 7, 5, 3, and 1. Each data point represents the average correlation between the target stock and the top $k$ similar stocks identified by the respective model.}
\label{chap4_find_simstock_dtw}
\end{figure*}

The off-diagonal plots in \cref{chap4_find_simstock_corr} and \cref{chap4_find_simstock_dtw} represents the outcomes of identifying for similar stocks in exchanges different from the exchange of the query stock. Note that Peer is not available for this scenario, because most trading platforms do not provide information on similar stocks in other exchanges. The off-diagonal plots demonstrate that \texttt{SimStock} outperforms the baselines in all different exchanges scenarios, indicating its effectiveness in leveraging temporal domain generalization to identify similar stocks across different markets. For instance, in the US to SSE, US to SZSE and US to TSE scenarios, \texttt{SimStock} achieves correlations around 0.3 for TOP@1, surpassing the performance of Corr1, Corr2, and TS2VEC. Similarly, for the SSE to US, SSE to SZSE and SSE to TSE scenarios, SimStock exhibits higher correlations compared to the baselines, showcasing its ability to capture similarities between stocks in these closely related exchanges.

However, it is important to note that the overall performance of all models in the different exchanges scenarios is lower compared to the same exchange scenario. This can be attributed to the inherent differences in market dynamics, regulations, and economic factors across different exchanges. Despite these challenges, \texttt{SimStock} consistently outperforms the baselines in all cases, highlighting its robustness and adaptability in finding similar stocks across diverse market conditions.

The DTW results in the off-diagonal plots further support the superiority of \texttt{SimStock} in identifying similar stocks across exchanges. The DTW approach is a more suitable alternative to the correlation metric when considering different timestamps. \texttt{SimStock} consistently achieves lower DTW distances compared to the baselines, indicating its ability to capture the overall patterns and similarities in stock price movements, even when applied to different markets. These findings underline the potential of \texttt{SimStock} in enabling investors and analysts to identify similar stocks across different exchanges.

\clearpage

\clearpage

\subsubsection{Qualitative evaluation} \label{section_4_2_3} The similarity between stocks should not be measured only on time-series distances. Similar stocks may be in different industries, but they should have similar fundamental information. Unfortunately, however, it is almost impossible to measure such similarity in a quantitative way. Therefore, we provide a qualitative analysis of those results. 

To assess the effectiveness of SimStock in identifying fundamentally similar stocks, we conduct a nuanced analysis by examining the actual stocks recommended by each method for a diverse set of query stocks from various sectors (\cref{chap4_pair_stocks_name}). We compare the top@3 similar stocks identified by \texttt{SimStock} with those of the baseline methods, including TS2VEC, Corr1, Corr2, and Peer, to evaluate whether \texttt{SimStock} captures relevant business characteristics, financial attributes, and industry-specific similarities better than the baselines.

When we begin by examining the technology sector, \texttt{SimStock} identifies highly relevant similar stocks for query stocks such as Apple (AAPL), Microsoft (MSFT), and Meta (META). For AAPL, \texttt{SimStock}'s top@$k$ similar stocks include Microsoft (MSFT), Tyler Technologies (TYL), and Intuit (INTU), all well-established companies in the software and technology domain. In contrast, the baseline methods identify a mix of technology and non-technology companies, such as Amazon (AMZN) for TS2VEC and Corr2, and Thermo Fisher Scientific (TMO) for Corr1. Similarly, for MSFT, \texttt{SimStock} identifies Cadence Design Systems (CDNS), Manhattan Associates (MANH), and Tyler Technologies (TYL) as similar stocks, while the baselines include manufacturing companies like Danaher (DHR) for Corr1. 

Next, we analyze the healthcare sector, where \texttt{SimStock} identifies relevant similar stocks for query stocks such as Pfizer (PFE) and Centene (CNC). For PFE, \texttt{SimStock}'s top similar stocks include BioNTech (BNTX), Moderna (MRNA), and Johnson \& Johnson (JNJ), all major players in the pharmaceutical and biotechnology industries. The baseline methods, however, identify less relevant stocks, such as NantKwest (NKNG) for TS2VEC and Metrocity Bankshares (MCBS) for Corr1. For CNC, \texttt{SimStock} identifies other health insurance and financial services companies like Bank of Montreal (BMO) and Morgan Stanley (MS), while the baselines include less related stocks, such as Humana (HUM) for TS2VEC and Stericycle (SRCL) for Corr2.

In the energy sector, \texttt{SimStock} identifies industry-specific similar stocks for the query stock Exxon-Mobil (XOM), such as Marathon Oil (MRO), Cenovus Energy (CVE), and Hess Corporation (HES). The baseline methods also identify some relevant energy stocks, such as MRO for TS2VEC and Murphy Oil Corporation (MUR) for Corr1 and Marathon Petroleum (MPC) for Corr2. That is, all methods successfully identified stocks that are similar to XOM.

Finally, we examine the financial sector, where \texttt{SimStock} identifies relevant similar stocks for query stocks such as Wells Fargo (WFC) and Visa (V). For WFC, \texttt{SimStock}'s top similar stocks include Bank of America (BAC), Fifth Third Bancorp (FITB), and F.N.B. Corporation (FNB), all well-known banking institutions. Similar, the baseline methods also identify relevant banking stocks, such as BAC for Corr2, JPMorgan Chase (JPM) for TS2VEC and Berkshire Hills Bancorp (BHLH) for Corr1. For V, SimStock identifies Mastercard (MA), Stifel Financial (SF), and Intercontinental Hotels Group (IHG) as similar stocks, while the baselines include less related stocks, such as Planet Fitness (PLNT) for Corr2.

\begin{table}[!h]
\centering
    \input{Tables/chap4_pair_stocks_name}
    \caption{Top@3 similar stocks identified by \texttt{SimStock} and baseline methods (TS2VEC, Corr1, Corr2, and Peer) for a diverse set of query stocks from the technology, healthcare, energy, and financial sectors. }
    \label{chap4_pair_stocks_name}
\end{table}

The above provides further reassurance that \texttt{SimStock} robustly identifies relevant companies as similar to each other in a way intuitive to most human market participants.

\subsection{Application to pairs trading} In this section, we investigate the practical application of the similar stocks identified by \texttt{SimStock} in the context of pairs trading. Pairs trading is a market-neutral trading strategy that exploits the relative mispricing between two highly correlated securities \citep{gatev2006pairs}. The strategy involves simultaneously buying the relatively underpriced security and selling the relatively overpriced security, with the expectation that the prices will converge to their long-term equilibrium. We assess the profitability of pairs trading using the top@$k$ similar stocks identified by \texttt{SimStock} and compare its performance with benchmark models, namely TS2VEC, Corr1, and Corr2. We added Cointegration method (Coint) \citep{nelson1982trends, enders2004cointegration} to the benchmark model, which selected 3 stocks from the US exchange that exhibited the strongest cointegration relationships, as determined by their $p$-values in descending order.

\subsubsection{Pairs trading procedure} We employ the price ratio approach for pairs trading, which involves constructing a spread between two securities by computing the ratio of their prices. The spread is then normalized by its historical mean and standard deviation to obtain the Z-score, which measures the deviation of the current spread from its historical average in terms of standard deviations. The Z-score is calculated as follows:

\begin{equation}
\begin{aligned}
PR_t &= P_{q,t} / P_{s,t} \\
Z_t &=  (\mu_{1}^{PR_{t}} - \mu_{2}^{PR_{t}}) / \sigma_{2}^{PR_{t}}
\end{aligned}
\end{equation}
where $P_{q,t}$ and $P_{s,t}$ are the prices of the query stock and similar stock in the pair at time $t$, and $\mu_1$, $\mu_2$, and $\sigma_2$ are the rolling mean and standard deviation of the price ratio up to time $t$, calculated using lookback periods of $L_1$ and $L_2$ trading days, respectively.

Trading signals are generated based on the Z-score, with the opening of a long-short position when the Z-score exceeds a pre-specified threshold of $\pm 1.25 \text{ SD}$, indicating a significant deviation from the long-term equilibrium. Specifically, when $Z_t > 1.25 \text{ SD}$, we sell the overpriced security and buy the underpriced security. Conversely, when $Z_t < -1.25 \text{ SD}$, we buy the overpriced security and sell the underpriced security. The position is closed when the Z-score reverts back within the range $[-0.5 \text{ SD}, 0.5 \text{ SD}]$, suggesting price convergence. To limit potential losses, a stop-loss \citep{hwang2023stop} mechanism is implemented, whereby positions are closed if the portfolio value drops below a predetermined threshold $L_{\text{stop}}$ of 500 USD.

To determine optimal values for the hyperparameters $L_{1}$ and $L_{2}$, we perform a grid search over the training period from 2022-01-01 to 2022-12-31. Different $L_1$ and $L_2$ values are determined for each query stock paired with each of its top 3 similar stocks identified by \texttt{SimStock} and the benchmark models (TS2VEC, Corr1, Corr2), resulting in 3 pairs per query stock per model. Pairs trading is applied independently to each pair, with overall performance for each query stock and model calculated as the average wealth across the 3 pairs, starting with an initial trading capital of 10,000 USD. For all trades executed in this study, a fixed proportional transaction cost of 10 basis points (0.1\%) is applied.

\subsubsection{Evaluation of pairs trading performance} \cref{chap4_pair_stocks_wealth_mdd} presents the average terminal wealth and maximum drawdown (MDD) achieved by applying pairs trading to the top@3 similar stocks identified by \texttt{SimStock} and the benchmark models (TS2VEC, Corr1, Corr2, Coint) for each query stock. The results of pair trading between query stock and TOP@1 stock can be found in Appendix B.

\texttt{SimStock} achieves the highest average terminal wealth for 6 out of the 12 query stocks (AAP, CMG, MSFT, WFC, BA and CVS). Coint performs best for 2 query stocks (XOM and PFE), while Corr1, Corr2 and TS2VEC each perform best for 1 query stock (MA, AMZN and V respectively). For the remaining query stocks, \texttt{SimStock} ranks second in terms of wealth for 4 cases (V, XOM, META and MA). 

Regarding maximum drawdown, \texttt{SimStock} exhibits the lowest MDD for 7 out of the 12 query stocks (AAPL, CMG, MSFT, WFC, PFE, META and MA), indicating better risk management. TS2VEC, Corr1, Corr2 and Coint achieve the lowest MDD for 1 query stock each (AMZN, V, CVS and XOM respectively). For the remaining 5 query stocks (V, XOM, AMZN, BA and CVS), \texttt{SimStock} has the second-lowest MDD. The ability to generate higher returns while maintaining lower drawdowns suggests the robustness of the pairs identified by \texttt{SimStock}.

It is important to note that TS2VEC, Corr2, and Coint failed to generate buy/sell signals for certain stocks. Specifically, TS2VEC and Coint were unable to generate signals for AAPL and WFC, while Corr2 failed to do so for WFC. SimStock, on the other hand, generated a buy/sell signal for one out of the three stocks, specifically META. 

The results demonstrate the superior performance of \texttt{SimStock} in identifying profitable pairs for trading compared to the benchmark models. \texttt{SimStock}'s ability to achieve the highest average terminal wealth for the majority of query stocks while maintaining the lowest maximum drawdown for most cases highlights its effectiveness in capturing meaningful similarities between stocks that translate into successful pairs trading strategies.

\begin{table}[!h]
\centering
    \input{Tables/chap_4_pair_trading_top3}

    \begin{flushleft}
        \textit{Note:  NaN** values in both the terminal wealth and MDD indicate that the method failed to generate buy/sell signals for all three stocks in the pair. NaN* values only in the standard deviation indicate that the method failed to generate buy/sell signals for two out of the three stocks in the pair. For all other values, all method generated buy/sell signals for all three stocks in the pair.}
    \end{flushleft}
    \caption{Average terminal wealth (first row) and maximum drawdown (MDD) (second row) achieved by applying pairs trading to the top@3 similar stocks identified by SimStock, TS2VEC, Corr1, Corr2, coint for each query stock. he best performing method is highlighted in bold, while the second best is in underline. }
    \label{chap4_pair_stocks_wealth_mdd}
\end{table}

\subsection{Application to index tracking of thematic ETFs} Thematic ETFs have gained popularity among retail investors who wish to invest in specific themes or trends, such as innovative technologies. These ETFs are composed of assets that share a common theme, providing a contrast to traditional sector-based classifications, which can be slow and ambiguous in reflecting technological advancements and innovations. As thematic ETFs aim to identify innovative tech companies, many investors are interested in tracking these ETFs.

In this section, we compare the performance of \texttt{SimStock} and other baselines in identifying stocks to track four popular thematic ETFs. Our goal is to find similar stocks from the US, SSE, SZSE, and TSE exchanges that can effectively track the performance of these thematic ETFs. In other words, we use the thematic ETFs as queries to find similar stocks in each model.

For this analysis, we selected the four popular thematic ETFs: ARK Innovation ETF (ARKK), First Trust Cloud Computing ETF (SKYY), Global X Robotics \& AI ETF (BOTZ), and Global X Lithium \& Battery Tech ETF (LIT). The in-sample and out-of-sample periods are the same as those used in \cref{section_4_1}.

\subsubsection{Evaluation metrics} To evaluate the performance of \texttt{SimStock} and the baselines in tracking thematic ETFs, we first find the top $k$ similar stocks based on the thematic ETF queries. We then create equal-weighted portfolios using these similar stocks to track the thematic ETFs. We use two metrics to assess the tracking performance: Tracking Error (TE) and Tracking Error Volatility (TEV). \citep{focardi2004methodology}.

The end-of-the-period tracking error (TE) is calculated as follows:

\begin{equation}
\text{TE} = \sqrt{ \frac{1}{n} \sum_{j=1}^{n} (R^{I}_{j} - R^{P}_{j})^2 }
\end{equation}
where $R^{I}_{j}$ and $R^{P}_{j}$ are the cumulative returns of the query ETF and the tracking portfolio at period $j$, respectively, and $n$ is the number of periods.

Tracking Error Volatility (TEV) measures the volatility of the difference in returns between the tracking portfolio and the target ETF. It is calculated as the standard deviation of the difference in returns:

\begin{equation}
\text{TEV} = \sqrt{ \frac{1}{n} \sum_{j=1}^{n} (R^{I}_{j} - R^{P}_{j} - \mathbb{E}[{R^{I} - R^{P}}])^2 }
\end{equation}
where $\mathbb{E}[{R^{I} - R^{P}}]$ is the mean difference in returns between the target ETF and the tracking portfolio over the $n$ periods. For each thematic ETF, we find the top@$k$ similar stocks on each exchange (US, SSE, SZSE, TSE) using \texttt{SimStock} and the baseline methods. We then create equal-weighted portfolios using these similar stocks to track the performance of the thematic ETFs.

\subsubsection{Quantitative evaluation} In this subsection, we quantitatively evaluate the performance of \texttt{SimStock} and the baseline methods in tracking the four thematic ETFs. \cref{chap4_Tracking_RMSE_ETFs} report the tracking error (TE) for each ETF using the top@$k$ similar stocks identified by each method across the four exchanges. The best performing method for each $k$ and exchange is highlighted in bold, while the second best is in underline. The results of the evaluation with Tracking Error Volatility (TEV) can be found in Appendix B.

A clear pattern emerges from the results: \texttt{SimStock} consistently achieves the best (bold) or second lowest (underline) tracking error for the majority of $k$ values across three of the four ETFs (ARKK, SKYY, and BOTZ) and exchanges. This strong performance demonstrates that the similar stocks identified by \texttt{SimStock} are highly effective at tracking these thematic ETFs.  In contrast to \texttt{SimStock}, while the Corr1 and Corr2 baselines show reasonable performance in some cases, they lack consistency across different scenarios. TS2VEC, on the other hand, generally exhibits higher tracking errors, indicating that the stocks it identifies do not track the ETFs as closely.

Notably, \texttt{SimStock}'s performance on the US exchange is particularly impressive for ARKK, SKYY, and BOTZ. It substantially outperforms the other methods, especially for lower values of $k$ (e.g., top 10-20 similar stocks), suggesting that \texttt{SimStock} excels at identifying a concentrated set of the most relevant stocks for tracking these ETF themes. \texttt{SimStock}'s ability to identify highly relevant stocks in this market, even with a smaller number of stocks (low top@$k$ values), is a testament to its effectiveness and precision in capturing the key components driving these thematic ETFs. While the performance gap between \texttt{SimStock} and the other methods is less pronounced on the other exchanges, \texttt{SimStock} still achieves the best or second-best performance in most cases for ARKK, SKYY, and BOTZ. However, \texttt{SimStock}'s performance in tracking the Global X Lithium \& Battery Tech ETF (LIT) is relatively weaker compared to the other ETFs.

\begin{figure*}[!htbp]
\centering
\includegraphics[width=\textwidth]{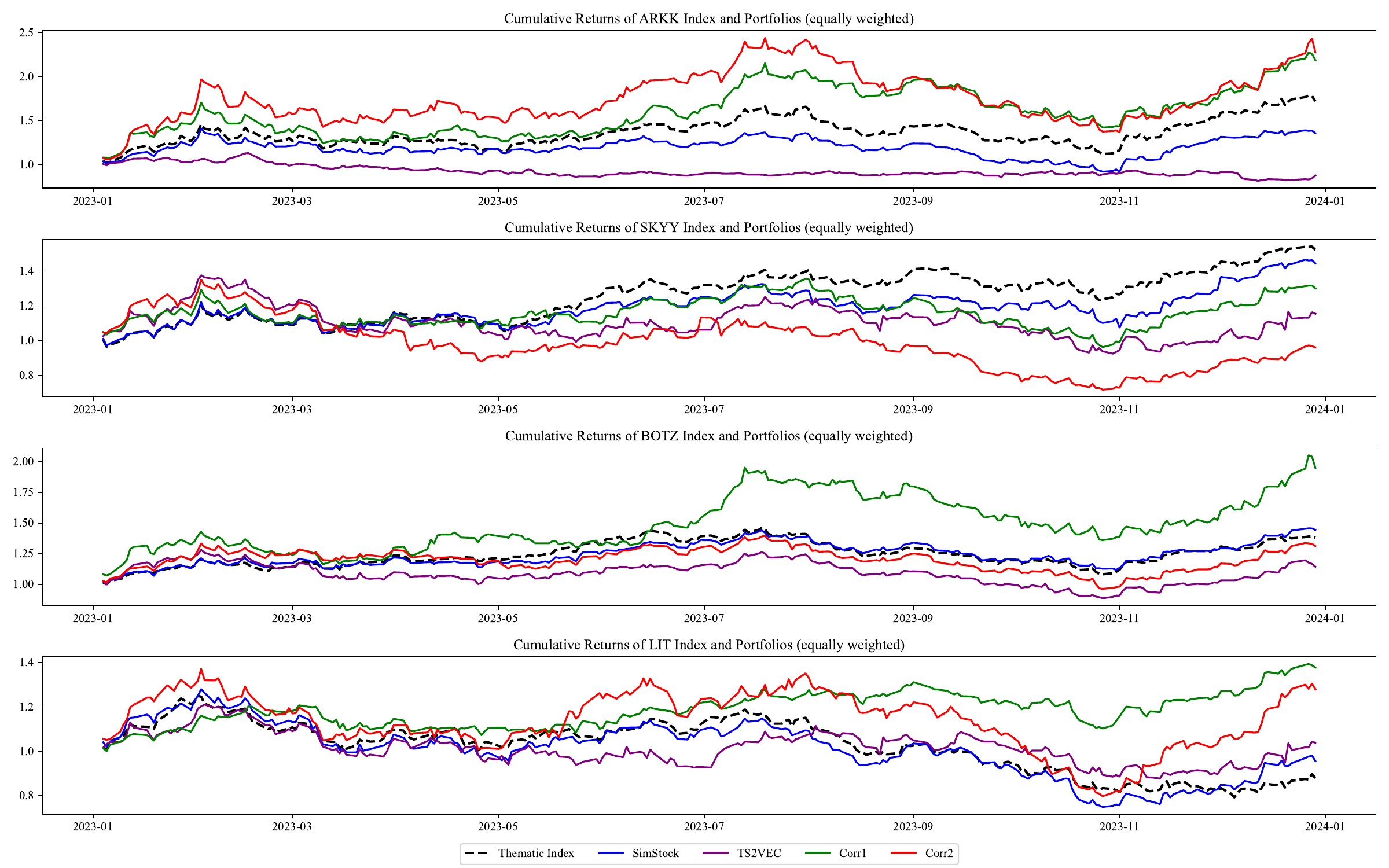}
\caption{Cumulative return curves of the four thematic ETFs (ARKK, SKYY, BOTZ, and LIT) and their corresponding tracking portfolios constructed using the top 10 similar stocks identified by \texttt{SimStock} and the baseline methods (TS2VEC, Corr1, and Corr2) from the US exchange. The closer a portfolio's curve follows the respective ETF curve (dotted black line), the better the tracking performance. }
\label{chap4_cumulative_retun_curves_for_index}
\end{figure*}

In addition to the quantitative evaluation using tracking error and tracking error volatility, we can also visually assess the performance of \texttt{SimStock} and the baseline methods in tracking the thematic ETFs by examining their cumulative return curves. \cref{chap4_cumulative_retun_curves_for_index} showcases the cumulative return curves for each thematic ETF (ARKK, SKYY, BOTZ, and LIT) and the corresponding tracking portfolios created using the top@10 similar stocks identified by each method on the US exchange.

The cumulative return curves provide a clear visual representation of how closely the tracking portfolios match the performance of their respective thematic ETFs over time. For ARKK, SKYY, and BOTZ, the \texttt{SimStock} portfolio (shown in blue) closely follows the trajectory of the thematic ETF (shown in dotted black), indicating that the top 10 stocks identified by \texttt{SimStock} effectively capture the performance of these ETFs. In contrast, the portfolios based on the other methods (TS2VEC in purple, Corr1 in green, and Corr2 in red) exhibit larger deviations from the ETF curves, suggesting that their selected stocks are less effective at tracking the ETFs.

The cumulative return curves represent \texttt{SimStock}'s ability to identify a concentrated set of stocks that effectively track the performance of thematic ETFs, particularly for ARKK, SKYY, and BOTZ, based on the top@10 similar stocks from the US exchange. 

\clearpage
\begin{table*}[!htbp]
    \input{Tables/chap4_Tracking_RMSE_ARKK}
    \vspace{0.5cm}
    
    \input{Tables/chap4_Tracking_RMSE_SKYY}

    \vspace{0.5cm}
    
    \input{Tables/chap4_Tracking_RMSE_BOTZ}
    \vspace{0.5cm}

\input{Tables/chap4_Tracking_RMSE_LIT}

    \caption{Tracking errors (TE) of \texttt{SimStock} and baseline methods for tracking the performance of four thematic ETFs (ARKK, SKYY, BOTZ, and LIT) using top k similar stocks from the US, SSE, SZSE, and TSE exchanges. The best performing method for each $k$ and exchange is highlighted in bold, while the second best is in underline.}
    \label{chap4_Tracking_RMSE_ETFs}
\end{table*}
\clearpage

\subsubsection{Qualitative evaluation} For the ARK Innovation ETF (ARKK), which focuses on disruptive innovation, \texttt{SimStock} identifies stocks of companies involved in cutting-edge technologies such as 3D computing (MLTS, DDD), a fintech (SQ), and artificial intelligence (SHOP) company. Similarly, for the First Trust Cloud Computing ETF (SKYY), \texttt{SimStock} finds stocks of leading cloud computing (NET, DBX, TYL, GWRE and GLOB) and software-as-a-service (INTU, ESTC and PAYC) providers. The qualitative assessment of \texttt{SimStock}'s similar stock selections for all thematic ETFs can be found in Appendix C.

\subsection{Application to portfolio optimization} In this section, we investigate whether \texttt{SimStock} embeddings can enhance portfolio optimization. Specifically, we construct the correlation matrix using the \texttt{SimStock} embedding as a similarity measure, and use it as an input for portfolio optimization. We compare the portfolio performance using the SimStock embedding with other covariance estimators. Notice that \texttt{SimStock} is designed to identify stocks with similar behavior, but it does not actively search for stocks that exhibit completely different dynamics compared to a given stock. This may limit its ability to ensure optimal portfolio diversification. While our approach uses deep learning to estimate the covariance matrix ($\Sigma$) for portfolio optimization, it differs from existing deep learning-based portfolio optimization methods. Most existing deep learning approaches for portfolio optimization focus on predicting optimal portfolio weights \citep{zhang2021universal, uysal2023end} or estimating portfolio parameters, particularly the expected returns ($\mu$) \citep{butler2023integrating, costa2023distributionally}. To the best of our knowledge, no prior deep learning-based portfolio research has specifically targeted the estimation of the covariance matrix ($\Sigma$). Our study aims to fill this gap by proposing a deep learning method to estimate $\Sigma$ and comparing its performance with conventional covariance estimators in traditional portfolio optimization.

\subsubsection{Optimization formulation}  We consider the Mean-Variance Optimization (MVO) framework \citep{markowits1952portfolio} for portfolio optimization. The MVO approach determines the optimal asset weights ($\omega_i$) in a portfolio, subject to constraints on risk and return, assuming the expected return ($\mu_i$), variance ($\sigma^2_{ii}$), and covariance ($\sigma_{ij}$) of each asset are known. This study adopts the experimental design and methodology established by \citep{gerber2021gerber}. We randomly sampled the asset classes 100 times with 10, 30 and 50 stocks each from the S\&P500 and JPX Prime 150.

The long-only MVO problem, accounting for transaction costs, is formulated as:

\begin{equation}
\begin{aligned}
\text{Maximize: } & w^T \mu - \psi 1^T |w - w_0| \\
\text{Subject to: } & w^T \Sigma w \leq \sigma^2_\text{target}, \quad w^T 1 = 0, \quad 0 \leq w_k \leq 1, \forall k = {1, 2, \cdots, N}.
\end{aligned}
\end{equation}%

The objective function seeks to maximize the portfolio's expected return ($\omega^T \mu$) while minimizing transaction costs, represented by the term $\psi 1^T |\omega - \omega_0|$. Here, $\psi$ denotes the fixed proportional transaction cost, set to 10 basis points (0.1\%) in this study, and $\omega_0$ represents the previous portfolio weights. By including this term, we penalize deviations from the previous weights, effectively controlling portfolio turnover and reducing trading costs.

The optimization is subject to three constraints: (1) the portfolio variance must not exceed a predetermined risk target ($\sigma^2_\text{target}$), (2) the sum of portfolio weights must equal unity, and (3) individual asset weights must be non-negative and not exceed one, prohibiting short selling. By solving this constrained optimization problem, we obtain the optimal portfolio weights $\omega^*$ for a given risk target $\sigma_\text{target}$ and transaction cost $\psi$. The set of optimal portfolios across various risk levels constitutes the efficient frontier, with each point representing the highest achievable return for a specific level of risk. 

We estimate the expected return $\mu_{ti}$ for asset $i$ at time $t$ using the sample mean of its historical returns over a $T$-month lookback window. For the covariance matrix $\Sigma$, we compare the performance of historical covariance (HC), shrinkage method (SM) \citep{ledoit2004honeyorigin}, Gerber statistic (GS) \citep{gerber2021gerber}, TS(TS2VEC) \citep{yue2022ts2vec} embedding and \texttt{SimStock} embedding (SS).

\subsubsection{Backtesting procedure} 
To evaluate the performance of the different covariance estimators in the context of portfolio optimization, we employ the following backtesting procedure. Starting from Jan 2021, at the beginning of each month, we use the monthly returns of the current asset universe over the previous $T = 12$ months to estimate the expected return vector $\mu$ and the covariance matrix $\Sigma$.

We then apply a quadratic optimizer to solve for the optimal portfolio weights $\omega^*$ given a specific risk target $\sigma_\text{target}$ for MVO and the optimal portfolio weights $w^*$ that minimize the portfolio variance for MVP. The previous portfolio is rebalanced according to the new optimal weights, and this optimized portfolio is held for one month. At the end of the month, the realized portfolio return is calculated as $w^{*T} \tilde{r}$, where $\tilde{r}$ represents the vector of realized asset returns for that month.

This process is repeated by rolling the in-sample period forward by one month and computing the updated efficient portfolio for the next month. The rolling-window approach allows the portfolios to adapt to structural changes in the market and mitigates data mining bias. As the initialization of the first portfolio requires two years of monthly returns, our performance evaluation covers the period from Jan 2022 to Feb 2024.

\subsubsection{Efficient frontiers} 
\begin{figure}
\centering
\includegraphics[width=\textwidth]{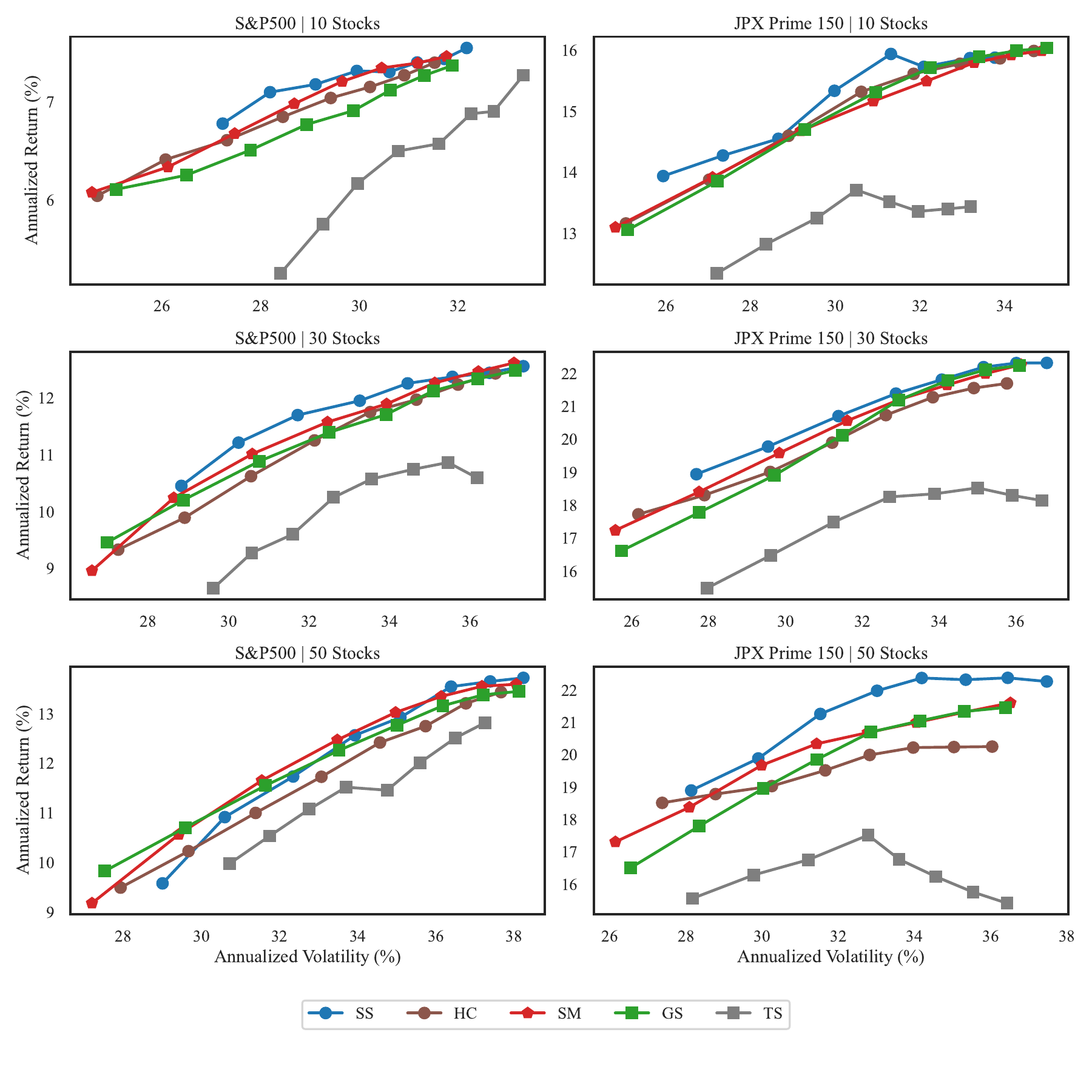}
\caption{Ex-post efficient frontiers displaying annualized return and volatility of portfolios optimized for different risk targets. The black vertical dotted lines represent the average volatility of the S\&P500 and JPX Prime 150, respectively.}
\label{chap4_ex_eff}
\end{figure}

In this subsection, to evaluate the performance of portfolios optimized using the \texttt{SimStock} embedding covariance matrix ($\Sigma_\text{SS}$), we compare their ex-post efficient frontiers with those of portfolios constructed using traditional covariance estimators, namely the historical covariance (HC), shrinkage method (SM), Gerber statistic (GS) and TS2VEC (TS), denoted as $\Sigma_\text{HC}$, $\Sigma_\text{SM}$, $\Sigma_\text{GS}$ and $\Sigma_\text{TS}$, respectively. To obtain the covariance matrix $\Sigma_\text{SS}$ using the TS2VEC and \texttt{SimStock} embedding, we first compute the pairwise L2 distances between the embeddings of each stock pair. These distances are then scaled to the range [-1, 1] using max-min normalization, which brings the stocks closer to 1 in the correlation matrix for smaller L2 distances, indicating greater similarity between stocks. The resulting normalized distances are used to populate the correlation matrix, represents the correlation coefficient between stocks $i$ and $j$ as estimated by embeddings. We then compute the covariance matrix $\Sigma_\text{TS}$ and $\Sigma_\text{SS}$. The analysis is conducted on portfolios of varying sizes (10, 30, and 50 stocks) from both the S\&P500 and JPX Prime 150 universes, across different risk targets. 

As shown in \cref{chap4_ex_eff}, the \texttt{SimStock} embedding approach slightly outperforms the other covariance estimators across all portfolio sizes and risk levels in both the S\&P500 and JPX Prime 150 universes. The portfolios constructed using $\Sigma_\text{SS}$ achieve higher annualized returns for a given level of annualized volatility compared to those based on $\Sigma_\text{HC}$, $\Sigma_\text{SM}$, $\Sigma_\text{GS}$ and $\Sigma_\text{TS}$. This outperformance is slightly more evident for smaller portfolio sizes (10 and 30 stocks), where the \texttt{SimStock} embedding method demonstrates an advantage in capturing meaningful relationships between assets, leading to superior risk-adjusted performance. While our approach did not aim to take into account the correlation coefficients between all stocks, these results suggest that \texttt{SimStock} has the potential to slightly improve portfolio optimization results in portfolio construction settings. The results showing robustness to potential other choices to construct the covariance from \texttt{SimStock} can be found in Appendix F.

The underperformance of TS2VEC is particularly noticeable when compared to the \texttt{SimStock} (SS) embedding approach, which consistently outperforms all other methods. The gap between the TS and SS curves is most evident for the smaller portfolio sizes (10 and 30 stocks), suggesting that the TS2VEC embeddings may struggle to capture the nuanced relationships between assets in these more concentrated portfolios.

\subsubsection{Detailed investment performances} This section compares the performance of five portfolio construction methods (Simstock embedding, Historical Covariance, Shrinkage Method, Gerber Statistic, and TS2VEC embedding) for MVO, using 30-stock portfolios from the S\&P 500 and JPX Prime 150 universes. See Appendix E for results for 10 and 50 stocks.

\cref{chap4_table_snp500} and \cref{chap4_table_tse150} present the performance metrics for the target risk optimizations in the S\&P500 and JPX Prime 150 universes, respectively, at four different risk target levels (24\%, 27\%, 30\%, and 33\%). See Appendix E for results for 10 and 50 stocks.

In the S\&P500 universe (\cref{chap4_table_snp500}), the Simstock embedding (SS) method produced the highest returns at all risk target levels while maintaining risk measures comparable to the other methods. Returns and risk measures increased for all methods as the risk target rose. The Historical Covariance (HC), Shrinkage Method (SM), and Gerber Statistic (GS) methods performed similarly to each other, but slightly worse than SS, particularly at higher risk levels. For example, at the 24\% risk target, SS achieved an 11.31\% arithmetic return compared to 10.04\%, 9.61\%, and 10.14\% for HC, SM, and GS respectively. At the 33\% risk target, SS's return increased to 13.33\%, while HC, SM and GS saw returns of 12.50\%, 12.85\% and 12.69\%. The TS2VEC embedding (TS) method consistently underperformed the others, with a 9.40\% return at the 24\% risk level and 11.49\% at the 33\% level, below all other methods.

The JPX Prime 150 universe (\cref{chap4_table_tse150}) showed similar patterns, with SS producing the best returns, followed by HC, GS, SM and TS in that order across risk levels. SS achieved a 20.12\% return at the 24\% risk target and 22.47\% at the 33\% target. In contrast, TS had the lowest returns at 14.60\% and 18.45\% for those risk levels. The HC, GS, and SM methods produced returns that were higher than TS but lower than SS at each corresponding risk level.

Despite potentially higher volatility which is evident in the ex-post efficient frontiers in \cref{chap4_ex_eff}, the performance of the SS method highlights its potential as a meaningful and effective stock embedding technique for portfolio optimization. 

\clearpage
\begin{sidewaystable}
\renewcommand{\arraystretch}{1}
\input{portfolio_figures/snp_30_ts}
\caption{This table presents the performance metrics for four portfolio construction methods in the S\&P500: Simstock embedding (SS), historical covariance (HC), shrinkage method (SM), Gerber statistic (GS) and TS2VEC embedding (TS). The portfolios were optimized for four different risk target levels: 24\%, 27\%, 30\%, and 33\%. The performance was evaluated over the full testing period from January 2022 to February 2024. The 3-month U.S. Treasury Bill rate was used as the risk-free rate for performance calculations. Transaction costs were modeled as 10 basis points of the traded volume for each rebalancing event.}
\label{chap4_table_snp500}
\vspace{20mm}
\input{portfolio_figures/tse_30_ts}
\caption{This table presents the performance metrics for four portfolio construction methods in the JPX Prime 150: Simstock embedding(SS), historical covariance (HC), shrinkage method (SM), Gerber statistic (GS) and TS2VEC embedding (TS). The portfolios were optimized for four different risk target levels: 24\%, 27\%, 30\%, and 33\%. The performance was evaluated over the full testing period from January 2022 to February 2024. The 3-month U.S. Treasury Bill rate was used as the risk-free rate for performance calculations. Transaction costs were modeled as 10 basis points of the traded volume for each rebalancing event.}
\label{chap4_table_tse150}
\end{sidewaystable}
\clearpage

\subsubsection{Characteristics of \texttt{SimStock} similarity matrix} To investigate what distinguishes the \texttt{SimStock} embedding method from other portfolio optimization methods, we can compare the correlation matrices generated by different methodologies. Specifically, we propose the following mapping:

\begin{equation}
 \frac{\left\|\  \text{MD}-\text{RC}^{\text{future}}   \right\|_{\text{F}} }{\left\|\  \text{MD}-\text{RC}^{\text{past}}   \right\|_{\text{F}}} \le 1
\end{equation}

Here, MD refers to the correlation matrix obtained using a specific methodology (e.g., SS, SM, GS or TS), while $\text{RC}^{\text{future}}$ and $\text{RC}^{\text{past}}$ represent the realized correlation matrices for the future and past periods, respectively.

If the value of this mapping is less than or close to 1, it implies that $\left\|\  \text{MD}-\text{RC}^{\text{future}}   \right\|_{\text{F}} \le \left\|\  \text{MD}-\text{RC}^{\text{past}}   \right\|_{\text{F}}$. This means that the current methodology is tracking the future correlation matrix well, as the Frobenius distance between the method's correlation matrix and the future realized correlation matrix is smaller than the distance between the method's correlation matrix and the past correlation matrix.

\begin{figure}[h!]
\centering
\includegraphics[width=\textwidth]{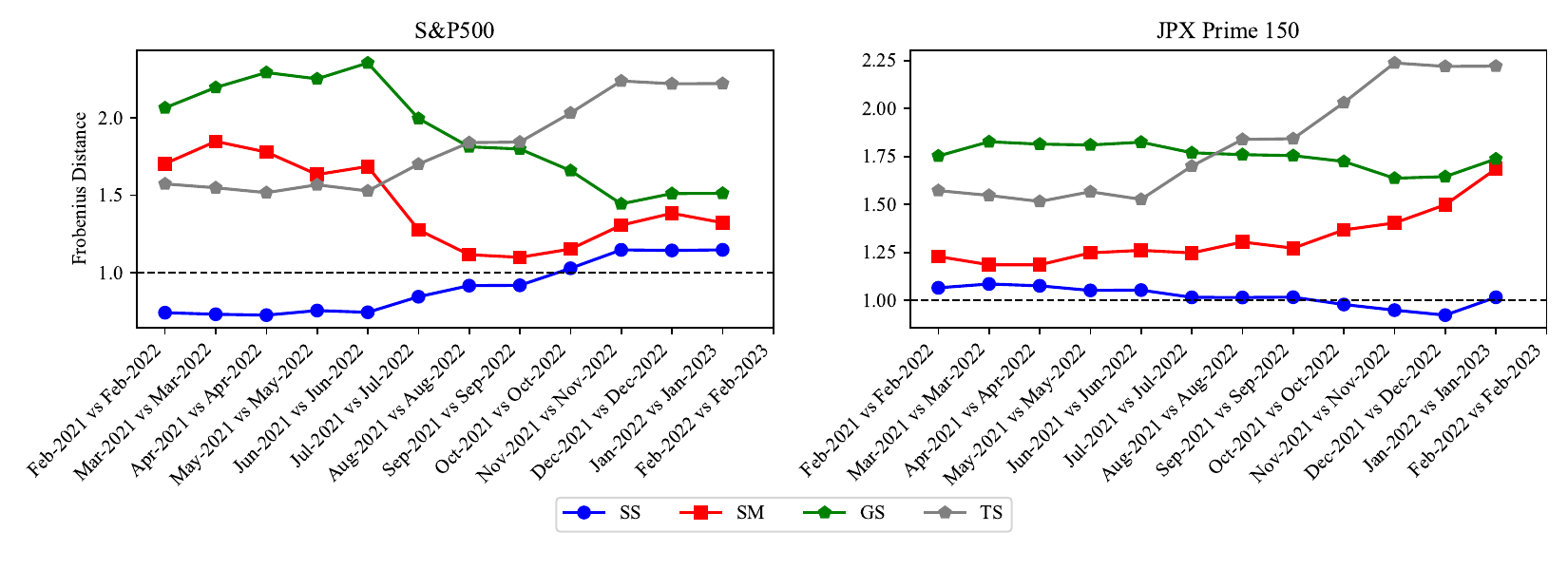}
\caption{Comparison of Frobenius distances between correlation matrices generated by different methodologies (SS: SimStock embedding, SM: shrinkage method, GS: Gerber statistic) and realized correlation matrices for consecutive years in the S\&P500 and JPX Prime 150 universes. Lower values indicate better tracking of the future correlation matrix. The SS method consistently exhibits lower Frobenius distances compared to the SM , GS and TS methods across most month pairs in both universes, suggesting its superior ability to capture future asset correlations.}
\label{chap4_corr_comparison}
\end{figure}

\cref{chap4_corr_comparison} illustrates the Frobenius distances for different methodologies in the S\&P500 and JPX Prime 150 universes. The figure compares the correlation matrices between the same months in consecutive years (e.g., Feb-2021 vs Feb-2022, Mar-2021 vs Mar-2022, etc.). For example, ``Feb-2021 vs Feb-2022'' means that the correlation series for the period from February 2021 to January 2022 is compared to the correlation series for the period from February 2022 to January 2023, representing the current year and the future year, respectively.

For the S\&P500, the Frobenius distances for the SimStock embedding method (SS) are consistently lower than those of the shrinkage method (SM), Gerber statistic (GS) and TS2VEC embedding (TS) across all month pairs. This suggests that the SS method is better at tracking the future correlation matrix compared to the SM and GS methods.

Similarly, in the JPX Prime 150 universe, the SS method generally exhibits lower Frobenius distances than the SM, GS and TS methods for most month pairs. This indicates that the SS method is more effective at capturing the future correlation structure of the assets compared to the other methodologies.

The ability of the SimStock embedding method to better track the future correlation matrix can be attributed to its unique approach, temporal representation learning, which effectively captures the time-varying patterns in stock price data. This improved understanding of the underlying asset dynamics could enable the SS method to generate correlation matrices that more closely resemble the future realized correlations, thereby providing an advantage edge in portfolio optimization.

%% file: Tables/chap4_prepro.tex
\fontsize{8}{9}\selectfont{
\begin{tabular}{ll}
\hline
\multicolumn{1}{c}{\textbf{Price features}} & \multicolumn{1}{c}{\textbf{Description}} \\ \hline
$z_{\text{Open}}$                           & $\text{Open}_{t}/\text{Close}_{t} -1$ \\
$z_{\text{High}}$                           & $\text{High}_{t}/\text{Close}_{t} -1$ \\
$z_{\text{Low}}$                            & $\text{Low}_{t}/\text{Close}_{t} -1$ \\
$z_{\text{Close}}$                          & $\text{Close}_{t}/\text{Close}_{t-1} -1$ \\
$z_{\text{Volume}}$                         & $\text{Volume}_{t}/\text{Volume}_{t-1} -1$ \\ \hline
\end{tabular}}

%% file: Tables/chap4_pair_stocks_name.tex
\fontsize{6}{7}\selectfont{
\begin{tabular}{ccccccccccc}
\hline
\multicolumn{1}{l}{\multirow{2}{*}{\textbf{\begin{tabular}[c]{@{}l@{}}Query\\ Stock\end{tabular}}}} & \multicolumn{1}{l}{\multirow{2}{*}{\textbf{Method}}} & \multicolumn{3}{l}{\textbf{TOP@3 similar stocks}} & \multicolumn{1}{l}{} & \multicolumn{1}{l}{\multirow{2}{*}{\textbf{\begin{tabular}[c]{@{}l@{}}Query\\ Stock\end{tabular}}}} & \multicolumn{1}{l}{\multirow{2}{*}{\textbf{Method}}} & \multicolumn{3}{l}{\textbf{TOP@3 similar stocks}} \\
\multicolumn{1}{l}{}                                                                                & \multicolumn{1}{l}{}                                 & \textbf{First}  & \textbf{Second} & \textbf{Third} & \textbf{}            & \multicolumn{1}{l}{}                                                                                & \multicolumn{1}{l}{}                                 & \textbf{First}  & \textbf{Second} & \textbf{Third} \\ \cline{1-5} \cline{7-11} 
\multirow{5}{*}{AAPL}                                                                               & SimStock                                             & MSFT            & TYL            & INTU           &                      & \multirow{5}{*}{PFE}                                                                                & SimStock                                             & BNTX            & MRNA           & JNJ            \\
                                                                                                    & TS2VEC                                               & AMZN            & WTM            & AMD            &                      &                                                                                                     & TS2VEC                                               & NKNG            & DJCO           & ESGR           \\
                                                                                                    & Corr1                                                & TMO             & SNPS           & CNDS           &                      &                                                                                                     & Corr1                                                & ICL             & MCBS           & PTSI           \\
                                                                                                    & Corr2                                                & GLOB            & AMZN           & TYL            &                      &                                                                                                     & Corr2                                                & RMR             & NVS            & BUD            \\
                                                                                                    & Peer                                                 & MSFT            & NVDA           & ASML           &                      &                                                                                                     & Peer                                                 & LLY             & ABBV           & NVO            \\ \cline{1-5} \cline{7-11} 
\multirow{5}{*}{CMG}                                                                                & SimStock                                             & AMZN            & MANH           & MSFT           &                      & \multirow{5}{*}{AMZN}                                                                               & SimStock                                             & CMG             & INTU           & MANH           \\
                                                                                                    & TS2VEC                                               & NVR             & USLM           & NEU            &                      &                                                                                                     & TS2VEC                                               & AAPL            & F              & PLPC           \\
                                                                                                    & Corr1                                                & DHR             & DSGX           & PCTY           &                      &                                                                                                     & Corr1                                                & ACMR            & ADBE           & FIVN           \\
                                                                                                    & Corr2                                                & PAYC            & PCTY           & MANH           &                      &                                                                                                     & Corr2                                                & TEAM            & LYV            & GENE           \\
                                                                                                    & Peer                                                 & HLT             & RACE           & AZO            &                      &                                                                                                     & Peer                                                 & TSLA            & BKNG           & SBUX           \\ \cline{1-5} \cline{7-11} 
\multirow{5}{*}{MSFT}                                                                               & SimStock                                             & CDNS            & MANH           & TYL            &                      & \multirow{5}{*}{BA}                                                                                 & SimStock                                             & IVZ             & SPR            & UAA            \\
                                                                                                    & TS2VEC                                               & GOOG            & GOOGL          & MA             &                      &                                                                                                     & TS2VEC                                               & LPL             & NNI            & FCX            \\
                                                                                                    & Corr1                                                & DHR             & TMO            & FAST           &                      &                                                                                                     & Corr1                                                & NCLH            & SPR            & SOHO           \\
                                                                                                    & Corr2                                                & GOOGL           & GOOGL          & DAVA           &                      &                                                                                                     & Corr2                                                & RVSB            & ENVA           & STT            \\
                                                                                                    & Peer                                                 & AAPL            & NVDA           & ASML           &                      &                                                                                                     & Peer                                                 & NOC             & CNI            & WM             \\ \cline{1-5} \cline{7-11} 
\multirow{5}{*}{WFC}                                                                                & SimStock                                             & BAC             & FITB           & FNB            &                      & \multirow{5}{*}{META}                                                                               & SimStock                                             & SPOT            & PYPL           & FORM           \\
                                                                                                    & TS2VEC                                               & JPM             & C              & MA             &                      &                                                                                                     & TS2VEC                                               & UHAL            & MAR            & MSFT           \\
                                                                                                    & Corr1                                                & BHLH            & WNEB           & RVSB           &                      &                                                                                                     & Corr1                                                & CVNA            & GREE           & INDP           \\
                                                                                                    & Corr2                                                & BAC             & WBS            & CFG            &                      &                                                                                                     & Corr2                                                & SKYW            & JAGX           & CTHR           \\
                                                                                                    & Peer                                                 & CHTR            & NTES           & ATVI           &                      &                                                                                                     & Peer                                                 & MCD             & LOW            & TM             \\ \cline{1-5} \cline{7-11} 
\multirow{5}{*}{V}                                                                                  & SimStock                                             & MA              & SF             & IHG            & \multicolumn{1}{l}{} & \multirow{5}{*}{MA}                                                                                 & SimStock                                             & V               & BKNG           & IHG            \\
                                                                                                    & TS2VEC                                               & MA              & MSFT           & KO             & \multicolumn{1}{l}{} &                                                                                                     & TS2VEC                                               & V               & KO             & NUE            \\
                                                                                                    & Corr1                                                & MA              & TDY            & ROP            & \multicolumn{1}{l}{} &                                                                                                     & Corr1                                                & V               & TDY            & FICO           \\
                                                                                                    & Corr2                                                & MA              & PLNT           & RELX           & \multicolumn{1}{l}{} &                                                                                                     & Corr2                                                & V               & GES            & RTO            \\
                                                                                                    & Peer                                                 & MA              & ADBE           & CSCO           & \multicolumn{1}{l}{} &                                                                                                     & Peer                                                 & JPM             & BAC            & V              \\ \cline{1-5} \cline{7-11} 
\multirow{5}{*}{XOM}                                                                                & SimStock                                             & MRO             & CVE            & HES            & \multicolumn{1}{l}{} & \multirow{5}{*}{CVS}                                                                                & SimStock                                             & CNC             & BMO            & MS             \\
                                                                                                    & TS2VEC                                               & MRO             & CVX            & NUE            & \multicolumn{1}{l}{} &                                                                                                     & TS2VEC                                               & HUM             & VNR            & BKNG           \\
                                                                                                    & Corr1                                                & MUR             & MRO            & EOG            & \multicolumn{1}{l}{} &                                                                                                     & Corr1                                                & CCB             & CHRD           & RJF            \\
                                                                                                    & Corr2                                                & MPC             & HES            & ERF            & \multicolumn{1}{l}{} &                                                                                                     & Corr2                                                & SRCL            & MLM            & NMFC           \\
                                                                                                    & Peer                                                 & CVX             & SHEL           & TTE            & \multicolumn{1}{l}{} &                                                                                                     & Peer                                                 & ANTM            & MDT            & GSK            \\ \cline{1-5} \cline{7-11} 
\end{tabular}}

%% file: Tables/chap_4_pair_trading_top3.tex
\useunder{\uline}{\ul}{}
\fontsize{7}{8}\selectfont{
\begin{tabular}{cccccc}
\hline
\multirow{2}{*}{\textbf{\begin{tabular}[c]{@{}c@{}}Query\\ Stock\end{tabular}}} & \multicolumn{5}{c}{\textbf{Wealth}}                                                                                                                                                                \\
                                                                                & \textbf{SimStock}                           & \textbf{TS2VEC}                 & \textbf{Corr1}                       & \textbf{Corr2}                       & \textbf{Coint}                       \\ \hline
AAPL                                                                            & \textbf{959.96 \tiny{$\pm$473.76}}          & $\text{NaN}^{**}$               & 234.53 \tiny{$\pm$1164.22}           & 915.16 \tiny{$\pm$338.61}            & $\text{NaN}^{**}$                    \\
CMG                                                                             & \textbf{545.81 \tiny{$\pm$722.54}}          & 282.17 \tiny{$\pm$713.52}       & -1069.43 \tiny{$\pm$721.61}          & -1098.98 \tiny{$\pm$892.15}          & -856.49 \tiny{$\pm$2964.75}          \\
MSFT                                                                            & \textbf{754.51 \tiny{$\pm$69.73}}           & 497.55 \tiny{$\pm$784.46}       & 477.31 \tiny{$\pm$1650.20}           & -305.99 \tiny{$\pm$2111.35}          & 454.68 \tiny{$\pm$857.63}            \\
WFC                                                                             & \textbf{567.80 \tiny{$\pm$169.84}}          & -777.72 \tiny{$\pm$1123.20}     & 390.84 \tiny{$\pm$736.03}            & $\text{NaN}^{**}$                    & -1309.26 \tiny{$\pm$2141.77}         \\
V                                                                               & {\ul 354.24 \tiny$\pm$116.60}               & 23.66 \tiny{$\pm$226.45}        & 329.23 \tiny{$\pm$99.14}             & \textbf{406.00 \tiny{$\pm$164.55}}   & 240.39 \tiny{$\pm$1067.06}           \\
XOM                                                                             & {\ul 403.26 \tiny$\pm$265.92}               & 364.02 \tiny{$\pm$248.26}       & -1.24 \tiny{$\pm$166.4}              & 13.54 \tiny{$\pm$1014.40}            & \textbf{1184.00 \tiny{$\pm$2526.09}} \\
PFE                                                                             & -413.02 \tiny{$\pm$674.64}                  & 113.17 \tiny{$\pm$2265.54}      & {\ul 192.04 \tiny$\pm$1656.18}       & -162.91 \tiny{$\pm$1543.09}          & \textbf{419.17 \tiny{$\pm$210.61}}   \\
AMZN                                                                            & 120.90 \tiny{$\pm$244.16}                   & {\ul 380.26 \tiny$\pm$1304.23}  & -597.30 \tiny{$\pm$1002.48}          & \textbf{2045.12 \tiny{$\pm$2088.20}} & -1183.58 \tiny{$\pm$3522.64}         \\
BA                                                                              & \textbf{572.25 \tiny{$\pm$2255.81}}         & 14.50 \tiny{$\pm$850.40}        & -1209.89 \tiny{$\pm$657.74}          & -652.71 \tiny{$\pm$1337.71}          & 143.38 \tiny{$\pm$724.86}            \\
META                                                                            & {\ul 1343.46 \tiny$\pm$ $\text{NaN}^*$}       & -597.33 \tiny{$\pm$1276.49}     & \textbf{1819.52 \tiny{$\pm$1901.72}} & -1694.84 \tiny{$\pm$3247.40}         & 326.92 \tiny{$\pm$1266.86}           \\
MA                                                                              & {\ul 123.06 \tiny$\pm$66.54}                & -98.50 \tiny{$\pm$267.16}       & \textbf{246.91 \tiny{$\pm$69.16}}    & -245.95 \tiny{$\pm$954.56}           & -576.08 \tiny{$\pm$893.63}           \\
CVS                                                                             & \textbf{1101.66 \tiny{$\pm$536.96}}         & -618.20 \tiny{$\pm$753.38}      & 981.67 \tiny{$\pm$753.47}            & 794.88 \tiny{$\pm$370.08}            & -209.44 \tiny{$\pm$1313.68}          \\ \hline
\multicolumn{1}{l}{}                                                            & \multicolumn{1}{l}{}                        & \multicolumn{1}{l}{}            & \multicolumn{1}{l}{}                 & \multicolumn{1}{l}{}                 & \multicolumn{1}{l}{}                 \\ \hline
\multirow{2}{*}{\textbf{\begin{tabular}[c]{@{}c@{}}Query\\ Stock\end{tabular}}} & \multicolumn{5}{c}{\textbf{Maximum Drawdown (\%)}}                                                                                                                                                 \\
                                                                                & \textbf{SimStock}                           & \textbf{TS2VEC}                 & \textbf{Corr1}                       & \textbf{Corr2}                       & \textbf{Coint}                       \\ \hline
AAPL                                                                            & \textbf{-1.91 \tiny{$\pm$0.67}}             & $\text{NaN}^{**}$               & -6.05 \tiny{$\pm$3.47}               & {\ul -3.4 \tiny$\pm$2.00}            & $\text{NaN}^{**}$                    \\
CMG                                                                             & \textbf{-2.98 \tiny{$\pm$2.54}}             & {\ul -6.37 \tiny$\pm$2.22}      & -14.96 \tiny{$\pm$6.38}              & -15.52 \tiny{$\pm$10.36}             & -20.03 \tiny{$\pm$15.47}             \\
MSFT                                                                            & \textbf{-3.79 \tiny{$\pm$0.21}}             & {\ul -5.66 \tiny$\pm$0.24}      & -7.12 \tiny{$\pm$4.72}               & -12.89 \tiny{$\pm$12.54}             & -13.23 \tiny{$\pm$2.96}              \\
WFC                                                                             & \textbf{-2.2 \tiny{$\pm$3.77}}              & -9.16 \tiny{$\pm$8.29}          & {\ul -5.75 \tiny$\pm$4.53}           & $\text{NaN}^{**}$                    & -23.95 \tiny{$\pm$13.75}             \\
V                                                                               & {\ul -2.22 \tiny$\pm$2.55}                  & -2.38 \tiny{$\pm$2.77}          & \textbf{-0.43 \tiny{$\pm$0.75}}      & -3.17 \tiny{$\pm$4.47}               & -7.46 \tiny{$\pm$5.87}               \\
XOM                                                                             & {\ul -3.38 \tiny$\pm$0.44}                  & \textbf{-3.02 \tiny{$\pm$2.95}} & -4.71 \tiny{$\pm$1.42}               & -6.61 \tiny{$\pm$7.06}               & -9.29 \tiny{$\pm$7.58}               \\
PFE                                                                             & \textbf{-7.84 \tiny{$\pm$7.41}}             & -10.03 \tiny{$\pm$13.49}        & -9.99 \tiny{$\pm$9.43}               & -10.43 \tiny{$\pm$5.06}              & {\ul -9.03 \tiny$\pm$5.56}           \\
AMZN                                                                            & {\ul -5.05 \tiny$\pm$4.99}                  & \textbf{-4.67 \tiny{$\pm$2.33}} & -15.12 \tiny{$\pm$9.38}              & -8.96 \tiny{$\pm$4.15}               & -31.90 \tiny{$\pm$13.12}             \\
BA                                                                              & {\ul -12.09 \tiny$\pm$11.57}                & \textbf{-5.87 \tiny{$\pm$2.13}} & -15.46 \tiny{$\pm$8.62}              & -12.86 \tiny{$\pm$3.26}              & -14.43 \tiny{$\pm$7.32}              \\
META                                                                            & \textbf{-4.91 \tiny{$\pm$$\text{NaN}^{*}$}} & {\ul -8.75 \tiny$\pm$8.98}      & -11.58 \tiny{$\pm$4.95}              & -33.98 \tiny{$\pm$25.84}             & -25.40 \tiny{$\pm$11.24}             \\
MA                                                                              & \textbf{-2.18 \tiny{$\pm$2.51}}             & -5.01 \tiny{$\pm$5.19}          & {\ul -2.58 \tiny$\pm$2.72}           & -6.85 \tiny{$\pm$7.23}               & -11.51 \tiny{$\pm$4.66}              \\
CVS                                                                             & {\ul -3.01 \tiny$\pm$1.84}                  & -9.08 \tiny{$\pm$8.32}          & -4.58 \tiny{$\pm$4.59}               & \textbf{-2.92 \tiny{$\pm$1.37}}      & -19.85 \tiny{$\pm$17.02}             \\ \hline
\end{tabular}}

%% file: Tables/chap4_Tracking_RMSE_ARKK.tex
\useunder{\uline}{\ul}{}
\centering
\fontsize{7}{8}\selectfont{
\begin{tabular}{cccccccc}
\hline
\multicolumn{1}{l}{}                           &                           & \multicolumn{6}{c}{\textbf{ARK Innovation ETF (ARKK)}}                                                                                                                                                                                                          \\
\multicolumn{1}{l}{\multirow{-2}{*}{\textbf{Exchange}}} & \multirow{-2}{*}{\textbf{Methods}} & \textbf{TOP@10}                              & \textbf{TOP@15}                              & \textbf{TOP@20}                        & \textbf{TOP@25}                        & \textbf{TOP@30}                        & \textbf{TOP@35}                        \\ \hline
                                               & Corr1                     & {\color[HTML]{000000} {\ul \textbf{0.2828}}} & {\color[HTML]{000000} {\ul \textbf{0.1363}}} & {\color[HTML]{000000} \textbf{0.0746}} & {\color[HTML]{000000} \textbf{0.0822}} & {\color[HTML]{000000} \textbf{0.0649}} & {\color[HTML]{000000} \textbf{0.0592}} \\
                                               & Corr2                     & 0.4264                                       & 0.2431                                       & 0.1132                                 & {\color[HTML]{000000} {\ul 0.1455}}    & {\color[HTML]{000000} {\ul 0.0963}}    & {\color[HTML]{000000} {\ul 0.1149}}    \\
                                               & TS2VEC                    & 0.4752                                       & 0.5772                                       & 0.5321                                 & 0.4887                                 & 0.3723                                 & 0.3046                                 \\
\multirow{-4}{*}{US}                           & SimStock                  & {\color[HTML]{000000} \textbf{0.1934}}       & {\color[HTML]{000000} \textbf{0.0764}}       & {\color[HTML]{000000} {\ul 0.0975}}    & 0.1785                                 & 0.1822                                 & 0.2414                                 \\ \hline
                                               & Corr1                     & 0.4014                                       & 0.4244                                       & 0.4191                                 & 0.4250                                 & 0.4333                                 & 0.4365                                 \\
                                               & Corr2                     & {\color[HTML]{000000} \textbf{0.2753}}       & {\color[HTML]{000000} \textbf{0.2651}}       & {\color[HTML]{000000} \textbf{0.2739}} & {\color[HTML]{000000} {\ul 0.2883}}    & {\color[HTML]{000000} \textbf{0.2681}} & {\color[HTML]{000000} \textbf{0.2747}} \\
                                               & TS2VEC                    & 0.3813                                       & 0.3815                                       & 0.3748                                 & 0.4037                                 & 0.4030                                 & 0.4142                                 \\
\multirow{-4}{*}{SSE}                          & SimStock                  & {\color[HTML]{000000} {\ul 0.3064}}          & {\color[HTML]{000000} {\ul 0.3133}}          & {\color[HTML]{000000} {\ul 0.2778}}    & {\color[HTML]{000000} \textbf{0.2774}} & {\color[HTML]{000000} {\ul 0.2754}}    & {\color[HTML]{000000} {\ul 0.3069}}    \\ \hline
                                               & Corr1                     & 0.3745                                       & 0.3843                                       & 0.3607                                 & 0.4006                                 & 0.3843                                 & 0.3923                                 \\
                                               & Corr2                     & {\color[HTML]{000000} {\ul 0.2302}}          & {\color[HTML]{000000} {\ul 0.2063}}          & {\color[HTML]{000000} {\ul 0.2138}}    & {\color[HTML]{000000} \textbf{0.2208}} & {\color[HTML]{000000} {\ul 0.2380}}    & {\color[HTML]{000000} {\ul 0.2547}}    \\
                                               & TS2VEC                    & 0.2793                                       & 0.2843                                       & 0.3132                                 & 0.3074                                 & 0.2993                                 & 0.3010                                 \\
\multirow{-4}{*}{SZSE}                         & SimStock                  & {\color[HTML]{000000} \textbf{0.1621}}       & {\color[HTML]{000000} \textbf{0.1643}}       & {\color[HTML]{000000} \textbf{0.1959}} & {\color[HTML]{000000} 0.2252}          & {\color[HTML]{000000} \textbf{0.2354}} & {\color[HTML]{000000} \textbf{0.2272}} \\ \hline
                                               & Corr1                     & 0.3965                                       & 0.3214                                       & 0.2694                                 & 0.2533                                 & 0.2591                                 & 0.2573                                 \\
                                               & Corr2                     & {\color[HTML]{000000} {\ul 0.1458}}          & {\color[HTML]{000000} {\ul 0.1707}}          & {\color[HTML]{000000} {\ul 0.1850}}    & {\color[HTML]{000000} {\ul 0.1846}}    & {\color[HTML]{000000} {\ul 0.1869}}    & {\color[HTML]{000000} {\ul 0.2002}}    \\
                                               & TS2VEC                    & 0.2712                                       & 0.2940                                       & 0.2574                                 & 0.2856                                 & 0.2841                                 & 0.2790                                 \\
\multirow{-4}{*}{TSE}                          & SimStock                  & {\color[HTML]{000000} \textbf{0.1440}}       & {\color[HTML]{000000} \textbf{0.1322}}       & {\color[HTML]{000000} \textbf{0.1310}} & {\color[HTML]{000000} \textbf{0.1328}} & {\color[HTML]{000000} \textbf{0.1262}} & {\color[HTML]{000000} \textbf{0.1321}} \\ \hline
\end{tabular}}

%% file: Tables/chap4_Tracking_RMSE_SKYY.tex
\useunder{\uline}{\ul}{}
\centering
\fontsize{7}{8}\selectfont{
\begin{tabular}{cccccccc}
\hline
\multicolumn{1}{l}{}                                    &                                    & \multicolumn{6}{c}{\textbf{First Trust Cloud Computing ETF (SKYY)}}                                                                                                                                                                                 \\
\multicolumn{1}{l}{\multirow{-2}{*}{\textbf{Exchange}}} & \multirow{-2}{*}{\textbf{Methods}} & \textbf{\textbf{TOP@10}}               & \textbf{\textbf{TOP@15}}               & \textbf{\textbf{TOP@20}}               & \textbf{\textbf{TOP@25}}               & \textbf{\textbf{TOP@30}}               & \textbf{\textbf{TOP@35}}               \\ \hline
                                                        & Corr1                              & {\color[HTML]{000000} {\ul 0.1502}}    & {\color[HTML]{000000} {\ul 0.1968}}    & {\color[HTML]{000000} 0.2064}          & {\color[HTML]{000000} {\ul 0.1937}}    & {\color[HTML]{000000} {\ul 0.1310}}    & {\color[HTML]{000000} {\ul 0.1235}}    \\
                                                        & Corr2                              & {\color[HTML]{000000} 0.3643}          & {\color[HTML]{000000} 0.2941}          & {\color[HTML]{000000} 0.2130}          & {\color[HTML]{000000} 0.2137}          & {\color[HTML]{000000} 0.1673}          & {\color[HTML]{000000} 0.1482}          \\
                                                        & TS2VEC                             & {\color[HTML]{000000} 0.2525}          & {\color[HTML]{000000} 0.2370}          & {\color[HTML]{000000} {\ul 0.2104}}    & {\color[HTML]{000000} 0.2377}          & {\color[HTML]{000000} 0.2262}          & {\color[HTML]{000000} 0.2263}          \\
\multirow{-4}{*}{US}                                    & SimStock                           & {\color[HTML]{000000} \textbf{0.0939}} & {\color[HTML]{000000} \textbf{0.0366}} & {\color[HTML]{000000} \textbf{0.0645}} & {\color[HTML]{000000} \textbf{0.0814}} & {\color[HTML]{000000} \textbf{0.0603}} & {\color[HTML]{000000} \textbf{0.0596}} \\ \hline
                                                        & Corr1                              & {\color[HTML]{000000} 0.2990}          & {\color[HTML]{000000} 0.3471}          & {\color[HTML]{000000} 0.3612}          & {\color[HTML]{000000} 0.3715}          & {\color[HTML]{000000} 0.3757}          & {\color[HTML]{000000} 0.3872}          \\
                                                        & Corr2                              & {\color[HTML]{000000} 0.1865}          & {\color[HTML]{000000} 0.1961}          & {\color[HTML]{000000} 0.1941}          & {\color[HTML]{000000} \textbf{0.1554}} & {\color[HTML]{000000} \textbf{0.1647}} & {\color[HTML]{000000} \textbf{0.1760}} \\
                                                        & TS2VEC                             & {\color[HTML]{000000} 0.2939}          & {\color[HTML]{000000} 0.3475}          & {\color[HTML]{000000} 0.3857}          & {\color[HTML]{000000} 0.3616}          & {\color[HTML]{000000} 0.3572}          & {\color[HTML]{000000} 0.3240}          \\
\multirow{-4}{*}{SSE}                                   & SimStock                           & {\color[HTML]{000000} \textbf{0.1607}} & {\color[HTML]{000000} \textbf{0.1291}} & {\color[HTML]{000000} \textbf{0.1688}} & {\color[HTML]{000000} {\ul 0.1615}}    & {\color[HTML]{000000} {\ul 0.1748}}    & {\color[HTML]{000000} {\ul 0.1912}}    \\ \hline
                                                        & Corr1                              & {\color[HTML]{000000} 0.4157}          & {\color[HTML]{000000} 0.4014}          & {\color[HTML]{000000} 0.3893}          & {\color[HTML]{000000} 0.3915}          & {\color[HTML]{000000} 0.3771}          & {\color[HTML]{000000} 0.3603}          \\
                                                        & Corr2                              & {\color[HTML]{000000} {\ul 0.1329}}    & {\color[HTML]{000000} \textbf{0.1390}} & {\color[HTML]{000000} \textbf{0.1532}} & {\color[HTML]{000000} \textbf{0.1617}} & {\color[HTML]{000000} {\ul 0.1679}}    & {\color[HTML]{000000} {\ul 0.1771}}    \\
                                                        & TS2VEC                             & {\color[HTML]{000000} 0.2388}          & {\color[HTML]{000000} 0.2789}          & {\color[HTML]{000000} 0.2616}          & {\color[HTML]{000000} 0.2598}          & {\color[HTML]{000000} 0.2690}          & {\color[HTML]{000000} 0.2499}          \\
\multirow{-4}{*}{SZSE}                                  & SimStock                           & {\color[HTML]{000000} \textbf{0.1230}} & {\color[HTML]{000000} {\ul 0.1543}}    & {\color[HTML]{000000} {\ul 0.1731}}    & {\color[HTML]{000000} {\ul 0.1651}}    & {\color[HTML]{000000} \textbf{0.1559}} & {\color[HTML]{000000} \textbf{0.1455}} \\ \hline
                                                        & Corr1                              & {\color[HTML]{000000} 0.1363}          & {\color[HTML]{000000} 0.1822}          & {\color[HTML]{000000} 0.1756}          & {\color[HTML]{000000} 0.2212}          & {\color[HTML]{000000} 0.2206}          & {\color[HTML]{000000} 0.2300}          \\
                                                        & Corr2                              & {\color[HTML]{000000} {\ul 0.1118}}    & {\color[HTML]{000000} {\ul 0.1489}}    & {\color[HTML]{000000} {\ul 0.1232}}    & {\color[HTML]{000000} {\ul 0.1389}}    & {\color[HTML]{000000} {\ul 0.1321}}    & {\color[HTML]{000000} {\ul 0.1261}}    \\
                                                        & TS2VEC                             & {\color[HTML]{000000} 0.2220}          & {\color[HTML]{000000} 0.2395}          & {\color[HTML]{000000} 0.2384}          & {\color[HTML]{000000} 0.2276}          & {\color[HTML]{000000} 0.2290}          & {\color[HTML]{000000} 0.2369}          \\
\multirow{-4}{*}{TSE}                                   & SimStock                           & {\color[HTML]{000000} \textbf{0.0940}} & {\color[HTML]{000000} \textbf{0.0733}} & {\color[HTML]{000000} \textbf{0.0734}} & {\color[HTML]{000000} \textbf{0.0660}} & {\color[HTML]{000000} \textbf{0.0593}} & {\color[HTML]{000000} \textbf{0.0691}} \\ \hline
\end{tabular}}

%% file: Tables/chap4_Tracking_RMSE_BOTZ.tex
\useunder{\uline}{\ul}{}

\centering
\fontsize{7}{8}\selectfont{
\begin{tabular}{cccccccc}
\hline
\multicolumn{1}{l}{}                                    &                                    & \multicolumn{6}{c}{\textbf{Global X Robotics \& AI ETF (BOTZ)}}                                                                                                                                                                                      \\
\multicolumn{1}{l}{\multirow{-2}{*}{\textbf{Exchange}}} & \multirow{-2}{*}{\textbf{Methods}} & \textbf{\textbf{TOP@10}}               & \textbf{\textbf{TOP@15}}               & \textbf{\textbf{TOP@20}}               & \textbf{\textbf{TOP@25}}               & \textbf{\textbf{TOP@30}}               & \textbf{\textbf{TOP@35}}               \\ \hline
                                                        & Corr1                              & 0.2847                                 & 0.1581                                 & {\color[HTML]{000000} {\ul 0.0981}}    & {\color[HTML]{000000} {\ul 0.0881}}    & {\color[HTML]{000000} {\ul 0.0784}}    & {\color[HTML]{000000} {\ul 0.0897}}    \\
                                                        & Corr2                              & {\color[HTML]{000000} {\ul 0.0946}}    & {\color[HTML]{000000} {\ul 0.1526}}    & 0.1249                                 & 0.1216                                 & 0.1448                                 & 0.1373                                 \\
                                                        & TS2VEC                             & 0.1890                                 & 0.3221                                 & 0.3368                                 & 0.2763                                 & 0.2833                                 & 0.3247                                 \\
\multirow{-4}{*}{US}                                    & SimStock                           & {\color[HTML]{000000} \textbf{0.0364}} & {\color[HTML]{000000} \textbf{0.0647}} & {\color[HTML]{000000} \textbf{0.0946}} & {\color[HTML]{000000} \textbf{0.0910}} & {\color[HTML]{000000} \textbf{0.0890}} & {\color[HTML]{000000} \textbf{0.0818}} \\ \hline
                                                        & Corr1                              & 0.2912                                 & 0.2870                                 & 0.2831                                 & 0.2738                                 & 0.2745                                 & 0.2815                                 \\
                                                        & Corr2                              & {\color[HTML]{000000} \textbf{0.0863}} & {\color[HTML]{000000} \textbf{0.1205}} & {\color[HTML]{000000} \textbf{0.1198}} & {\color[HTML]{000000} \textbf{0.1179}} & {\color[HTML]{000000} \textbf{0.1220}} & {\color[HTML]{000000} \textbf{0.1382}} \\
                                                        & TS2VEC                             & 0.2512                                 & 0.2535                                 & 0.2638                                 & 0.2955                                 & 0.2646                                 & 0.2696                                 \\
\multirow{-4}{*}{SSE}                                   & SimStock                           & {\color[HTML]{000000} {\ul 0.1471}}    & {\color[HTML]{000000} {\ul 0.1520}}    & {\color[HTML]{000000} {\ul 0.1514}}    & {\color[HTML]{000000} {\ul 0.1573}}    & {\color[HTML]{000000} {\ul 0.1509}}    & {\color[HTML]{000000} {\ul 0.1283}}    \\ \hline
                                                        & Corr1                              & 0.2259                                 & 0.2063                                 & 0.2329                                 & 0.2466                                 & 0.2645                                 & 0.2763                                 \\
                                                        & Corr2                              & {\color[HTML]{000000} {\ul 0.1016}}    & {\color[HTML]{000000} \textbf{0.1142}} & {\color[HTML]{000000} \textbf{0.1143}} & {\color[HTML]{000000} \textbf{0.1144}} & {\color[HTML]{000000} \textbf{0.1242}} & {\color[HTML]{000000} \textbf{0.1169}} \\
                                                        & TS2VEC                             & 0.1842                                 & 0.1516                                 & 0.1527                                 & 0.1387                                 & 0.1376                                 & 0.1694                                 \\
\multirow{-4}{*}{SZSE}                                  & SimStock                           & {\color[HTML]{000000} \textbf{0.0812}} & {\color[HTML]{000000} {\ul 0.1289}}    & {\color[HTML]{000000} {\ul 0.1171}}    & {\color[HTML]{000000} {\ul 0.1246}}    & {\color[HTML]{000000} {\ul 0.1331}}    & {\color[HTML]{000000} {\ul 0.1366}}    \\ \hline
                                                        & Corr1                              & 0.1646                                 & 0.1420                                 & 0.1457                                 & 0.1454                                 & 0.1411                                 & 0.1236                                 \\
                                                        & Corr2                              & {\color[HTML]{000000} {\ul 0.0951}}    & {\color[HTML]{000000} {\ul 0.0916}}    & {\color[HTML]{000000} {\ul 0.0863}}    & {\color[HTML]{000000} {\ul 0.0844}}    & {\color[HTML]{000000} {\ul 0.0910}}    & {\color[HTML]{000000} {\ul 0.0895}}    \\
                                                        & TS2VEC                             & 0.2155                                 & 0.2343                                 & 0.1943                                 & 0.1952                                 & 0.1959                                 & 0.1843                                 \\
\multirow{-4}{*}{TSE}                                   & SimStock                           & {\color[HTML]{000000} \textbf{0.0639}} & {\color[HTML]{000000} \textbf{0.0730}} & {\color[HTML]{000000} \textbf{0.0712}} & {\color[HTML]{000000} \textbf{0.0797}} & {\color[HTML]{000000} \textbf{0.0802}} & {\color[HTML]{000000} \textbf{0.0764}} \\ \hline
\end{tabular}}

%% file: Tables/chap4_Tracking_RMSE_LIT.tex
\centering
\fontsize{7}{8}\selectfont{
\begin{tabular}{cccccccc}
\hline
\multicolumn{1}{l}{}                                    &                                    & \multicolumn{6}{c}{\textbf{Global X Lithium \& Battery Tech ETF (LIT)}}                                                                                                                                                                              \\
\multicolumn{1}{l}{\multirow{-2}{*}{\textbf{Exchange}}} & \multirow{-2}{*}{\textbf{Methods}} & \textbf{\textbf{TOP@10}}               & \textbf{\textbf{TOP@15}}               & \textbf{\textbf{TOP@20}}               & \textbf{\textbf{TOP@25}}               & \textbf{\textbf{TOP@30}}               & \textbf{\textbf{TOP@35}}               \\ \hline
                                                        & Corr1                              & 0.2167                                 & 0.2819                                 & 0.2512                                 & 0.2638                                 & 0.2732                                 & 0.2465                                 \\
                                                        & Corr2                              & 0.1424                                 & 0.2337                                 & 0.2234                                 & 0.2195                                 & 0.2512                                 & 0.1669                                 \\
                                                        & TS2VEC                             & {\color[HTML]{000000} {\ul 0.0838}}    & {\color[HTML]{000000} {\ul 0.1094}}    & {\color[HTML]{000000} {\ul 0.0859}}    & {\color[HTML]{000000} {\ul 0.0924}}    & {\color[HTML]{000000} {\ul 0.1080}}    & {\color[HTML]{000000} {\ul 0.1070}}    \\
\multirow{-4}{*}{US}                                    & SimStock                           & {\color[HTML]{000000} \textbf{0.0446}} & {\color[HTML]{000000} \textbf{0.0557}} & {\color[HTML]{000000} \textbf{0.0647}} & {\color[HTML]{000000} \textbf{0.0666}} & {\color[HTML]{000000} \textbf{0.0694}} & {\color[HTML]{000000} \textbf{0.0703}} \\ \hline
                                                        & Corr1                              & 0.1756                                 & 0.1671                                 & 0.1485                                 & 0.1431                                 & 0.1354                                 & 0.1294                                 \\
                                                        & Corr2                              & 0.1159                                 & {\color[HTML]{000000} {\ul 0.1061}}    & {\color[HTML]{000000} {\ul 0.0983}}    & {\color[HTML]{000000} {\ul 0.0923}}    & {\color[HTML]{000000} {\ul 0.0899}}    & {\color[HTML]{000000} {\ul 0.0905}}    \\
                                                        & TS2VEC                             & {\color[HTML]{000000} \textbf{0.0715}} & {\color[HTML]{000000} \textbf{0.0798}} & {\color[HTML]{000000} \textbf{0.0772}} & {\color[HTML]{000000} \textbf{0.0790}} & {\color[HTML]{000000} \textbf{0.0817}} & {\color[HTML]{000000} \textbf{0.0803}} \\
\multirow{-4}{*}{SSE}                                   & SimStock                           & {\color[HTML]{000000} {\ul 0.1003}}    & 0.1355                                 & 0.1084                                 & 0.1009                                 & 0.1161                                 & 0.1211                                 \\ \hline
                                                        & Corr1                              & {\color[HTML]{000000} {\ul 0.1562}}    & {\color[HTML]{000000} {\ul 0.1431}}    & {\color[HTML]{000000} {\ul 0.1676}}    & {\color[HTML]{000000} {\ul 0.1676}}    & {\color[HTML]{000000} {\ul 0.1700}}    & {\color[HTML]{000000} {\ul 0.1730}}    \\
                                                        & Corr2                              & 0.2638                                 & 0.2566                                 & 0.2073                                 & 0.1906                                 & 0.1838                                 & 0.1690                                 \\
                                                        & TS2VEC                             & {\color[HTML]{000000} \textbf{0.1504}} & {\color[HTML]{000000} \textbf{0.1236}} & {\color[HTML]{000000} \textbf{0.1138}} & {\color[HTML]{000000} \textbf{0.1090}} & {\color[HTML]{000000} \textbf{0.0997}} & {\color[HTML]{000000} \textbf{0.0962}} \\
\multirow{-4}{*}{SZSE}                                  & SimStock                           & 0.2861                                 & 0.3187                                 & 0.2833                                 & 0.2892                                 & 0.2734                                 & 0.2412                                 \\ \hline
                                                        & Corr1                              & {\color[HTML]{000000} {\ul 0.1839}}    & {\color[HTML]{000000} {\ul 0.2430}}    & {\color[HTML]{000000} {\ul 0.2477}}    & {\color[HTML]{000000} {\ul 0.2769}}    & {\color[HTML]{000000} {\ul 0.2600}}    & {\color[HTML]{000000} {\ul 0.2242}}    \\
                                                        & Corr2                              & 0.3249                                 & 0.3030                                 & 0.2580                                 & 0.2457                                 & 0.2826                                 & 0.2662                                 \\
                                                        & TS2VEC                             & {\color[HTML]{000000} \textbf{0.1563}} & {\color[HTML]{000000} \textbf{0.1537}} & {\color[HTML]{000000} \textbf{0.1642}} & {\color[HTML]{000000} \textbf{0.1727}} & {\color[HTML]{000000} \textbf{0.1502}} & {\color[HTML]{000000} \textbf{0.1768}} \\
\multirow{-4}{*}{TSE}                                   & SimStock                           & 0.2557                                 & 0.2794                                 & 0.2763                                 & 0.2866                                 & 0.2861                                 & 0.2735                                 \\ \hline
\end{tabular}}

%% file: portfolio_figures/snp_30_ts.tex
\fontsize{6}{9}\selectfont{
\begin{tabular}{r|ccccc|ccccc|ccccc|ccccc}
\hline
S\&P500 | 30 Stocks   & \multicolumn{5}{c|}{Target Volatility (24\%)} & \multicolumn{5}{c|}{Target Volatility (27\%)} & \multicolumn{5}{c|}{Target Volatility (30\%)} & \multicolumn{5}{c}{Target Volatility (33\%)} \\ \hline
Covariance Method     & SS      & HC      & SM     & GS     & TS     & SS      & HC      & SM     & GS     & TS     & SS      & HC     & SM      & GS     & TS     & SS      & HC     & SM     & GS     & TS     \\ \hline
Arithmetic Return (\%) & 11.31   & 10.04   & 9.61   & 10.14  & 9.40   & 12.22   & 10.75   & 11.02  & 11.05  & 10.51  & 12.84   & 11.66  & 12.03   & 11.97  & 10.86  & 13.33   & 12.50  & 12.85  & 12.69  & 11.49  \\ 
Geometric Return (\%)  & 10.39   & 9.30    & 8.97   & 9.43   & 8.19   & 11.158  & 9.86    & 10.21  & 10.17  & 9.43   & 11.60   & 10.57  & 10.967  & 10.83  & 9.56   & 11.88   & 11.18  & 11.51  & 11.32  & 10.37  \\ 
Cumulative Return (\%) & 37.36   & 33.07   & 31.60  & 33.40  & 26.64  & 40.54   & 35.40   & 36.63  & 36.49  & 31.04  & 42.64   & 38.44  & 39.930  & 39.61  & 31.53  & 44.18   & 41.27  & 42.59  & 42.05  & 34.46  \\ 
Annualized SD (\%)     & 28.82   & 27.27   & 26.61  & 27.00  & 29.54  & 30.24   & 28.92   & 28.66  & 28.90  & 30.24  & 31.72   & 30.58  & 30.623  & 30.79  & 31.37  & 33.25   & 32.17  & 32.49  & 32.54  & 32.26  \\ 
Annualized Skewness   & -0.12   & -0.15   & -0.12  & -0.122 & -0.17  & -0.14   & -0.17   & -0.16  & -0.16  & -0.21  & -0.18   & -0.20  & -0.191  & -0.20  & -0.21  & -0.22   & -0.22  & -0.22  & -0.23  & -0.24  \\ 
Annualized Kurtosis   & 3.17    & 3.22    & 3.23   & 3.161  & 2.81   & 3.22    & 3.28    & 3.29   & 3.24   & 2.86   & 3.27    & 3.33   & 3.343   & 3.28   & 2.88   & 3.27    & 3.33   & 3.33   & 3.28   & 2.93   \\ 
Maximum Drawdown (\%)  & -24.90  & -23.96  & -23.47 & -23.59 & -25.59 & -25.65  & -25.36  & -24.57 & -25.04 & -25.93 & -26.69  & -26.44 & -25.971 & -26.57 & -26.64 & -28.20  & -27.63 & -27.57 & -28.08 & -26.47 \\ 
Monthly 95\% VaR (\%)   & -10.44  & -10.22  & -9.99  & -10.05 & -11.11 & -10.77  & -10.63  & -10.53 & -10.52 & -11.38 & -11.19  & -10.95 & -10.921 & -11.01 & -11.69 & -11.72  & -11.40 & -11.48 & -11.56 & -12.2  \\
Sharpe Ratio          & 0.44    & 0.40    & 0.39   & 0.42   & 0.31   & 0.46    & 0.41    & 0.43   & 0.43   & 0.31   & 0.46    & 0.42   & 0.447   & 0.43   & 0.35   & 0.45    & 0.43   & 0.44   & 0.43   & 0.36   \\ 
Annualized Turnover   & 8.68    & 8.39    & 8.49   & 8.36   & 7.92   & 8.69    & 8.48    & 8.56   & 8.49   & 8.04   & 8.73    & 8.54   & 8.592   & 8.56   & 7.99   & 8.67    & 8.57   & 8.51   & 8.54   & 7.86   \\ \hline
\end{tabular}}

%% file: portfolio_figures/tse_30_ts.tex
\fontsize{6}{9}\selectfont{
\begin{tabular}{r|ccccc|ccccc|ccccc|ccccc}
\hline
JPX Prime 150 | 30 Stocks & \multicolumn{5}{c|}{Target Volatility (24\%)} & \multicolumn{5}{c|}{Target Volatility (27\%)} & \multicolumn{5}{c|}{Target Volatility (30\%)} & \multicolumn{5}{c}{Target Volatility (33\%)} \\ \hline
Covariance Method         & SS      & HC      & SM     & GS     & TS     & SS      & HC      & SM     & GS     & TS     & SS      & HC      & SM     & GS     & TS     & SS      & HC     & SM     & GS     & TS     \\ \hline
Arithmetic Return (\%)     & 20.12   & 19.02   & 18.61  & 17.80  & 14.60  & 20.85   & 19.78   & 20.02  & 19.20  & 15.67  & 21.70   & 20.65   & 21.40  & 20.51  & 16.83  & 22.47   & 21.70  & 22.50  & 21.96  & 18.45  \\ 
Geometric Return (\%)      & 18.63   & 17.72   & 17.24  & 16.62  & 12.96  & 19.24   & 18.31   & 18.40  & 17.78  & 14.25  & 19.96   & 19.01   & 19.58  & 18.90  & 15.19  & 20.62   & 19.90  & 20.57  & 20.14  & 16.79  \\ 
Cumulative Return (\%)     & 71.15   & 66.53   & 64.60  & 61.55  & 44.16  & 74.32   & 69.56   & 70.25  & 67.23  & 49.12  & 77.95   & 73.23   & 76.27  & 72.90  & 52.85  & 81.67   & 77.87  & 81.37  & 79.30  & 59.32  \\ 
Annualized SD (\%)         & 26.83   & 26.16   & 25.56  & 25.71  & 27.19  & 28.51   & 27.88   & 27.74  & 27.74  & 28.39  & 29.99   & 29.59   & 29.83  & 29.69  & 29.62  & 31.33   & 31.20  & 31.59  & 31.46  & 30.66  \\ 
Annualized Skewness       & 0.28    & 0.16    & 0.17   & 0.13   & 0.11   & 0.32    & 0.17    & 0.21   & 0.18   & 0.15   & 0.34    & 0.19    & 0.23   & 0.21   & 0.16   & 0.32    & 0.19   & 0.23   & 0.20   & 0.13   \\ 
Annualized Kurtosis       & 3.37    & 2.94    & 2.99   & 2.92   & 2.69   & 3.43    & 3.02    & 3.08   & 3.01   & 2.67   & 3.49    & 3.06    & 3.17   & 3.09   & 2.73   & 3.51    & 3.12   & 3.22   & 3.16   & 2.72   \\ 
Maximum Drawdown (\%)      & -19.17  & -19.63  & -19.52 & -19.44 & -21.87 & -20.37  & -20.89  & -20.85 & -20.62 & -22.89 & -21.16  & -22.09  & -22.15 & -21.96 & -22.57 & -22.19  & -23.28 & -23.29 & -23.14 & -24.30 \\ 
Monthly 95\% VaR (\%)       & -8.52   & -8.89   & -8.49  & -8.82  & -9.86  & -9.00   & -9.46   & -9.20  & -9.36  & -9.92  & -9.41   & -9.95   & -9.79  & -9.90  & -10.29 & -9.79   & -10.46 & -10.31 & -10.40 & -10.75 \\ 
Sharpe Ratio              & 0.96    & 0.92    & 0.91   & 0.86   & 0.63   & 0.93    & 0.89    & 0.90   & 0.86   & 0.65   & 0.92    & 0.87    & 0.90   & 0.87   & 0.67   & 0.92    & 0.87   & 0.90   & 0.88   & 0.71   \\ 
Annualized Turnover       & 8.81    & 8.38    & 8.50   & 8.54   & 8.21   & 8.83    & 8.52    & 8.57   & 8.62   & 8.15   & 8.81    & 8.59    & 8.56   & 8.59   & 8.29   & 8.85    & 8.58   & 8.54   & 8.54   & 8.19   \\ \hline
\end{tabular}}

%% file: content/chap5_con.tex
\section{Conclusion}
In this paper, we examine \texttt{SimStock}, a novel temporal self-supervised learning framework that aims to learn robust and informative representations of financial time series data. By incorporating techniques from SSL and temporal domain generalization, \texttt{SimStock} captures the complex relationships between different financial assets while accounting for temporal shifts in the data distribution. The proposed dimension corruption method integrates temporal patterns into the corruption process, enabling \texttt{SimStock} to learn representations that are robust to noise and non-stationarity in the data. 

We conduct extensive experiments on four real-world benchmarks with thousands of stocks to demonstrate the effectiveness of \texttt{SimStock} in finding similar stocks. Our results show that \texttt{SimStock} consistently outperforms existing methods in both same exchange and different exchanges scenarios, achieving state-of-the-art performance in terms of accuracy and robustness. A qualitative analysis further highlights the ability of \texttt{SimStock} to identify stocks with similar fundamental characteristics and industry-specific similarities, simplifying the process of screening potential investment opportunities.

The practical utility of \texttt{SimStock} is demonstrated through various financial applications, including pairs trading, index tracking, and portfolio optimization. In the pairs trading experiment, using the similar stocks identified by \texttt{SimStock} leads to superior profitability compared to traditional methods. For index tracking, \texttt{SimStock} exhibits lower tracking errors and better alignment with the underlying themes of thematic ETFs, particularly for ARKK, SKYY, and BOTZ. Finally, in portfolio optimization, the \texttt{SimStock} embedding approach slightly outperforms traditional covariance estimators, achieving higher risk-adjusted returns across different portfolio sizes and risk levels.

However, our current analysis does not examine the stability of the embedding space over time. While recent generative models \citep{kim2024topp} and large language models (LLMs) \citep{rho2024language, nie2024survey} have investigated such issues, the case of continuous time series remains unexplored. To address this gap, future research could focus on a theoretical analysis of the embedding space in the context of financial time series. This analysis could potentially incorporate notions of regime change \citep{wood2023few} to better understand how the embedding space evolves over time and adapts to different market conditions. Also, this study primarily focused on time series data, utilizing sector information as the only static data input. Future research could expand on this approach by incorporating additional textual data, such as firm descriptions, news articles, and financial statements, to develop a more comprehensive representation model for stock data. This expanded approach has the potential to provide a deeper understanding of the complex dynamics between stocks.



%% file: Appendix/Appendix_A.tex
\section*{Appendix}
\section*{Appendix A: Ablation sutdy} 

In this appendix, we present an ablation study on the impact of different values of the mixing parameter $\lambda$ on the model's performance. We evaluate the model using the correlation coefficient and dynamic time warping (DTW) distance across four stock market indices (US, SSE, SZSE, and TSE). The results are presented in \cref{corr_lambda} and \cref{dtw_lambda}.

\cref{corr_lambda} shows that the highest correlation coefficients for each index are obtained with $\lambda = 0.7$, indicating the best performance in terms of the correlation between the model's predictions and the actual stock market data. Similarly, \cref{dtw_lambda} demonstrates that the lowest DTW distances for each index are achieved with $\lambda = 0.7$, suggesting the best performance in terms of the similarity between the model's predictions and the actual stock market data.

\setcounter{table}{0}
\renewcommand{\thetable}{A.\arabic{table}}
\begin{table*}[!htbp]
\centering
    \input{Appendix_Table/lambda_corr}
\caption{Correlation coefficient results for different values of $\lambda$ across four stock market indices. The best result for each index is in bold, while the second best is underlined.}
\label{corr_lambda}
\end{table*}

\setcounter{table}{1}
\begin{table*}[!htbp]
\centering
    \input{Appendix_Table/lambda_dtw}
\caption{DTW distance results for different values of $\lambda$ across four stock market indices. The best (smallest) result for each index is in bold, while the second best is underlined.}
\label{dtw_lambda}
\end{table*}

%% file: Appendix_Table/lambda_corr.tex
\useunder{\uline}{\ul}{}
\fontsize{6}{8}\selectfont{
\begin{tabular}{ccccccccccccc}
\cline{1-6} \cline{8-13}
\textbf{US}     & \textbf{TOP@1}  & \textbf{TOP@3}  & \textbf{TOP@5}  & \textbf{TOP@7}  & \textbf{TOP@9}  &  & \textbf{SSE} & \textbf{TOP@1}  & \textbf{TOP@3}  & \textbf{TOP@5}  & \textbf{TOP@7}  & \textbf{TOP@9}  \\ \cline{1-6} \cline{8-13} 
\textbf{$\lambda$=0.1}       & 0.7576          & 0.6179          & 0.5601          & 0.5267          & 0.5058          &  & \textbf{$\lambda$=0.1}     & 0.7193          & 0.5483          & 0.4811          & 0.4396          & 0.4137          \\
\textbf{$\lambda$=0.3}       & 0.7723          & 0.6212          & 0.5638          & 0.5283          & 0.5052          &  & \textbf{$\lambda$=0.3}     & 0.7618          & 0.6048          & 0.5502          & 0.5150          & 0.4890          \\
\textbf{$\lambda$=0.5}       & 0.7876          & 0.6580          & 0.6056          & 0.5744          & 0.5542          &  & \textbf{$\lambda$=0.5}     & 0.8029          & 0.6513          & 0.5894          & 0.5559          & 0.5314          \\
\textbf{$\lambda$=0.7}       & \textbf{0.8128} & \textbf{0.6905} & \textbf{0.6414} & \textbf{0.6119} & \textbf{0.5923} &  & \textbf{$\lambda$=0.7}     & \textbf{0.8088} & \textbf{0.6691} & \textbf{0.6207} & \textbf{0.5879} & \textbf{0.5633} \\
\textbf{$\lambda$=0.9}       & {\ul 0.8104}    & {\ul 0.6870}    & {\ul 0.6374}    & {\ul 0.6088}    & {\ul 0.5892}    &  & \textbf{$\lambda$=0.9}     & {\ul 0.8054}    & {\ul 0.6759}    & {\ul 0.6222}    & {\ul 0.5883}    & {\ul 0.5642}    \\ \cline{1-6} \cline{8-13} 
                      &                 &                 &                 &                 &                 &  &                     &                 &                 &                 &                 &                 \\ \cline{1-6} \cline{8-13} 
\textbf{SZSE} & \textbf{TOP@1}  & \textbf{TOP@3}  & \textbf{TOP@5}  & \textbf{TOP@7}  & \textbf{TOP@9}  &  & \textbf{TSE} & \textbf{TOP@1}  & \textbf{TOP@3}  & \textbf{TOP@5}  & \textbf{TOP@7}  & \textbf{TOP@9}  \\ \cline{1-6} \cline{8-13} 
\textbf{$\lambda$=0.1}       & 0.7548          & 0.6286          & 0.5707          & 0.5404          & 0.5173          &  & \textbf{$\lambda$=0.1}     & 0.6337          & 0.4802          & 0.4127          & 0.3805          & 0.3549          \\
\textbf{$\lambda$=0.3}       & 0.7252          & 0.5838          & 0.5222          & 0.4872          & 0.4624          &  & \textbf{$\lambda$=0.3}     & 0.6482          & 0.4871          & 0.4258          & 0.3926          & 0.3711          \\
\textbf{$\lambda$=0.5}       & 0.6888          & 0.5463          & 0.4880          & 0.4506          & 0.4281          &  & \textbf{$\lambda$=0.5}     & 0.6772          & 0.5203          & 0.4644          & 0.4347          & 0.4161          \\
\textbf{$\lambda$=0.7}       & \textbf{0.7951} & \textbf{0.6768} & \textbf{0.6298} & \textbf{0.5961} & \textbf{0.5737} &  & \textbf{$\lambda$=0.7}     & {\ul 0.7668}    & {\ul 0.6483}    & {\ul 0.5964}    & {\ul 0.5651}    & {\ul 0.5476}    \\
\textbf{$\lambda$=0.9}       & {\ul 0.7873}    & {\ul 0.6713}    & {\ul 0.6180}     & {\ul 0.5873}    & {\ul 0.5644}    &  & \textbf{$\lambda$=0.9}     & \textbf{0.7811} & \textbf{0.6644} & \textbf{0.6143} & \textbf{0.5862} & \textbf{0.5686} \\ \cline{1-6} \cline{8-13} 
\end{tabular}}

%% file: Appendix_Table/lambda_dtw.tex
\useunder{\uline}{\ul}{}
\fontsize{6}{8}\selectfont{
\begin{tabular}{ccccccccccccc}
\cline{1-6} \cline{8-13}
\textbf{US}     & \textbf{TOP@1}   & \textbf{TOP@3}   & \textbf{TOP@5}   & \textbf{TOP@7}   & \textbf{TOP@9}   &  & \textbf{SSE} & \textbf{TOP@1}   & \textbf{TOP@3}   & \textbf{TOP@5}   & \textbf{TOP@7}   & \textbf{TOP@9}   \\ \cline{1-6} \cline{8-13} 
\textbf{$\lambda$=0.1}       & 14.3211          & 22.0003          & 24.9902          & 26.6191          & 27.6328          &  & \textbf{$\lambda$=0.1}     & 13.9537          & 21.7294          & 24.7087          & 26.5307          & 27.7407          \\
\textbf{$\lambda$=0.3}       & 13.6924          & 21.9408          & 24.8712          & 26.6015          & 27.6861          &  & \textbf{$\lambda$=0.3}     & 12.6262          & 20.9044          & 23.637           & 25.2049          & 26.3574          \\
\textbf{$\lambda$=0.5}       & 13.1073          & 20.4898          & 23.3020          & 24.8633          & 25.8934          &  & \textbf{$\lambda$=0.5}     & 11.0089          & 19.5019          & 22.5511          & 24.0833          & 25.0604          \\
\textbf{$\lambda$=0.7}       & \textbf{12.2124} & \textbf{19.4629} & \textbf{22.1019} & \textbf{23.5986} & \textbf{24.6195} &  & \textbf{$\lambda$=0.7}     & \textbf{11.2685} & {\ul 19.009}     & \textbf{21.5380} & \textbf{23.1172} & \textbf{24.2082} \\
\textbf{$\lambda$=0.9}       & {\ul 12.3086}    & {\ul 19.5703}    & {\ul 22.2254}    & {\ul 23.6872}    & {\ul 24.6773}    &  & \textbf{$\lambda$=0.9}     & {\ul 11.5052}    & \textbf{18.752}  & {\ul 21.5877}    & {\ul 23.1732}    & {\ul 24.2119}    \\ \cline{1-6} \cline{8-13} 
                      &                  &                  &                  &                  &                  &  &                     &                  &                  &                  &                  &                  \\ \cline{1-6} \cline{8-13} 
\textbf{SZSE} & \textbf{TOP@1}   & \textbf{TOP@3}   & \textbf{TOP@5}   & \textbf{TOP@7}   & \textbf{TOP@9}   &  & \textbf{TSE} & \textbf{TOP@1}   & \textbf{TOP@3}   & \textbf{TOP@5}   & \textbf{TOP@7}   & \textbf{TOP@9}   \\ \cline{1-6} \cline{8-13} 
\textbf{$\lambda$=0.1}       & 13.2464          & 19.705           & 22.2434          & 23.6564          & 24.7304          &  & \textbf{$\lambda$=0.1}     & 19.4737          & 27.5245          & 31.0088          & 32.5306          & 33.7523          \\
\textbf{$\lambda$=0.3}       & 14.1484          & 21.0873          & 23.8357          & 25.2652          & 26.2770          &  & \textbf{$\lambda$=0.3}     & 18.5924          & 26.811           & 29.9655          & 31.7022          & 32.7707          \\
\textbf{$\lambda$=0.5}       & 15.0496          & 21.9856          & 24.4632          & 26.0794          & 27.0110          &  & \textbf{$\lambda$=0.5}     & 17.7066          & 26.0258          & 28.8961          & 30.4193          & 31.3618          \\
\textbf{$\lambda$=0.7}       & \textbf{12.0045} & \textbf{18.4121} & \textbf{20.7238} & \textbf{22.1577} & \textbf{23.1457} &  & \textbf{$\lambda$=0.7}     & {\ul 14.6179}    & {\ul 21.4850}    & {\ul 24.2952}    & {\ul 25.8992}    & {\ul 26.7530}    \\
\textbf{$\lambda$=0.9}       & {\ul 12.399}     & {\ul 18.4378}    & {\ul 21.0680}    & {\ul 22.4532}    & {\ul 23.3963}    &  & \textbf{$\lambda$=0.9}     & \textbf{14.1545} & \textbf{21.0788} & \textbf{23.8008} & \textbf{25.3369} & \textbf{26.2443} \\ \cline{1-6} \cline{8-13} 
\end{tabular}}

%% file: Appendix/Appendix_B.tex
\section*{Appendix B: Pairs trading results for query stock and TOP@1 similar Stock} 

In this section, we present the pairs trading results obtained by applying the trading strategy to the query stock and its top@1 similar stock identified by SimStock and the benchmark models (TS2VEC, Corr1, Corr2, and Coint).

\cref{table_top1_pair_} shows the terminal wealth and maximum drawdown (MDD) achieved for each query stock and its top@1 similar stock. The best-performing model for each query stock is highlighted in bold, while the second-best model is underlined.

SimStock achieves the highest terminal wealth for 3 out of the 12 query stocks (AAPL, CMG, and BA). Coint outperforms for 5 query stocks (V, XOM, AMZN, META and MA), while Corr1 outperforms for 2 query stocks (CVS and WFC for Corr1) and Corr2 achieves the highest wealth for PFE. TS2VEC achieves the highest wealth for MSFT. For the query stocks V and MA, SimStock, TS2VEC, Corr1, and Corr2 yield the same results.

In terms of maximum drawdown, SimStock achieves the lowest MDD for 8 out of the 12 query stocks (CMG, MSFT, WFC, V, BA, META, MA and CVS). TS2VEC exhibits the lowest MDD for 4 query stocks (AAPL, V, AMZN and MA), while Corr1 outperforms for 2 query stocks (V and MA). Corr2 achieves the lowest MDD for WFC, V and MA, and Coint outperforms for XOM.

TS2VEC fails to generate buy/sell signals for the query stock PFE and WFC, resulting in NaN values for wealth and MDD. Similarly, Corr1 fails to generate signals for one of the occurrences of the MSFT query stock. These instances show limitations in the consistency of the benchmark models in identifying tradable opportunities.

SimStock achieves competitive performance in terms of terminal wealth and MDD for a significant number of query stocks, indicating its effectiveness in capturing meaningful similarities that translate into profitable trading strategies.

\setcounter{table}{0}
\renewcommand{\thetable}{B.\arabic{table}}
\begin{table*}[!htbp]
\centering

    \input{Appendix_Table/Pairs_trading_wealth_mdd_top1}

\caption{ Pairs trading results for the query stock and its top@1 similar stock identified by SimStock, TS2VEC, Corr1, Corr2 and Coint. The table presents the terminal wealth and maximum drawdown (MDD) achieved for each query stock and method combination. The best-performing method for each query stock is highlighted in bold, while the second-best method is underlined. NaN values indicate instances where the method failed to generate buy/sell signals.}
\label{table_top1_pair_}
\end{table*}

%% file: Appendix_Table/Pairs_trading_wealth_mdd_top1.tex
\useunder{\uline}{\ul}{}

\fontsize{7}{8}\selectfont{
\begin{tabular}{cccccc|ccccc}
\hline
\multirow{2}{*}{\textbf{\begin{tabular}[c]{@{}c@{}}Query\\ Stock\end{tabular}}} & \multicolumn{5}{c|}{\textbf{Wealth}}                                                          & \multicolumn{5}{c}{\textbf{Maximum Drawdown (\%)}}                                     \\
                                                                                & \textbf{SimStock} & \textbf{TS2VEC}  & \textbf{Corr1}   & \textbf{Corr2}   & \textbf{Coint}  & \textbf{SimStock} & \textbf{TS2VEC} & \textbf{Corr1} & \textbf{Corr2} & \textbf{Coint}  \\ \hline
AAPL                                                                            & \textbf{1506.88}  & 1302.67          & -797.21          & 766.51           & NaN             & {\ul -2.20}       & \textbf{-2.11}  & -8.55          & -5.70          & NaN             \\
CMG                                                                             & \textbf{372.84}   & -541.61          & -1203.23         & -1502.99         & -59.03          & \textbf{-5.18}& {\ul -8.53}& -13.60           & -21.10         & -8.63           \\
MSFT                                                                            & 802.21& \textbf{1275.91} & 847.28& 509.61           & {\ul 1209.71}& \textbf{-3.82}    & {\ul -5.41}     & -7.59          & -5.89          & -10.76          \\
WFC                                                                             & {\ul 751.44}      & NaN              & \textbf{1236.13} & {\ul 751.44}     & -2725.64        & \textbf{-0.04}    & NaN     & -6.82          & \textbf{-0.04} &    -31.51          \\
V                                                                               & {\ul 240.25}      & {\ul 240.25}     & {\ul 240.25}     & {\ul 240.25}     & \textbf{691.54} & \textbf{0.00}     & \textbf{0.00}   & \textbf{0.00}  & \textbf{0.00}  & {\ul -2.77}     \\
XOM                                                                             & {\ul 113.28}      & {\ul 113.28}     & -192.12           & -1147.63          & \textbf{1102.32}& {\ul -3.07}& {\ul -3.07}&  -5.47& -14.73         & \textbf{-1.39}  \\
PFE                                                                             & -309.74           & NaN              & 296.56           & \textbf{1499.64} & {\ul 442.02}    & -7.48& NaN             & -9.35& \textbf{-5.61}& {\ul -7.02}\\
AMZN                                                                            & 372.84            & {\ul 1224.03}    & -1741.79         & 660.63           & \textbf{2622.29}& {\ul -5.18}       & \textbf{-2.12}  & -22.62         & -5.58         & -27.15          \\
BA                                                                              & \textbf{1320.62}  & 117.68            & -920.97          & {\ul 891.91}      & 93.45           & \textbf{-3.66}    & {\ul -6.70}     & -11.95         & -9.10          & -22.77          \\
META                                                                            & 532.60            & -2046.25         & {\ul 1617.05}    & 1060.88          & \textbf{1788.35}& \textbf{-1.29}    & -17.95          & -16.03            & {\ul -12.04}    & -26.08          \\
MA                                                                              & {\ul 197.88}      & {\ul 197.88}     & {\ul 197.88}     & {\ul 197.88}     & \textbf{405.87} & \textbf{0.00}     & \textbf{0.00}   & \textbf{0.00}  & \textbf{0.00}  & {\ul -6.13}     \\
CVS                                                                             & {\ul 727.14}      & -1421.22         & \textbf{1486.60} & 438.36            & -1638.88        & \textbf{-1.38}    & -18.53& -9.88         & {\ul -4.44}& -39.44          \\ \hline
\end{tabular}}

%% file: Appendix/Appendix_C.tex
\section*{Appendix C: Application to index tracking of thematic ETFs}

\subsection*{Qualitative evaluation}  In this subsection, to provide a comprehensive qualitative evaluation, we examine the overlapping and non-overlapping tickers for each model and ETF combination, along with their respective industries, as illustrated in \cref{chap4_index_overlapping}. This analysis enables us to assess the extent to which the models capture the underlying themes of the ETFs. The experimental setup involves selecting 10 stocks to track each ETF on the US exchange, as shown in \cref{chap4_cumulative_retun_curves_for_index}.

For the ARKK ETF, which focuses on disruptive innovation, \texttt{SimStock} identifies several relevant stocks across various sectors. The overlapping tickers include ROKU (Roku Inc., a company that provides streaming media devices and services), SHOP (Shopify Inc., an e-commerce platform provider), and SQ (Square Inc., a financial technology company). The non-overlapping tickers selected by \texttt{SimStock} also align with the disruptive innovation theme, including companies from 3D printing (MTLS: Materialise NV; DDD: 3D Systems Corp.), biotechnology (EDIT: Editas Medicine Inc.), and e-commerce (MELI: MercadoLibre Inc.). In contrast, the baseline models (TS2VEC, and Corr1) either have no overlapping tickers or include stocks from less relevant industries, such as insurance (WTM: White Mountains Insurance Group Ltd.), sports (MANU: Manchester United plc), and real estate (FLJ: FLJ Group Ltd.).

For the SKYY ETF, which tracks the cloud computing industry, \texttt{SimStock} identifies NET (Cloudflare Inc., a content delivery network and DDoS mitigation company), cloud storage (DBX: Dropbox Inc.), financial software (INTU: Intuit Inc.), Elasticsearch (ESTC: Elastic N.V.) and PAYC (Paycom Software Inc., a provider of human capital management software) as overlapping tickers. The non-overlapping stocks selected by \texttt{SimStock} also show a strong connection to the cloud computing theme such as software compnay (TYL: Tyler Technologies Inc.; GWRE: Guidewire Software, Inc.; GLOB: Globant S.A.). The baseline models, however, include stocks from a wider range of industries that are less directly related to cloud computing, such as automotive (F: Ford Motor Co.), retail (W: Wayfair Inc.), and real estate (LAMR: Lamar Advertising Co.), biotechnology (LIFE: aTyr Pharma Inc.; VBIV: VBI Vaccines Inc.), and agriculture (VFF: Village Farms International Inc.).

For the BOTZ ETF, which focuses on robotics and artificial intelligence, \texttt{SimStock} identifies NVDA (NVIDIA Corp.) as an overlapping ticker. While the non-overlapping stocks selected by \texttt{SimStock} include some companies from other sectors, such as asset management (AMG: Affiliated Managers Group Inc.; GAM: General American Investors Co. Inc.) and building materials (MLM: Martin Marietta Materials Inc.), the model also identifies several companies from the semiconductor industry (QRVO: Qorvo Inc.; ASML: ASML Holding N.V.; LSCC: Lattice Semiconductor Corp.), which is a key enabler of robotics and AI. The baseline models, TS2VEC and Corr1, include stocks from a diverse range of industries that are less directly related to the robotics and AI theme, such as animal healthcare (IDXX: IDEXX Laboratories Inc.), insurance (AEG: Aegon N.V.), and electric vehicles (HYLN: Hyliion Holdings Corp.).

For the LIT ETF, which tracks the lithium and battery tech industry, \texttt{SimStock} identifies three overlapping tickers: ALB (Albemarle Corp.), ALTM (Altus Midstream Co.) and LAAC (Lithium Americas (Argentina) Corp), both of which are involved in lithium mining. However, the non-overlapping stocks selected by \texttt{SimStock} come from various other industries, such as asset management (AB: AllianceBernstein Holding L.P.; APAM: Artisan Partners Asset Management Inc.; BLK: BlackRock Inc.; KKR: KKR \& Co. Inc.) and software (BB: BlackBerry Ltd.), which are less directly related to the lithium and battery tech theme. The baseline models also struggle to identify relevant stocks for this ETF, with no overlapping tickers and selections from industries such as insurance (WTM: White Mountains Insurance Group Ltd.; ERIE: Erie Indemnity Co.; ASR: Grupo Aeroportuario del Sureste, S. A. B. de C. V.), automotive (F: Ford Motor Co.; SUZ: Suzuki Motor Corp.), and engineering \& construction (MYRG: MYR Group Inc.; FRHC: Freedom Holding Corp.).

This qualitative analysis represent that \texttt{SimStock}'s top@10 stocks for each ETF generally align better with the respective themes compared to the baseline models, particularly for the ARKK, SKYY, and BOTZ ETFs.  Howevere, the baseline models, TS2VEC and Corr1, include stocks from a diverse range of industries that are less directly related to the specific themes of the ETFs. Corr2 shows some improvement over the other baselines, with a few overlapping tickers for ARKK and BOTZ, but it still lags behind \texttt{SimStock} in terms of capturing the underlying themes of the ETFs.

\setcounter{figure}{0}
\renewcommand{\thefigure}{C.\arabic{figure}}
\begin{figure*}[!htbp]
\centering
\includegraphics[width=\textwidth]{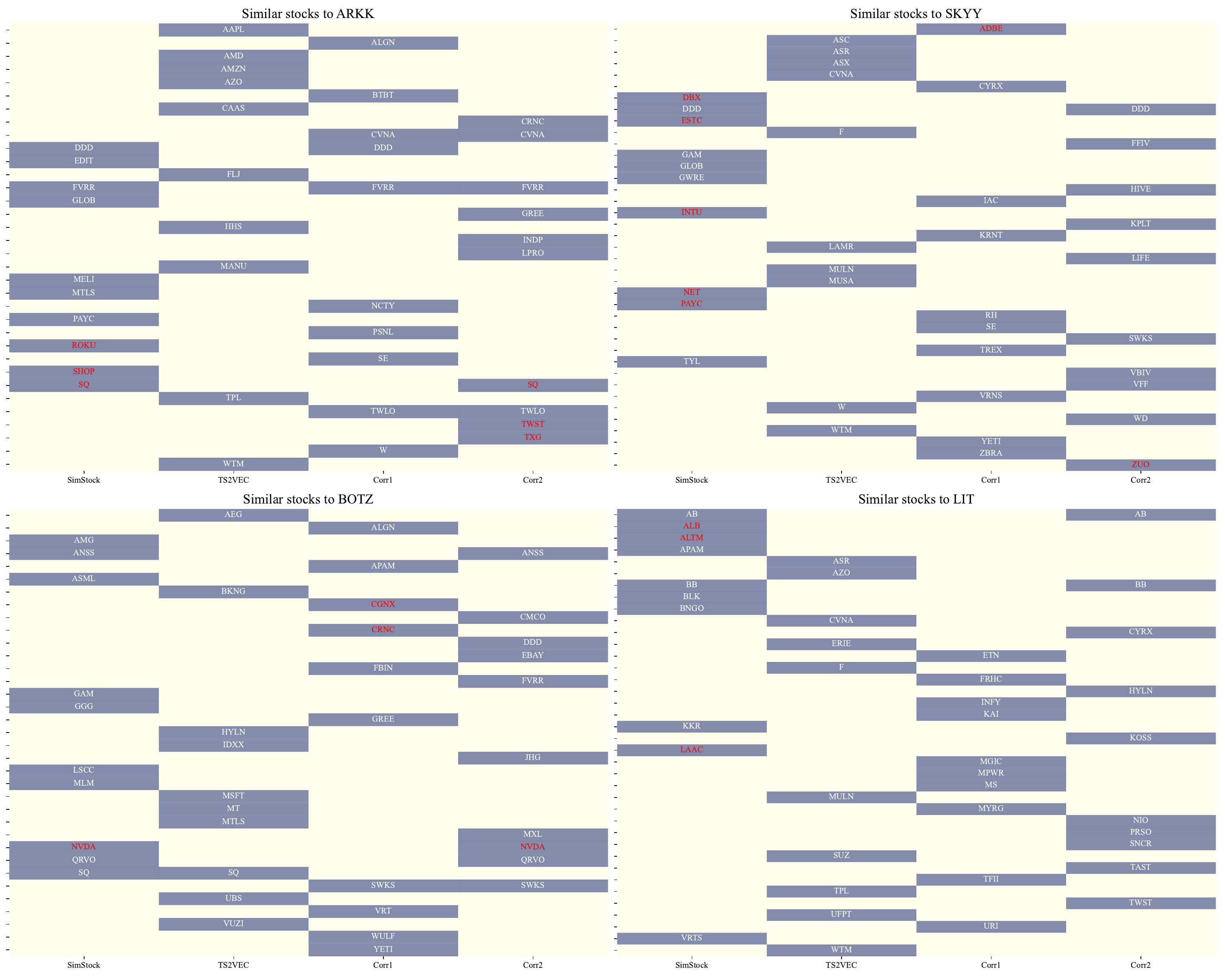}
\caption{Top@10 stocks selected by each model to track the four thematic ETFs (ARKK, SKYY, BOTZ, and LIT) on the US exchange. The overlapping tickers between the model selections and the actual ETF holdings are shown in red, while the non-overlapping tickers are shown in white. SimStock demonstrates a higher number of overlapping tickers with the ETF holdings compared to the baseline models, particularly for ARKK, SKYY, and BOTZ, indicating its ability to identify stocks that align with the underlying themes of these ETFs.}
\label{chap4_index_overlapping}
\end{figure*}

%% file: Appendix/Appendix_D.tex
\section*{Appendix D: Tracking Error Volatility (TEV) Results}

In this section, we present the Tracking Error Volatility (TEV) results to complement the Tracking Error (TE) analysis discussed in the main text. TEV measures the volatility of the difference in returns between the tracking portfolio and the target ETF, providing information into the stability and consistency of the tracking performance.

We computed the TEV for each combination of the four thematic ETFs (ARK Innovation ETF (ARKK), First Trust Cloud Computing ETF (SKYY), Global X Robotics \& AI ETF (BOTZ), and Global X Lithium \& Battery Tech ETF (LIT)) and the four exchanges (US, SSE, SZSE, and TSE). The TEV values were calculated for the top k similar stocks (k = 10, 15, 20, 25, 30, 35) identified by \texttt{SimStock} and the baseline methods (Corr1, Corr2, TS2VEC).

\cref{Table:B.1} present the TEV results for ARKK, SKYY, BOTZ, and LIT, respectively. In each table, the best performing method for each k and exchange is highlighted in bold, while the second best is in underline. The TEV results are consistent with the findings from the TE analysis in the main text. \texttt{SimStock} achieves the lowest (bold) or second lowest (underline) TEV values for the majority of k values and exchanges, particularly for ARKK, SKYY, and BOTZ. This indicates that the similar stocks identified by \texttt{SimStock} not only closely track the performance of these ETFs but also exhibit lower volatility in the tracking error, resulting in more stable tracking portfolios.

The performance of \texttt{SimStock} in terms of both TE and TEV demonstrates its effectiveness in identifying stocks that can accurately and consistently track the performance of thematic ETFs, especially for ARKK, SKYY, and BOTZ. The lower TEV values achieved by \texttt{SimStock} suggest that its tracking portfolios are less prone to large deviations from the target ETF returns, providing investors with more reliable and stable tracking options.

\setcounter{table}{0}
\renewcommand{\thetable}{D.\arabic{table}}
\begin{table*}[!htbp]

    \input{Appendix_Table/index_tracking_ARKK_TEV}
    
    \vspace{0.5cm}
    
    \input{Appendix_Table/index_tracking_SKYY_TEV}

    \vspace{0.5cm}
    
    \input{Appendix_Table/index_tracking_BOTZ_TEV}
    
    \vspace{0.5cm}
    
    \input{Appendix_Table/index_tracking_LIT_TEV}

\caption{Tracking errors volatility(TEV) of \texttt{SimStock} and baseline methods for tracking the performance of four thematic ETFs (ARKK, SKYY, BOTZ, and LIT) using top k similar stocks from the US, SSE, SZSE, and TSE exchanges. The best performing method for each k and exchange is highlighted in black, while the second best is in underscore.}

\label{Table:B.1}

\end{table*}

\clearpage

%% file: Appendix_Table/index_tracking_ARKK_TEV.tex
\useunder{\uline}{\ul}{}
\centering
\fontsize{7}{8}\selectfont{
\begin{tabular}{cccccccc}
\hline
\multicolumn{1}{l}{{\color[HTML]{000000} }}                                    & {\color[HTML]{000000} }                                   & \multicolumn{6}{c}{\textbf{ARK Innovation ETF (ARKK)}}                                                                                                                                                                                                                                                    \\
\multicolumn{1}{l}{\multirow{-2}{*}{{\color[HTML]{000000} \textbf{Exchange}}}} & \multirow{-2}{*}{{\color[HTML]{000000} \textbf{Methods}}} & {\color[HTML]{000000} \textbf{\textbf{TOP@10}}} & {\color[HTML]{000000} \textbf{\textbf{TOP@15}}} & {\color[HTML]{000000} \textbf{\textbf{TOP@20}}} & {\color[HTML]{000000} \textbf{\textbf{TOP@25}}} & {\color[HTML]{000000} \textbf{\textbf{TOP@30}}} & {\color[HTML]{000000} \textbf{\textbf{TOP@35}}} \\ \hline
                                                                               & Corr1                                                     & {\color[HTML]{000000} \textbf{0.1524}}          & {\color[HTML]{000000} {\ul 0.0893}}             & {\color[HTML]{000000} \textbf{0.0720}}          & {\color[HTML]{000000} \textbf{0.0630}}          & {\color[HTML]{000000} \textbf{0.0540}}          & {\color[HTML]{000000} \textbf{0.0474}}          \\
                                                                               & Corr2                                                     & {\color[HTML]{000000} {\ul 0.1669}}             & 0.1141                                          & 0.1018                                          & {\color[HTML]{000000} {\ul 0.1061}}             & {\color[HTML]{000000} {\ul 0.0894}}             & {\color[HTML]{000000} {\ul 0.1042}}             \\
                                                                               & TS2VEC                                                    & 0.1884                                          & 0.2225                                          & 0.1852                                          & 0.1701                                          & 0.1460                                          & 0.1292                                          \\
\multirow{-4}{*}{US}                                                           & SimStock                                                  & 0.1934                                          & {\color[HTML]{000000} \textbf{0.0764}}          & {\color[HTML]{000000} {\ul 0.0975}}             & 0.1785                                          & 0.1822                                          & 0.2414                                          \\ \hline
                                                                               & Corr1                                                     & 0.1747                                          & 0.1864                                          & 0.1820                                          & 0.1824                                          & 0.1833                                          & 0.1839                                          \\
                                                                               & Corr2                                                     & {\color[HTML]{000000} {\ul 0.1513}}             & {\color[HTML]{000000} \textbf{0.1414}}          & {\color[HTML]{000000} {\ul 0.1417}}             & {\color[HTML]{000000} {\ul 0.1435}}             & {\color[HTML]{000000} {\ul 0.1394}}             & {\color[HTML]{000000} \textbf{0.1406}}          \\
                                                                               & TS2VEC                                                    & 0.1633                                          & 0.1618                                          & 0.1527                                          & 0.1629                                          & 0.1637                                          & 0.1655                                          \\
\multirow{-4}{*}{SSE}                                                          & SimStock                                                  & {\color[HTML]{000000} \textbf{0.1414}}          & {\color[HTML]{000000} {\ul 0.1419}}             & {\color[HTML]{000000} \textbf{0.1352}}          & {\color[HTML]{000000} \textbf{0.1366}}          & {\color[HTML]{000000} \textbf{0.1358}}          & {\color[HTML]{000000} {\ul 0.1429}}             \\ \hline
                                                                               & Corr1                                                     & 0.1747                                          & 0.1824                                          & 0.1709                                          & 0.1845                                          & 0.1800                                          & 0.1815                                          \\
                                                                               & Corr2                                                     & {\color[HTML]{000000} \textbf{0.1399}}          & {\color[HTML]{000000} \textbf{0.1376}}          & {\color[HTML]{000000} \textbf{0.1355}}          & {\color[HTML]{000000} \textbf{0.1382}}          & {\color[HTML]{000000} \textbf{0.1394}}          & {\color[HTML]{000000} {\ul 0.1394}}             \\
                                                                               & TS2VEC                                                    & 0.1495                                          & 0.1456                                          & 0.1489                                          & 0.1463                                          & 0.1410                                          & 0.1415                                          \\
\multirow{-4}{*}{SZSE}                                                         & SimStock                                                  & {\color[HTML]{000000} {\ul 0.1436}}             & {\color[HTML]{000000} {\ul 0.1411}}             & {\color[HTML]{000000} {\ul 0.1410}}             & {\color[HTML]{000000} {\ul 0.1449}}             & {\color[HTML]{000000} {\ul 0.1397}}             & {\color[HTML]{000000} \textbf{0.1350}}          \\ \hline
                                                                               & Corr1                                                     & 0.1969                                          & 0.1759                                          & 0.1430                                          & 0.1388                                          & 0.1387                                          & 0.1358                                          \\
                                                                               & Corr2                                                     & {\color[HTML]{000000} {\ul 0.1165}}             & {\color[HTML]{000000} {\ul 0.1129}}             & {\color[HTML]{000000} {\ul 0.1213}}             & {\color[HTML]{000000} {\ul 0.1172}}             & {\color[HTML]{000000} {\ul 0.1149}}             & {\color[HTML]{000000} {\ul 0.1141}}             \\
                                                                               & TS2VEC                                                    & 0.1309                                          & 0.1335                                          & 0.1262                                          & 0.1279                                          & 0.1312                                          & 0.1251                                          \\
\multirow{-4}{*}{TSE}                                                          & SimStock                                                  & {\color[HTML]{000000} \textbf{0.1117}}          & {\color[HTML]{000000} \textbf{0.1116}}          & {\color[HTML]{000000} \textbf{0.1118}}          & {\color[HTML]{000000} \textbf{0.1125}}          & {\color[HTML]{000000} \textbf{0.1076}}          & {\color[HTML]{000000} \textbf{0.1074}}          \\ \hline
\end{tabular}}

%% file: Appendix_Table/index_tracking_SKYY_TEV.tex
\useunder{\uline}{\ul}{}
\centering
\fontsize{7}{8}\selectfont{
\begin{tabular}{cccccccc}
\hline
\multicolumn{1}{l}{}                                    &                                    & \multicolumn{6}{c}{\textbf{First Trust Cloud Computing ETF (SKYY)}}                                                                                                                                                                                 \\
\multicolumn{1}{l}{\multirow{-2}{*}{\textbf{Exchange}}} & \multirow{-2}{*}{\textbf{Methods}} & \textbf{\textbf{TOP@10}}               & \textbf{\textbf{TOP@15}}               & \textbf{\textbf{TOP@20}}               & \textbf{\textbf{TOP@25}}               & \textbf{\textbf{TOP@30}}               & \textbf{\textbf{TOP@35}}               \\ \hline
                                                        & Corr1                              & {\color[HTML]{000000} {\ul 0.1150}}    & {\color[HTML]{000000} {\ul 0.1406}}    & {\color[HTML]{000000} {\ul 0.1379}}    & {\color[HTML]{000000} {\ul 0.1245}}    & {\color[HTML]{000000} {\ul 0.1058}}    & {\color[HTML]{000000} {\ul 0.0993}}    \\
                                                        & Corr2                              & 0.2410                                 & 0.2298                                 & 0.2011                                 & 0.1916                                 & 0.1586                                 & 0.1455                                 \\
                                                        & TS2VEC                             & 0.1849                                 & 0.1655                                 & 0.1407                                 & 0.1397                                 & 0.1387                                 & 0.1330                                 \\
\multirow{-4}{*}{US}                                    & SimStock                           & {\color[HTML]{000000} \textbf{0.0616}} & {\color[HTML]{000000} \textbf{0.0364}} & {\color[HTML]{000000} \textbf{0.0561}} & {\color[HTML]{000000} \textbf{0.0653}} & {\color[HTML]{000000} \textbf{0.0541}} & {\color[HTML]{000000} \textbf{0.0521}} \\ \hline
                                                        & Corr1                              & 0.1699                                 & 0.1929                                 & 0.1993                                 & 0.2014                                 & 0.2035                                 & 0.2134                                 \\
                                                        & Corr2                              & {\color[HTML]{000000} {\ul 0.1490}}    & {\color[HTML]{000000} {\ul 0.1386}}    & {\color[HTML]{000000} {\ul 0.1303}}    & {\color[HTML]{000000} {\ul 0.1118}}    & {\color[HTML]{000000} {\ul 0.1133}}    & {\color[HTML]{000000} {\ul 0.1188}}    \\
                                                        & TS2VEC                             & 0.1897                                 & 0.1975                                 & 0.2151                                 & 0.2074                                 & 0.2013                                 & 0.1871                                 \\
\multirow{-4}{*}{SSE}                                   & SimStock                           & {\color[HTML]{000000} \textbf{0.1106}} & {\color[HTML]{000000} \textbf{0.0921}} & {\color[HTML]{000000} \textbf{0.1128}} & {\color[HTML]{000000} \textbf{0.1072}} & {\color[HTML]{000000} \textbf{0.1074}} & {\color[HTML]{000000} \textbf{0.1167}} \\ \hline
                                                        & Corr1                              & 0.2248                                 & 0.2183                                 & 0.2123                                 & 0.2156                                 & 0.2106                                 & 0.2042                                 \\
                                                        & Corr2                              & {\color[HTML]{000000} \textbf{0.1103}} & {\color[HTML]{000000} {\ul 0.1143}}    & {\color[HTML]{000000} {\ul 0.1164}}    & {\color[HTML]{000000} {\ul 0.1122}}    & {\color[HTML]{000000} {\ul 0.1212}}    & {\color[HTML]{000000} {\ul 0.1239}}    \\
                                                        & TS2VEC                             & 0.1642                                 & 0.1796                                 & 0.1684                                 & 0.1689                                 & 0.1702                                 & 0.1594                                 \\
\multirow{-4}{*}{SZSE}                                  & SimStock                           & {\color[HTML]{000000} \textbf{0.1103}} & {\color[HTML]{000000} \textbf{0.1130}} & {\color[HTML]{000000} \textbf{0.1157}} & {\color[HTML]{000000} \textbf{0.1105}} & {\color[HTML]{000000} \textbf{0.1047}} & {\color[HTML]{000000} \textbf{0.1020}} \\ \hline
                                                        & Corr1                              & 0.1308                                 & 0.1363                                 & 0.1239                                 & 0.1429                                 & 0.1450                                 & 0.1472                                 \\
                                                        & Corr2                              & {\color[HTML]{000000} \textbf{0.0723}} & {\color[HTML]{000000} {\ul 0.0846}}    & {\color[HTML]{000000} {\ul 0.0762}}    & {\color[HTML]{000000} {\ul 0.0871}}    & {\color[HTML]{000000} {\ul 0.0818}}    & {\color[HTML]{000000} {\ul 0.0810}}    \\
                                                        & TS2VEC                             & 0.1066                                 & 0.1142                                 & 0.1153                                 & 0.1136                                 & 0.1108                                 & 0.1215                                 \\
\multirow{-4}{*}{TSE}                                   & SimStock                           & {\color[HTML]{000000} {\ul 0.0842}}    & {\color[HTML]{000000} \textbf{0.0703}} & {\color[HTML]{000000} \textbf{0.0704}} & {\color[HTML]{000000} \textbf{0.0640}} & {\color[HTML]{000000} \textbf{0.0570}} & {\color[HTML]{000000} \textbf{0.0625}} \\ \hline
\end{tabular}}

%% file: Appendix_Table/index_tracking_BOTZ_TEV.tex
\useunder{\uline}{\ul}{}
\centering
\fontsize{7}{8}\selectfont{
\begin{tabular}{cccccccc}
\hline
\multicolumn{1}{l}{}                                    &                                    & \multicolumn{6}{c}{\textbf{Global X Robotics \& AI ETF (BOTZ)}}                                                                                                                                                                                      \\
\multicolumn{1}{l}{\multirow{-2}{*}{\textbf{Exchange}}} & \multirow{-2}{*}{\textbf{Methods}} & \textbf{\textbf{TOP@10}}               & \textbf{\textbf{TOP@15}}               & \textbf{\textbf{TOP@20}}               & \textbf{\textbf{TOP@25}}               & \textbf{\textbf{TOP@30}}               & \textbf{\textbf{TOP@35}}               \\ \hline
                                                        & Corr1                              & 0.1599                                 & 0.1030                                 & {\color[HTML]{000000} {\ul 0.0744}}    & {\color[HTML]{000000} {\ul 0.0758}}    & {\color[HTML]{000000} {\ul 0.0729}}    & {\color[HTML]{000000} {\ul 0.0774}}    \\
                                                        & Corr2                              & {\color[HTML]{000000} {\ul 0.0807}}    & {\color[HTML]{000000} {\ul 0.0971}}    & 0.0872                                 & 0.0892                                 & 0.0951                                 & 0.0865                                 \\
                                                        & TS2VEC                             & 0.1023                                 & 0.1464                                 & 0.1458                                 & 0.1218                                 & 0.1221                                 & 0.1338                                 \\
\multirow{-4}{*}{US}                                    & SimStock                           & {\color[HTML]{000000} \textbf{0.0359}} & {\color[HTML]{000000} \textbf{0.0360}} & {\color[HTML]{000000} \textbf{0.0465}} & {\color[HTML]{000000} \textbf{0.0535}} & {\color[HTML]{000000} \textbf{0.0551}} & {\color[HTML]{000000} \textbf{0.0548}} \\ \hline
                                                        & Corr1                              & 0.1326                                 & 0.1354                                 & 0.1379                                 & 0.1337                                 & 0.1315                                 & 0.1348                                 \\
                                                        & Corr2                              & {\color[HTML]{000000} {\ul 0.0810}}    & {\color[HTML]{000000} {\ul 0.0846}}    & {\color[HTML]{000000} {\ul 0.0814}}    & {\color[HTML]{000000} \textbf{0.0819}} & {\color[HTML]{000000} {\ul 0.0846}}    & {\color[HTML]{000000} {\ul 0.0887}}    \\
                                                        & TS2VEC                             & 0.1274                                 & 0.1204                                 & 0.1196                                 & 0.1315                                 & 0.1251                                 & 0.1225                                 \\
\multirow{-4}{*}{SSE}                                   & SimStock                           & {\color[HTML]{000000} \textbf{0.0664}} & {\color[HTML]{000000} \textbf{0.0745}} & {\color[HTML]{000000} \textbf{0.0760}} & {\color[HTML]{000000} {\ul 0.0828}}    & {\color[HTML]{000000} \textbf{0.0782}} & {\color[HTML]{000000} \textbf{0.0755}} \\ \hline
                                                        & Corr1                              & 0.1236                                 & 0.1143                                 & 0.1193                                 & 0.1250                                 & 0.1333                                 & 0.1369                                 \\
                                                        & Corr2                              & {\color[HTML]{000000} {\ul 0.0910}}    & 0.0884                                 & {\color[HTML]{000000} {\ul 0.0868}}    & {\color[HTML]{000000} {\ul 0.0836}}    & {\color[HTML]{000000} {\ul 0.0837}}    & {\color[HTML]{000000} \textbf{0.0808}} \\
                                                        & TS2VEC                             & 0.0914                                 & {\color[HTML]{000000} \textbf{0.0812}} & {\color[HTML]{000000} \textbf{0.0855}} & {\color[HTML]{000000} \textbf{0.0815}} & {\color[HTML]{000000} \textbf{0.0804}} & {\color[HTML]{000000} {\ul 0.0890}}    \\
\multirow{-4}{*}{SZSE}                                  & SimStock                           & {\color[HTML]{000000} \textbf{0.0808}} & {\color[HTML]{000000} {\ul 0.0875}}    & 0.1083                                 & 0.1030                                 & 0.1044                                 & 0.0991                                 \\ \hline
                                                        & Corr1                              & {\color[HTML]{000000} {\ul 0.0688}}    & {\color[HTML]{000000} {\ul 0.0680}}    & {\color[HTML]{000000} {\ul 0.0729}}    & {\color[HTML]{000000} \textbf{0.0757}} & {\color[HTML]{000000} \textbf{0.0724}} & {\color[HTML]{000000} \textbf{0.0643}} \\
                                                        & Corr2                              & 0.0773                                 & 0.0890                                 & 0.0837                                 & 0.0830                                 & 0.0864                                 & 0.0818                                 \\
                                                        & TS2VEC                             & 0.0970                                 & 0.0926                                 & 0.0834                                 & 0.0861                                 & 0.0804                                 & 0.0814                                 \\
\multirow{-4}{*}{TSE}                                   & SimStock                           & {\color[HTML]{000000} \textbf{0.0604}} & {\color[HTML]{000000} \textbf{0.0730}} & {\color[HTML]{000000} \textbf{0.0711}} & {\color[HTML]{000000} {\ul 0.0792}}    & {\color[HTML]{000000} {\ul 0.0794}}    & {\color[HTML]{000000} {\ul 0.0764}}    \\ \hline
\end{tabular}}

%% file: Appendix_Table/index_tracking_LIT_TEV.tex
\useunder{\uline}{\ul}{}
\centering
\fontsize{7}{8}\selectfont{
\begin{tabular}{cccccccc}
\hline
\multicolumn{1}{l}{}                                    &                                    & \multicolumn{6}{c}{\textbf{Global X Lithium \& Battery Tech ETF (LIT)}}                                                                                                                                                                              \\
\multicolumn{1}{l}{\multirow{-2}{*}{\textbf{Exchange}}} & \multirow{-2}{*}{\textbf{Methods}} & \textbf{\textbf{TOP@10}}               & \textbf{\textbf{TOP@15}}               & \textbf{\textbf{TOP@20}}               & \textbf{\textbf{TOP@25}}               & \textbf{\textbf{TOP@30}}               & \textbf{\textbf{TOP@35}}               \\ \hline
                                                        & Corr1                              & 0.1548                                 & 0.1744                                 & 0.1605                                 & 0.1708                                 & 0.1805                                 & 0.1706                                 \\
                                                        & Corr2                              & 0.0935                                 & 0.1440                                 & 0.1450                                 & 0.1433                                 & 0.1677                                 & 0.1246                                 \\
                                                        & TS2VEC                             & {\color[HTML]{000000} {\ul 0.0819}}    & {\color[HTML]{000000} {\ul 0.1091}}    & {\color[HTML]{000000} {\ul 0.0858}}    & {\color[HTML]{000000} {\ul 0.0869}}    & {\color[HTML]{000000} {\ul 0.1051}}    & {\color[HTML]{000000} {\ul 0.0996}}    \\
\multirow{-4}{*}{US}                                    & SimStock                           & {\color[HTML]{000000} \textbf{0.0389}} & {\color[HTML]{000000} \textbf{0.0513}} & {\color[HTML]{000000} \textbf{0.0501}} & {\color[HTML]{000000} \textbf{0.0492}} & {\color[HTML]{000000} \textbf{0.0560}} & {\color[HTML]{000000} \textbf{0.0536}} \\ \hline
                                                        & Corr1                              & {\color[HTML]{000000} {\ul 0.0840}}    & {\color[HTML]{000000} \textbf{0.0752}} & {\color[HTML]{000000} \textbf{0.0729}} & {\color[HTML]{000000} \textbf{0.0701}} & {\color[HTML]{000000} \textbf{0.0724}} & {\color[HTML]{000000} \textbf{0.0714}} \\
                                                        & Corr2                              & 0.0843                                 & 0.0837                                 & 0.0804                                 & 0.0843                                 & 0.0883                                 & 0.0901                                 \\
                                                        & TS2VEC                             & {\color[HTML]{000000} \textbf{0.0702}} & {\color[HTML]{000000} {\ul 0.0781}}    & {\color[HTML]{000000} {\ul 0.0764}}    & {\color[HTML]{000000} {\ul 0.0788}}    & {\color[HTML]{000000} {\ul 0.0817}}    & {\color[HTML]{000000} {\ul 0.0803}}    \\
\multirow{-4}{*}{SSE}                                   & SimStock                           & 0.0945                                 & 0.1193                                 & 0.1031                                 & 0.0942                                 & 0.1060                                 & 0.1114                                 \\ \hline
                                                        & Corr1                              & {\color[HTML]{000000} \textbf{0.0565}} & {\color[HTML]{000000} \textbf{0.0567}} & {\color[HTML]{000000} \textbf{0.0617}} & {\color[HTML]{000000} \textbf{0.0609}} & {\color[HTML]{000000} \textbf{0.0622}} & {\color[HTML]{000000} \textbf{0.0679}} \\
                                                        & Corr2                              & 0.2148                                 & 0.1898                                 & 0.1643                                 & 0.1509                                 & 0.1512                                 & 0.1399                                 \\
                                                        & TS2VEC                             & {\color[HTML]{000000} {\ul 0.0629}}    & {\color[HTML]{000000} {\ul 0.0661}}    & {\color[HTML]{000000} {\ul 0.0704}}    & {\color[HTML]{000000} {\ul 0.0723}}    & {\color[HTML]{000000} {\ul 0.0773}}    & {\color[HTML]{000000} {\ul 0.0821}}    \\
\multirow{-4}{*}{SZSE}                                  & SimStock                           & 0.2200                                 & 0.2361                                 & 0.2102                                 & 0.2057                                 & 0.1963                                 & 0.1817                                 \\ \hline
                                                        & Corr1                              & {\color[HTML]{000000} {\ul 0.1522}}    & {\color[HTML]{000000} {\ul 0.1718}}    & {\color[HTML]{000000} {\ul 0.1704}}    & 0.1839                                 & {\color[HTML]{000000} {\ul 0.1713}}    & {\color[HTML]{000000} {\ul 0.1556}}    \\
                                                        & Corr2                              & 0.1768                                 & 0.1807                                 & 0.1603                                 & 0.1575                                 & 0.1851                                 & 0.1771                                 \\
                                                        & TS2VEC                             & {\color[HTML]{000000} \textbf{0.1488}} & {\color[HTML]{000000} \textbf{0.1372}} & {\color[HTML]{000000} \textbf{0.1444}} & {\color[HTML]{000000} \textbf{0.1447}} & {\color[HTML]{000000} \textbf{0.1318}} & {\color[HTML]{000000} \textbf{0.1426}} \\
\multirow{-4}{*}{TSE}                                   & SimStock                           & 0.1580                                 & 0.1750                                 & 0.1715                                 & {\color[HTML]{000000} {\ul 0.1808}}    & 0.1837                                 & 0.1772                                 \\ \hline
\end{tabular}}

%% file: Appendix/Appendix_E.tex
\section*{Appendix E: Application to Portfolio optimization I}

\subsection*{Performance Statistics of Major Stock Indices}

In this section, we presents the performance statistics of four major stock indices: S\&P 500, JPX Prime 150, SSE 100, and SZSE 100 in \cref{chap_exchange_des}. The evaluation period spans from January 2023 to Feb 2024. The S\&P 500 and JPX Prime 150 indices exhibited positive returns during this period, while the SSE 100 and SZSE 100 indices experienced negative returns.

The S\&P 500 index, which represents the U.S. stock market, generated an arithmetic return of 4.87\% and a geometric return of 2.71\%. The cumulative return over the evaluation period was 13.02\%. The annualized standard deviation of 30.84\% indicates the high volatility of the index. The maximum drawdown of -31.32\% highlights the substantial decline the index experienced during the period. The monthly 95\% Value at Risk (VaR) of -11.9\% indicates the potential loss that could be incurred with a 95\% confidence level. The Sharpe ratio of 0.06 suggests a low risk-adjusted return for the S\&P 500 index.

The JPX Prime 150 index, representing the Japanese stock market, outperformed the other indices with an arithmetic return of 13.69\% and a geometric return of 11.61\%. The cumulative return over the evaluation period was 49.59\%. The annualized standard deviation of 27.18\% was lower than that of the S\&P 500 index, indicating relatively lower volatility. The annualized skewness and kurtosis values of -0.04 and 2.89, respectively, were similar to those of the S\&P 500 index. The maximum drawdown of -23.73\% was less severe compared to the other indices. The monthly 95\% VaR of -9.94\% was also lower than that of the S\&P 500 index. The Sharpe ratio of 0.58 indicates a higher risk-adjusted return for the JPX Prime 150 index compared to the S\&P 500 index.

The SSE 100 and SZSE 100 indices, representing the Chinese stock market, experienced negative returns during the evaluation period. The SSE 100 index had an arithmetic return of -5.36\% and a geometric return of -6.99\%, resulting in a cumulative return of -11.16\%. The SZSE 100 index had an arithmetic return of -7.81\% and a geometric return of -9.86\%, resulting in a cumulative return of -21.02\%. Both indices exhibited higher volatility compared to the S\&P 500 and JPX Prime 150 indices, with annualized standard deviations of 39.13\% and 37.39\%, respectively. The annualized skewness and kurtosis values were close to zero and slightly higher than three, respectively, indicating a relatively symmetric return distribution with slightly heavier tails compared to a normal distribution. The maximum drawdowns of -45.08\% and -46.1\% for the SSE 100 and SZSE 100 indices, respectively, were more severe compared to the S\&P 500 and JPX Prime 150 indices. The monthly 95\% VaR values of -16.24\% and -16.16\% were also higher, indicating a higher potential loss. The Sharpe ratios of -0.35 and -0.48 for the SSE 100 and SZSE 100 indices, respectively, suggest poor risk-adjusted returns during the evaluation period.

\setcounter{table}{0}
\renewcommand{\thetable}{E.\arabic{table}}
\begin{table}[ht]
\centering
    \input{Appendix_Table/exchange_des}
    \caption{Performance statistics of major stock indices from January 2023 to February 2024. The table presents various return and risk measures for the S\&P 500, JPX Prime 150, SSE 100, and SZSE 100 indices. The S\&P 500 and JPX Prime 150 indices exhibited positive returns, while the SSE 100 and SZSE 100 indices experienced negative returns during the evaluation period.}
    \label{chap_exchange_des}
\end{table}

\subsection*{Performance Metrics for Portfolio Optimization with 10 and 50 Stocks}
This appendix presents the performance metrics for MVO with 10 and 50 stocks in the S\&P 500 and JPX Prime 150 universes. The four portfolio construction methods evaluated are Simstock Embedding (SS), Historical Covariance (HC), Shrinkage Method (SM), and Gerber Statistic (GS).

We investigate the performance of the four covariance estimation methods (SS, HC, SM, GS, TS) for constructing portfolios with 10 and 50 stocks in the S\&P 500 and JPX Prime 150 universes at different target risk levels (24\%, 27\%, 30\%, and 33\%).

Tables C.3 and \cref{chap4_table_tse150_10} present the performance metrics for 10-stock portfolios in the S\&P 500 and JPX Prime 150, respectively. In both universes, the SS method generally outperforms the other methods across all risk target levels, achieving higher returns while maintaining similar risk measures. As the risk target level increases, the returns and risk measures for all methods also increase.

Tables C.5 and \cref{chap4_table_tse150_50} show the performance metrics for 50-stock portfolios in the S\&P 500 and JPX Prime 150, respectively. Similar to the 10-stock portfolios, the SS method demonstrates superior performance compared to the other methods across all risk target levels in both universes. The HC method has the second-best performance, followed by the SM, GS and TS methods. As the risk target level increases, the returns and risk measures for all methods increase.

The consistent outperformance of the Simstock Embedding (SS) method in terms of returns for both 10-stock and 50-stock portfolios in the S\&P 500 and JPX Prime 150 universes highlights its potential as an effective stock embedding technique for portfolio optimization.

\clearpage
\begin{sidewaystable}
    \input{portfolio_figures/snp_10_ts}
    \caption{This table presents the performance metrics for four portfolio construction methods in the JPX Prime 150: Simstock embedding(SS), historical covariance (HC), shrinkage method (SM), and Gerber statistic (GS). The portfolios were optimized for four different risk target levels: 24\%, 27\%, 30\%, and 33\%. The performance was evaluated over the full testing period from January 2022 to February 2024. The 3-month U.S. Treasury Bill rate was used as the risk-free rate for performance calculations. Transaction costs were modeled as 10 basis points of the traded volume for each rebalancing event.}
    \label{chap4_table_snp500_10}
    
    \vspace{20mm}
    
    \input{portfolio_figures/tse_10_ts}
    \caption{This table presents the performance metrics for four portfolio construction methods in the JPX Prime 150: Simstock embedding(SS), historical covariance (HC), shrinkage method (SM), and Gerber statistic (GS). The portfolios were optimized for four different risk target levels: 24\%, 27\%, 30\%, and 33\%. The performance was evaluated over the full testing period from January 2022 to February 2024. The 3-month U.S. Treasury Bill rate was used as the risk-free rate for performance calculations. Transaction costs were modeled as 10 basis points of the traded volume for each rebalancing event.}
    
    \label{chap4_table_tse150_10}
\end{sidewaystable}

\clearpage
\begin{sidewaystable}

    \input{portfolio_figures/snp_50_ts}
\caption{This table presents the performance metrics for four portfolio construction methods in the JPX Prime 150: Simstock embedding(SS), historical covariance (HC), shrinkage method (SM), and Gerber statistic (GS). The portfolios were optimized for four different risk target levels: 24\%, 27\%, 30\%, and 33\%. The performance was evaluated over the full testing period from January 2022 to February 2024. The 3-month U.S. Treasury Bill rate was used as the risk-free rate for performance calculations. Transaction costs were modeled as 10 basis points of the traded volume for each rebalancing event.}
\label{chap4_table_snp_50}
    \vspace{20mm}
    
    \input{portfolio_figures/tse_50_ts}
\caption{This table presents the performance metrics for four portfolio construction methods in the JPX Prime 150: Simstock embedding(SS), historical covariance (HC), shrinkage method (SM), and Gerber statistic (GS). The portfolios were optimized for four different risk target levels: 24\%, 27\%, 30\%, and 33\%. The performance was evaluated over the full testing period from January 2022 to February 2024. The 3-month U.S. Treasury Bill rate was used as the risk-free rate for performance calculations. Transaction costs were modeled as 10 basis points of the traded volume for each rebalancing event.}
\label{chap4_table_tse150_50}
\end{sidewaystable}
\clearpage

%% file: Appendix_Table/exchange_des.tex
\centering
\fontsize{7}{8}\selectfont{
\begin{tabular}{rcccc}
\hline
Evaluation Method                          & \multicolumn{4}{c}{Exchange}                 \\ \hline
\multicolumn{1}{r|}{Covariance Method}     & S\&P500 & JPX Prime 150 & SSE 100 & SZSE 100 \\ \hline
\multicolumn{1}{r|}{Arithmetic Return (\%)} & 4.87    & 13.69         & -5.36   & -7.81    \\ \cline{1-1}
\multicolumn{1}{r|}{Geometric Return (\%)}  & 2.71    & 11.61         & -6.99   & -9.86    \\ \cline{1-1}
\multicolumn{1}{r|}{Cumulative Return (\%)} & 13.02   & 49.59         & -11.16  & -21.02   \\ \cline{1-1}
\multicolumn{1}{r|}{Annualized SD (\%)}     & 30.84   & 27.18         & 39.13   & 37.39    \\ \cline{1-1}
\multicolumn{1}{r|}{Annualized Skewness}   & -0.03   & -0.04         & 0.04    & 0.01     \\ \cline{1-1}
\multicolumn{1}{r|}{Annualized Kurtosis}   & 2.81    & 2.89          & 3.12    & 3.24     \\ \cline{1-1}
\multicolumn{1}{r|}{Maximum Drawdown (\%)}  & -31.32  & -23.73        & -45.08  & -46.1    \\ \cline{1-1}
\multicolumn{1}{r|}{Monthly 95\% VaR (\%)}   & -11.9   & -9.94         & -16.24  & -16.16   \\ \cline{1-1}
\multicolumn{1}{r|}{Sharpe Ratio}          & 0.06    & 0.58          & -0.35   & -0.48    \\ \hline
\end{tabular}}

%% file: portfolio_figures/snp_10_ts.tex
\fontsize{6}{9}\selectfont{
\begin{tabular}{r|ccccc|ccccc|ccccc|ccccc}
\hline
S\&P500 | 10 Stocks    & \multicolumn{5}{c|}{Target Volatility (24\%)} & \multicolumn{5}{c|}{Target Volatility (27\%)} & \multicolumn{5}{c|}{Target Volatility (30\%)} & \multicolumn{5}{c}{Target Volatility (33\%)} \\ \hline
Covariance Method     & SS      & HC      & SM      & GS    & TS     & SS      & HC      & SM     & GS     & TS     & SS      & HC      & SM     & GS     & TS     & SS      & HC     & SM     & GS     & TS     \\ \hline
Arithmetic Return (\%) & 7.94    & 6.85    & 6.84    & 6.97  & 5.28   & 8.38    & 7.35    & 7.27   & 7.27   & 6.02   & 8.56    & 7.71    & 7.8    & 7.71   & 7.63   & 8.84    & 8.12   & 8.27   & 8.13   & 7.63   \\ 
Geometric Return (\%)  & 6.78    & 6.05    & 6.08    & 6.11  & 3.63   & 7.1     & 6.41    & 6.34   & 6.25   & 4.84   & 7.18    & 6.61    & 6.68   & 6.51   & 6.5    & 7.31    & 6.85   & 6.98   & 6.77   & 6.50   \\ 
Cumulative Return (\%) & 24.63   & 22.11   & 22.11   & 22.32 & 11.29  & 26.26   & 23.73   & 23.49  & 23.19  & 15.24  & 26.79   & 24.73   & 25.05  & 24.39  & 20.78  & 27.52   & 25.71  & 26.3   & 25.53  & 20.78  \\ 
Annualized SD (\%)     & 27.22   & 24.68   & 24.57   & 25.06 & 27.33  & 28.18   & 26.06   & 26.11  & 26.49  & 28.32  & 29.11   & 27.31   & 27.47  & 27.79  & 29.55  & 29.94   & 28.44  & 28.67  & 28.92  & 29.55  \\ 
Annualized Skewness   & -0.06   & -0.06   & -0.03   & -0.03 & -0.09  & -0.08   & -0.07   & -0.06  & -0.06  & -0.11  & -0.09   & -0.08   & -0.07  & -0.08  & -0.12  & -0.10   & -0.08  & -0.08  & -0.08  & -0.12  \\ 
Annualized Kurtosis   & 3.04    & 3.06    & 3.04    & 3.04  & 2.98   & 3.09    & 3.08    & 3.07   & 3.07   & 2.95   & 3.12    & 3.12    & 3.11   & 3.11   & 2.93   & 3.15    & 3.15   & 3.15   & 3.15   & 2.93   \\ 
Maximum Drawdown (\%)  & -26.07  & -24.22  & -24.04  & -24.4 & -26.91 & -27.13  & -25.28  & -25.23 & -25.73 & -26.75 & -27.93  & -26.52  & -26.41 & -26.93 & -26.27 & -28.59  & -27.49 & -27.46 & -27.85 & -26.27 \\ 
Monthly 95\% VaR (\%)   & -10.24  & -9.43   & -9.39   & -9.57 & -10.9  & -10.62  & -9.92   & -9.93  & -10.06 & -11.21 & -11.00  & -10.30  & -10.36 & -10.49 & -11.54 & -11.28  & -10.7  & -10.76 & -10.85 & -11.54 \\
Sharpe Ratio          & 0.24    & 0.22    & 0.23    & 0.23  & 0.07   & 0.25    & 0.23    & 0.23   & 0.22   & 0.16   & 0.25    & 0.23    & 0.23   & 0.22   & 0.24   & 0.24    & 0.23   & 0.24   & 0.23   & 0.24   \\ 
Annualized Turnover   & 7.15    & 6.81    & 6.88    & 6.94  & 6.54   & 7.31    & 7.00    & 7.09   & 7.15   & 6.72   & 7.22    & 7.15    & 7.21   & 7.25   & 6.90   & 7.23    & 7.21   & 7.24   & 7.26   & 6.90   \\ \hline
\end{tabular}}

%% file: portfolio_figures/tse_10_ts.tex
\fontsize{6}{9}\selectfont{
\begin{tabular}{r|ccccc|ccccc|ccccc|ccccc}
\hline
JPM Prime 150 | 10 Stocks & \multicolumn{5}{c|}{Target Volatility (24\%)} & \multicolumn{5}{c|}{Target Volatility (27\%)} & \multicolumn{5}{c|}{Target Volatility (30\%)} & \multicolumn{5}{c}{Target Volatility (33\%)} \\ \hline
Covariance Method         & SS     & HC      & SM      & GS     & TS     & SS      & HC     & SM      & GS     & TS     & SS      & HC      & SM     & GS     & TS     & SS      & HC     & SM     & GS     & TS     \\ \hline
Arithmetic Return (\%)     & 15.63  & 14.41   & 14.51   & 14.4   & 12.64  & 16.19   & 15.43  & 15.67   & 15.55  & 13.39  & 16.59   & 16.43   & 16.74  & 16.7   & 13.03  & 17.64   & 17.4   & 17.42  & 17.55  & 13.61  \\ 
Geometric Return (\%)      & 13.94  & 13.17   & 13.1    & 13.06  & 11.70  & 14.28   & 13.88  & 13.92   & 13.86  & 12.36  & 14.55   & 14.6    & 14.68  & 14.71  & 11.48  & 15.34   & 15.32  & 15.17  & 15.32  & 13.09  \\ 
Cumulative Return (\%)     & 53     & 49.12   & 48.85   & 48.69  & 39.37  & 55.09   & 53.04  & 53.44   & 53.24  & 41.84  & 56.99   & 56.99   & 57.98  & 58.02  & 38.53  & 61.27   & 61.1   & 61.09  & 61.66  & 44.65  \\ 
Annualized SD (\%)         & 25.93  & 25.05   & 24.8    & 25.09  & 26.65  & 27.34   & 27.01  & 27.09   & 27.21  & 27.85  & 28.65   & 28.89   & 29.18  & 29.27  & 28.88  & 29.97   & 30.61  & 30.89  & 30.95  & 29.31  \\ 
Annualized Skewness       & 0.18   & 0.25    & 0.24    & 0.21   & 0.15   & 0.21    & 0.28   & 0.3     & 0.27   & 0.18   & 0.24    & 0.32    & 0.33   & 0.3    & 0.22   & 0.24    & 0.32   & 0.34   & 0.32   & 0.18   \\ 
Annualized Kurtosis       & 3.55   & 3.3     & 3.33    & 3.3    & 3.08   & 3.64    & 3.38   & 3.43    & 3.38   & 3.05   & 3.73    & 3.46    & 3.52   & 3.45   & 2.98   & 3.69    & 3.51   & 3.58   & 3.53   & 2.94   \\ 
Maximum Drawdown (\%)      & -21.1  & -20.52  & -20.56  & -20.88 & -22.53 & -22.37  & -22    & -22.29  & -22.6  & -22.68 & -23.38  & -23.38  & -23.89 & -24.09 & -23.89 & -24.44  & -24.78 & -25.24 & -25.36 & -24.67 \\ 
Monthly 95\% VaR (\%)       & -8.83  & -8.76   & -8.61   & -8.81  & -8.98  & -9.35   & -9.41  & -9.29   & -9.48  & -9.38  & -9.78   & -9.97   & -10    & -10.1  & -9.68  & -10.18  & -10.46 & -10.5  & -10.57 & -10.22 \\ 
Sharpe Ratio              & 0.68   & 0.65    & 0.66    & 0.65   & 0.52   & 0.66    & 0.64   & 0.64    & 0.64   & 0.50   & 0.64    & 0.63    & 0.63   & 0.63   & 0.47   & 0.64    & 0.63   & 0.62   & 0.62   & 0.53   \\ 
Annualized Turnover       & 7.28   & 7.56    & 7.47    & 7.53   & 7.47   & 7.45    & 7.67   & 7.59    & 7.63   & 7.32   & 7.57    & 7.71    & 7.64   & 7.65   & 7.48   & 7.75    & 7.68   & 7.63   & 7.61   & 7.40   \\ \hline
\end{tabular}}

%% file: portfolio_figures/snp_50_ts.tex
\fontsize{6}{9}\selectfont{
\begin{tabular}{r|ccccc|ccccc|ccccc|ccccc}
\hline
S\&P500 | 50 Stocks    & \multicolumn{5}{c|}{Target Volatility (24\%)} & \multicolumn{5}{c|}{Target Volatility (27\%)} & \multicolumn{5}{c|}{Target Volatility (30\%)} & \multicolumn{5}{c}{Target Volatility (33\%)} \\ \hline
Covariance Method     & SS      & HC      & SM     & GS     & TS     & SS      & HC      & SM     & GS     & TS     & SS      & HC     & SM      & GS     & TS     & SS      & HC     & SM     & GS     & TS     \\ \hline
Arithmetic Return (\%) & 10.41   & 10.27   & 9.82   & 10.54  & 10.65  & 11.86   & 11.15   & 11.38  & 11.59  & 11.63  & 12.82   & 12.1   & 12.7    & 12.67  & 12.66  & 13.87   & 13.03  & 13.78  & 13.57  & 13.17  \\ 
Geometric Return (\%)  & 9.59    & 9.5     & 9.19   & 9.84   & 8.94   & 10.92   & 10.24   & 10.57  & 10.71  & 10.39  & 11.74   & 11.01  & 11.66   & 11.56  & 11.26  & 12.57   & 11.73  & 12.47  & 12.27  & 11.32  \\ 
Cumulative Return (\%) & 34.15   & 33.49   & 32.06  & 34.49  & 29.29  & 39.14   & 36.27   & 37.5   & 38.06  & 34.54  & 42.3    & 39.41  & 41.93   & 41.71  & 37.75  & 45.6    & 42.45  & 45.44  & 44.76  & 37.95  \\ 
Annualized SD (\%)     & 29      & 27.93   & 27.2   & 27.53  & 30.31  & 30.6    & 29.67   & 29.42  & 29.58  & 31.09  & 32.34   & 31.38  & 31.54   & 31.64  & 32.28  & 33.92   & 33.07  & 33.47  & 33.52  & 33.13  \\ 
Annualized Skewness   & -0.17   & -0.22   & -0.19  & -0.18  & -0.17  & -0.2    & -0.25   & -0.24  & -0.23  & -0.21  & -0.23   & -0.27  & -0.27   & -0.26  & -0.19  & -0.26   & -0.29  & -0.31  & -0.29  & -0.19  \\ 
Annualized Kurtosis   & 3.12    & 3.28    & 3.32   & 3.19   & 2.83   & 3.18    & 3.35    & 3.36   & 3.27   & 2.84   & 3.24    & 3.37   & 3.4     & 3.31   & 2.87   & 3.24    & 3.36   & 3.38   & 3.3    & 2.97   \\ 
Maximum Drawdown (\%)  & -25.4   & -24.36  & -23.35 & -23.44 & -26.04 & -25.79  & -25.44  & -24.66 & -24.97 & -26.18 & -26.33  & -26.6  & -25.94  & -26.53 & -26.15 & -27.29  & -27.68 & -27.34 & -27.85 & -26.57 \\ 
Monthly 95\% VaR (\%)   & -10.96  & -10.36  & -10.15 & -10.18 & -11.38 & -11.28  & -10.79  & -10.78 & -10.82 & -11.83 & -11.64  & -11.2  & -11.3   & -11.29 & -12.22 & -12.06  & -11.68 & -11.8  & -11.79 & -12.40 \\ 
Sharpe Ratio          & 0.39    & 0.41    & 0.39   & 0.43   & 0.36   & 0.44    & 0.42    & 0.44   & 0.45   & 0.40   & 0.46    & 0.43   & 0.46    & 0.46   & 0.45   & 0.48    & 0.44   & 0.47   & 0.47   & 0.46   \\ 
Annualized Turnover   & 8.94    & 8.57    & 8.69   & 8.5    & 8.10   & 8.9     & 8.63    & 8.74   & 8.57   & 8.21   & 8.84    & 8.66   & 8.67    & 8.6    & 8.04   & 8.7     & 8.63   & 8.55   & 8.54   & 7.82   \\ \hline
\end{tabular}}

%% file: portfolio_figures/tse_50_ts.tex
\fontsize{6}{9}\selectfont{
\begin{tabular}{r|ccccc|ccccc|ccccc|ccccc}
\hline
JPM Prime 150 | 50 Stocks & \multicolumn{5}{c|}{Target Volatility (24\%)} & \multicolumn{5}{c|}{Target Volatility (27\%)} & \multicolumn{5}{c|}{Target Volatility (30\%)} & \multicolumn{5}{c}{Target Volatility (33\%)} \\ \hline
Covariance Method         & SS      & HC      & SM     & GS     & TS     & SS      & HC      & SM     & GS     & TS     & SS      & HC      & SM     & GS     & TS     & SS      & HC     & SM     & GS     & TS     \\ \hline
Arithmetic Return (\%)     & 20.3    & 19.7    & 18.6   & 17.55  & 15.69  & 21.3    & 20.03   & 19.79  & 18.99  & 15.32  & 22.74   & 20.34   & 21.2   & 20.22  & 15.46  & 23.49   & 20.93  & 21.93  & 21.25  & 15.84  \\ 
Geometric Return (\%)      & 18.9    & 18.52   & 17.31  & 16.51  & 14.00  & 19.89   & 18.79   & 18.38  & 17.8   & 14.16  & 21.27   & 19.04   & 19.68  & 18.96  & 13.82  & 21.98   & 19.52  & 20.34  & 19.86  & 14.74  \\ 
Cumulative Return (\%)     & 70.97   & 69.2    & 63.79  & 59.9   & 48.15  & 75.61   & 70.44   & 68.66  & 65.51  & 48.77  & 82.1    & 71.7    & 74.76  & 70.89  & 47.44  & 85.94   & 74.21  & 78.27  & 75.42  & 51.04  \\ 
Annualized SD (\%)         & 28.13   & 27.37   & 26.14  & 26.54  & 26.85  & 29.89   & 28.77   & 28.08  & 28.34  & 28.32  & 31.52   & 30.25   & 29.98  & 30.02  & 29.28  & 33.01   & 31.65  & 31.42  & 31.43  & 30.25  \\ 
Annualized Skewness       & -0.03   & 0.00    & -0.01  & -0.01  & -0.10  & 0.02    & -0.03   & 0.02   & 0.01   & -0.09  & 0.03    & -0.03   & 0.02   & -0.01  & -0.08  & 0.03    & -0.06  & -0.02  & -0.07  & -0.03  \\ 
Annualized Kurtosis       & 2.89    & 2.86    & 2.83   & 2.79   & 2.62   & 2.88    & 2.88    & 2.90   & 2.87   & 2.66   & 2.89    & 2.88    & 2.93   & 2.88   & 2.63   & 2.9     & 2.91   & 2.91   & 2.91   & 2.62   \\ 
Maximum Drawdown (\%)      & -21.47  & -20.57  & -20.47 & -20.64 & -23.59 & -22.41  & -21.99  & -21.58 & -21.61 & -23.34 & -22.99  & -23.12  & -22.54 & -22.45 & -24.94 & -23.79  & -24.39 & -23.67 & -23.67 & -25.09 \\ 
Monthly 95\% VaR (\%)       & -9.73   & -9.68   & -8.9   & -9.27  & -10.17 & -10.25  & -10.23  & -9.55  & -9.87  & -10.59 & -10.74  & -10.8   & -10.22 & -10.51 & -10.99 & -11.22  & -11.35 & -10.83 & -11.18 & -11.51 \\ 
Sharpe Ratio              & 0.93    & 0.93    & 0.91   & 0.85   & 0.68   & 0.93    & 0.9     & 0.91   & 0.86   & 0.64   & 0.96    & 0.87    & 0.92   & 0.88   & 0.62   & 0.95    & 0.85   & 0.91   & 0.88   & 0.62   \\ 
Annualized Turnover       & 8.99    & 8.77    & 8.98   & 8.96   & 8.67   & 9.00    & 8.91    & 9.08   & 8.99   & 8.57   & 8.91    & 9.01    & 9.00   & 8.93   & 8.76   & 8.86    & 8.98   & 8.97   & 8.90   & 8.80   \\ \hline
\end{tabular}}

%% file: Appendix/Appendix_F.tex
\section*{Appendix F: Application to Portfolio optimization II}
\subsection*{Robustness of SimStock covariance construction for portfolio optimization}

To assess the robustness of the SimStock embedding approach for constructing the covariance matrix in portfolio optimization, we compare the ex-post efficient frontiers of portfolios based on different choices for computing the covariance matrix from the embeddings. Specifically, we consider the following methods:
\begin{itemize}
    \item L2: Pairwise L2 distances between stock embeddings, scaled to the range [-1, 1] using max-min normalization (the method used in the main text).
    \item L1: Pairwise L1 distances between stock embeddings, scaled to the range [-1, 1] using max-min normalization.
    \item CORR: Pearson correlation coefficients between stock embeddings.
    \item CKA \citep{kornblith2019similarity} : Centered Kernel Alignment (CKA) with a linear kernel between stock embeddings.
\end{itemize}

\cref{chap4_ABL_eff} presents the ex-post efficient frontiers for portfolios of various sizes (10, 30, and 50 stocks) from both the S\&P500 and JPX Prime 150 universes, constructed using the different covariance matrix computation methods. \cref{chap4_ABL_eff} presents the ex-post efficient frontiers for portfolios of various sizes (10, 30, and 50 stocks) from both the S\&P500 and JPX Prime 150 universes, constructed using the different covariance matrix computation methods.

 The results show that the choice of method for constructing the covariance matrix from the SimStock embeddings does impact the overall portfolio optimization performance. In most cases, the L2 distance method (used in the main text) achieves the best performance, followed by the L1 distance, CKA, and CORR methods. The superiority of the L2 and L1 distance methods suggests that capturing the dissimilarity between stock embeddings is more effective for constructing the covariance matrix than directly using correlation coefficients or kernel alignment.

\setcounter{figure}{0}
\renewcommand{\thefigure}{F.\arabic{figure}}
\begin{figure}[h]
\centering
\includegraphics[width=\textwidth]{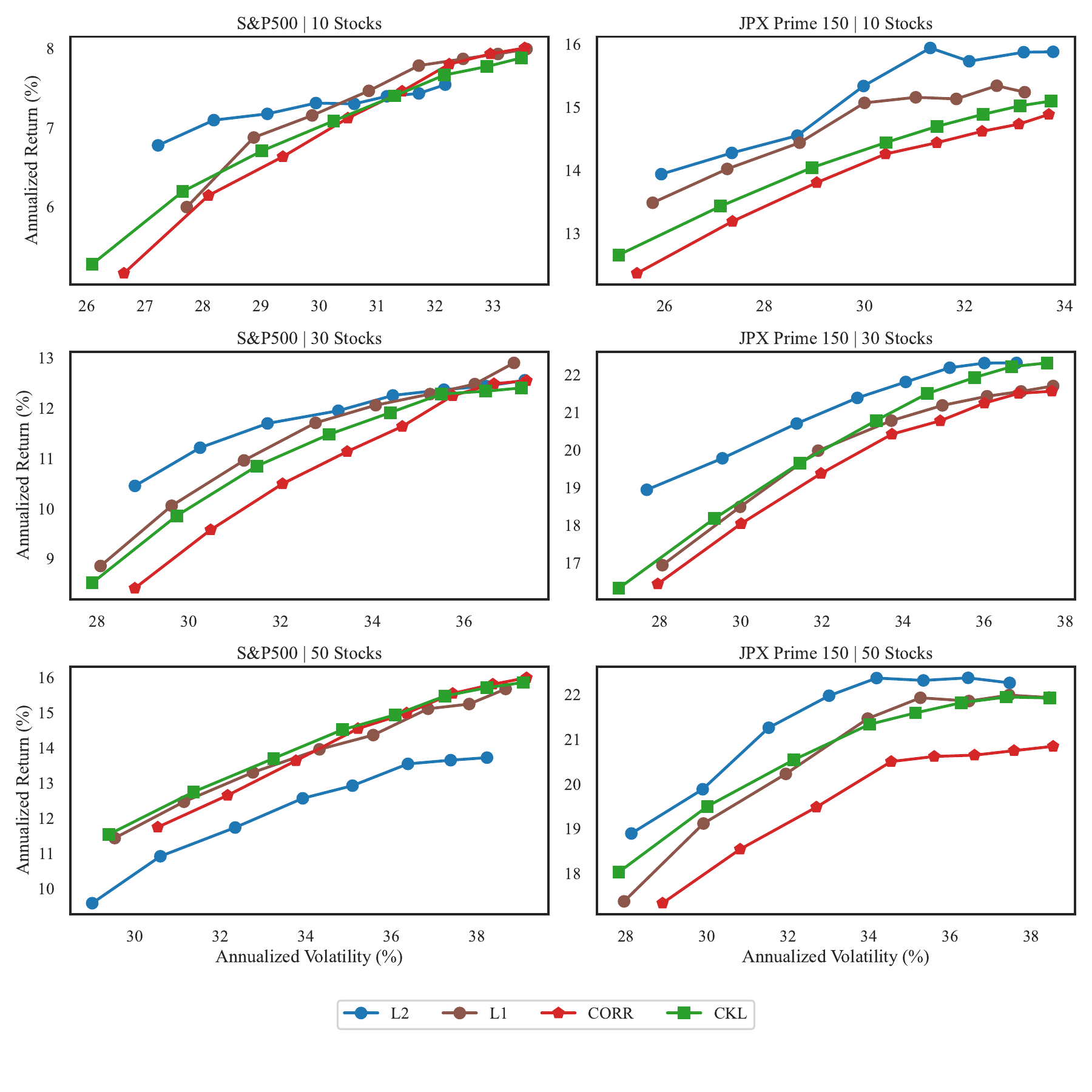}
\caption{Ex-post efficient frontiers displaying annualized return and volatility of portfolios optimized using different methods for computing the covariance matrix from SimStock embeddings.}
\label{chap4_ABL_eff}
\end{figure}

%% file: main.bbl
\begin{thebibliography}{84}
\providecommand{\natexlab}[1]{#1}
\providecommand{\noopsort}[1]{}
\providecommand{\printfirst}[2]{#1}
\providecommand{\singleletter}[1]{#1}
\providecommand{\switchargs}[2]{#2#1}

\bibitem[\protect\citeauthoryear{Ang and Bekaert}{2002}]{ang2002international}
Ang, A. and Bekaert, G., International asset allocation with regime shifts. {\itshape The review of financial studies}, 2002, \textbf{15}, 1137--1187.

\bibitem[\protect\citeauthoryear{Bai {\itshape{et~al.}}}{2022}]{bai2022temporal}
Bai, G., Ling, C. and Zhao, L., Temporal Domain Generalization with Drift-Aware Dynamic Neural Networks. {\itshape arXiv preprint arXiv:2205.10664}, 2022.

\bibitem[\protect\citeauthoryear{Balntas {\itshape{et~al.}}}{2016}]{balntas2016learning}
Balntas, V., Riba, E., Ponsa, D. and Mikolajczyk, K., Learning local feature descriptors with triplets and shallow convolutional neural networks.. In {\itshape Proceedings of the }{\itshape Bmvc}, Vol. ~1, p.~3, 2016.

\bibitem[\protect\citeauthoryear{Barberis}{2000}]{barberis2000investing}
Barberis, N., Investing for the long run when returns are predictable. {\itshape The Journal of Finance}, 2000, \textbf{55}, 225--264.

\bibitem[\protect\citeauthoryear{Berndt and Clifford}{1994}]{berndt1994using}
Berndt, D.J. and Clifford, J., Using dynamic time warping to find patterns in time series.. In {\itshape Proceedings of the }{\itshape KDD workshop}, Vol. ~10, pp. 359--370, 1994.

\bibitem[\protect\citeauthoryear{Butler and Kwon}{2023}]{butler2023integrating}
Butler, A. and Kwon, R.H., Integrating prediction in mean-variance portfolio optimization. {\itshape Quantitative Finance}, 2023, \textbf{23}, 429--452.

\bibitem[\protect\citeauthoryear{Caron {\itshape{et~al.}}}{2021}]{caron2021emerging}
Caron, M., Touvron, H., Misra, I., J{\'e}gou, H., Mairal, J., Bojanowski, P. and Joulin, A., Emerging properties in self-supervised vision transformers. In {\itshape Proceedings of the }{\itshape Proceedings of the IEEE/CVF international conference on computer vision}, pp. 9650--9660, 2021.

\bibitem[\protect\citeauthoryear{Chen {\itshape{et~al.}}}{2020}]{chen2020simple}
Chen, T., Kornblith, S., Norouzi, M. and Hinton, G., A simple framework for contrastive learning of visual representations. In {\itshape Proceedings of the }{\itshape International conference on machine learning}, pp. 1597--1607, 2020.

\bibitem[\protect\citeauthoryear{Choi and Kang}{2023}]{choi2023multi}
Choi, H. and Kang, P., Multi-task self-supervised time-series representation learning. {\itshape arXiv preprint arXiv:2303.01034}, 2023.

\bibitem[\protect\citeauthoryear{Chopra and Ziemba}{2013}]{chopra2013effect}
Chopra, V.K. and Ziemba, W.T., The effect of errors in means, variances, and covariances on optimal portfolio choice. In {\itshape Handbook of the fundamentals of financial decision making: Part I}, pp. 365--373, 2013, World Scientific.

\bibitem[\protect\citeauthoryear{Chung {\itshape{et~al.}}}{2022}]{chung2022effects}
Chung, M., Lee, Y., Kim, J.H., Kim, W.C. and Fabozzi, F.J., The effects of errors in means, variances, and correlations on the mean-variance framework. {\itshape Quantitative Finance}, 2022, \textbf{22}, 1893--1903.

\bibitem[\protect\citeauthoryear{Cont}{2001}]{cont2001empirical}
Cont, R., Empirical properties of asset returns: stylized facts and statistical issues. {\itshape Quantitative finance}, 2001, \textbf{1}, 223.

\bibitem[\protect\citeauthoryear{Costa and Iyengar}{2023}]{costa2023distributionally}
Costa, G. and Iyengar, G.N., Distributionally robust end-to-end portfolio construction. {\itshape Quantitative Finance}, 2023, \textbf{23}, 1465--1482.

\bibitem[\protect\citeauthoryear{Devlin {\itshape{et~al.}}}{2018}]{devlin2018bert}
Devlin, J., Chang, M.W., Lee, K. and Toutanova, K., Bert: Pre-training of deep bidirectional transformers for language understanding. {\itshape arXiv preprint arXiv:1810.04805}, 2018.

\bibitem[\protect\citeauthoryear{Enders}{2004}]{enders2004cointegration}
Enders, W., Cointegration and error-correction models. {\itshape Applied econometrics time series}, 2004, pp. 319--386.

\bibitem[\protect\citeauthoryear{Fama and French}{1993}]{fama1993common}
Fama, E.F. and French, K.R., Common risk factors in the returns on stocks and bonds. {\itshape Journal of financial economics}, 1993, \textbf{33}, 3--56.

\bibitem[\protect\citeauthoryear{Fama and French}{2015}]{fama2015five}
Fama, E.F. and French, K.R., A five-factor asset pricing model. {\itshape Journal of financial economics}, 2015, \textbf{116}, 1--22.

\bibitem[\protect\citeauthoryear{Fang {\itshape{et~al.}}}{2022}]{fang2022corrupted}
Fang, Y., Dong, L., Bao, H., Wang, X. and Wei, F., Corrupted image modeling for self-supervised visual pre-training. {\itshape arXiv preprint arXiv:2202.03382}, 2022.

\bibitem[\protect\citeauthoryear{Focardi and Fabozzi}{2004}]{focardi2004methodology}
Focardi, S.M. and Fabozzi, F.J., A methodology for index tracking based on time-series clustering. {\itshape Quantitative Finance}, 2004, \textbf{4}, 417--425.

\bibitem[\protect\citeauthoryear{Gatev {\itshape{et~al.}}}{2006}]{gatev2006pairs}
Gatev, E., Goetzmann, W.N. and Rouwenhorst, K.G., Pairs trading: Performance of a relative-value arbitrage rule. {\itshape The Review of Financial Studies}, 2006, \textbf{19}, 797--827.

\bibitem[\protect\citeauthoryear{Gerber {\itshape{et~al.}}}{2021}]{gerber2021gerber}
Gerber, S., Markowitz, H., Ernst, P., Miao, Y., Sargen, P. {\itshape et~al.}, The Gerber statistic: A robust co-movement measure for portfolio optimization. {\itshape No and Sargen, Paul, The Gerber Statistic: A Robust Co-Movement Measure for Portfolio Optimization (July 4, 2021)}, 2021.

\bibitem[\protect\citeauthoryear{Gorishniy {\itshape{et~al.}}}{2021}]{gorishniy2021revisiting}
Gorishniy, Y., Rubachev, I., Khrulkov, V. and Babenko, A., Revisiting deep learning models for tabular data. {\itshape Advances in Neural Information Processing Systems}, 2021, \textbf{34}, 18932--18943.

\bibitem[\protect\citeauthoryear{Grill {\itshape{et~al.}}}{2020}]{grill2020bootstrap}
Grill, J.B., Strub, F., Altch{\'e}, F., Tallec, C., Richemond, P., Buchatskaya, E., Doersch, C., Avila~Pires, B., Guo, Z., Gheshlaghi~Azar, M. {\itshape et~al.}, Bootstrap your own latent-a new approach to self-supervised learning. {\itshape Advances in neural information processing systems}, 2020, \textbf{33}, 21271--21284.

\bibitem[\protect\citeauthoryear{Guidolin and Timmermann}{2007}]{guidolin2007asset}
Guidolin, M. and Timmermann, A., Asset allocation under multivariate regime switching. {\itshape Journal of Economic Dynamics and Control}, 2007, \textbf{31}, 3503--3544.

\bibitem[\protect\citeauthoryear{He {\itshape{et~al.}}}{2020}]{he2020momentum}
He, K., Fan, H., Wu, Y., Xie, S. and Girshick, R., Momentum contrast for unsupervised visual representation learning. In {\itshape Proceedings of the }{\itshape Proceedings of the IEEE/CVF conference on computer vision and pattern recognition}, pp. 9729--9738, 2020.

\bibitem[\protect\citeauthoryear{Hochreiter and Schmidhuber}{1997}]{hochreiter1997long}
Hochreiter, S. and Schmidhuber, J., Long short-term memory. {\itshape Neural computation}, 1997, \textbf{9}, 1735--1780.

\bibitem[\protect\citeauthoryear{Hwang {\itshape{et~al.}}}{2023{\natexlab{a}}}]{hwang2023simstock}
Hwang, Y., Lee, J., Kim, D., Noh, S., Hong, J. and Lee, Y., SimStock: Representation Model for Stock Similarities. In {\itshape Proceedings of the }{\itshape Proceedings of the Fourth ACM International Conference on AI in Finance}, pp. 533--540, 2023{\natexlab{a}}.

\bibitem[\protect\citeauthoryear{Hwang {\itshape{et~al.}}}{2023{\natexlab{b}}}]{hwang2023stop}
Hwang, Y., Park, J., Lee, Y. and Lim, D.Y., Stop-loss adjusted labels for machine learning-based trading of risky assets. {\itshape Finance Research Letters}, 2023{\natexlab{b}}, \textbf{58}, 104285.

\bibitem[\protect\citeauthoryear{Jenni {\itshape{et~al.}}}{2020}]{jenni2020video}
Jenni, S., Meishvili, G. and Favaro, P., Video representation learning by recognizing temporal transformations. In {\itshape Proceedings of the }{\itshape Computer Vision--ECCV 2020: 16th European Conference, Glasgow, UK, August 23--28, 2020, Proceedings, Part XXVIII 16}, pp. 425--442, 2020.

\bibitem[\protect\citeauthoryear{Jiang {\itshape{et~al.}}}{2024}]{jiang2024mixtral}
Jiang, A.Q., Sablayrolles, A., Roux, A., Mensch, A., Savary, B., Bamford, C., Chaplot, D.S., Casas, D.d.l., Hanna, E.B., Bressand, F. {\itshape et~al.}, Mixtral of experts. {\itshape arXiv preprint arXiv:2401.04088}, 2024.

\bibitem[\protect\citeauthoryear{Jing and Tian}{2020}]{jing2020self}
Jing, L. and Tian, Y., Self-supervised visual feature learning with deep neural networks: A survey. {\itshape IEEE transactions on pattern analysis and machine intelligence}, 2020, \textbf{43}, 4037--4058.

\bibitem[\protect\citeauthoryear{Jobson and Korkie}{1980}]{jobson1980estimation}
Jobson, J.D. and Korkie, B., Estimation for Markowitz efficient portfolios. {\itshape Journal of the American Statistical Association}, 1980, \textbf{75}, 544--554.

\bibitem[\protect\citeauthoryear{Kan and Zhou}{2004}]{kan2004optimal}
Kan, R. and Zhou, G., Optimal estimation for economic gains: Portfolio choice with parameter uncertainty. {\itshape Washington University, St. Louis}, 2004.

\bibitem[\protect\citeauthoryear{Kan and Zhou}{2007}]{kan2007optimal}
Kan, R. and Zhou, G., Optimal portfolio choice with parameter uncertainty. {\itshape Journal of Financial and Quantitative Analysis}, 2007, \textbf{42}, 621--656.

\bibitem[\protect\citeauthoryear{Kim {\itshape{et~al.}}}{2021}]{kim2021mean}
Kim, J.H., Lee, Y., Kim, W.C. and Fabozzi, F.J., Mean-variance optimization for asset allocation. {\itshape Journal of Portfolio Management}, 2021, \textbf{47}, 24--40.

\bibitem[\protect\citeauthoryear{Kim {\itshape{et~al.}}}{2024}]{kim2024topp}
Kim, P.J., Jang, Y., Kim, J. and Yoo, J., TopP\&R: Robust Support Estimation Approach for Evaluating Fidelity and Diversity in Generative Models. {\itshape Advances in Neural Information Processing Systems}, 2024, \textbf{36}.

\bibitem[\protect\citeauthoryear{Kornblith {\itshape{et~al.}}}{2019}]{kornblith2019similarity}
Kornblith, S., Norouzi, M., Lee, H. and Hinton, G., Similarity of neural network representations revisited. In {\itshape Proceedings of the }{\itshape International conference on machine learning}, pp. 3519--3529, 2019.

\bibitem[\protect\citeauthoryear{Ledoit and Wolf}{2003}]{ledoit2003improved}
Ledoit, O. and Wolf, M., Improved estimation of the covariance matrix of stock returns with an application to portfolio selection. {\itshape Journal of empirical finance}, 2003, \textbf{10}, 603--621.

\bibitem[\protect\citeauthoryear{Ledoit and Wolf}{2004{\natexlab{a}}}]{ledoit2004honeyorigin}
Ledoit, O. and Wolf, M., Honey, I Shrunk the Sample Covariance Matrix. {\itshape Journal of Portfolio Management}, 2004{\natexlab{a}}, \textbf{30}, 110.

\bibitem[\protect\citeauthoryear{Ledoit and Wolf}{2004{\natexlab{b}}}]{ledoit2004well}
Ledoit, O. and Wolf, M., A well-conditioned estimator for large-dimensional covariance matrices. {\itshape Journal of multivariate analysis}, 2004{\natexlab{b}}, \textbf{88}, 365--411.

\bibitem[\protect\citeauthoryear{Lee {\itshape{et~al.}}}{2019}]{lee2019latent}
Lee, K., Chang, M.W. and Toutanova, K., Latent retrieval for weakly supervised open domain question answering. {\itshape arXiv preprint arXiv:1906.00300}, 2019.

\bibitem[\protect\citeauthoryear{Lee {\itshape{et~al.}}}{2023}]{lee2023overview}
Lee, Y., Thompson, J.R., Kim, J.H., Kim, W.C. and Fabozzi, F.A., An overview of machine learning for asset management. {\itshape The Journal of Portfolio Management}, 2023, \textbf{49}, 31--63.

\bibitem[\protect\citeauthoryear{Lim {\itshape{et~al.}}}{2019}]{lim2019enhancing}
Lim, B., Zohren, S. and Roberts, S., Enhancing time series momentum strategies using deep neural networks. {\itshape arXiv preprint arXiv:1904.04912}, 2019.

\bibitem[\protect\citeauthoryear{Lintner}{1975}]{lintner1975valuation}
Lintner, J., The valuation of risk assets and the selection of risky investments in stock portfolios and capital budgets. In {\itshape Stochastic optimization models in finance}, pp. 131--155, 1975, Elsevier.

\bibitem[\protect\citeauthoryear{Lo}{2017}]{lo2017adaptive}
Lo, A., {\itshape Adaptive markets: Financial evolution at the speed of thought}, 2017, Princeton University Press.

\bibitem[\protect\citeauthoryear{Lu {\itshape{et~al.}}}{2018}]{lu2018learning}
Lu, J., Liu, A., Dong, F., Gu, F., Gama, J. and Zhang, G., Learning under concept drift: A review. {\itshape IEEE transactions on knowledge and data engineering}, 2018, \textbf{31}, 2346--2363.

\bibitem[\protect\citeauthoryear{Mann {\itshape{et~al.}}}{2020}]{mann2020language}
Mann, B., Ryder, N., Subbiah, M., Kaplan, J., Dhariwal, P., Neelakantan, A., Shyam, P., Sastry, G., Askell, A., Agarwal, S. {\itshape et~al.}, Language models are few-shot learners. {\itshape arXiv preprint arXiv:2005.14165}, 2020.

\bibitem[\protect\citeauthoryear{Markowits}{1952}]{markowits1952portfolio}
Markowits, H.M., Portfolio selection. {\itshape Journal of Finance}, 1952, \textbf{7}, 71--91.

\bibitem[\protect\citeauthoryear{Markowitz}{1952}]{Markowitz1952}
Markowitz, H., Portfolio Selection. {\itshape The Journal of Finance}, 1952, \textbf{7}, 77--91.

\bibitem[\protect\citeauthoryear{Michaud}{1989}]{michaud1989markowitz}
Michaud, R.O., The Markowitz optimization enigma: Is ‘optimized’optimal?. {\itshape Financial analysts journal}, 1989, \textbf{45}, 31--42.

\bibitem[\protect\citeauthoryear{Nelson and Plosser}{1982}]{nelson1982trends}
Nelson, C.R. and Plosser, C.R., Trends and random walks in macroeconmic time series: some evidence and implications. {\itshape Journal of monetary economics}, 1982, \textbf{10}, 139--162.

\bibitem[\protect\citeauthoryear{Nie {\itshape{et~al.}}}{2024}]{nie2024survey}
Nie, Y., Kong, Y., Dong, X., Mulvey, J.M., Poor, H.V., Wen, Q. and Zohren, S., A Survey of Large Language Models for Financial Applications: Progress, Prospects and Challenges. {\itshape arXiv preprint arXiv:2406.11903}, 2024.

\bibitem[\protect\citeauthoryear{Pan and Yang}{2009}]{pan2009survey}
Pan, S.J. and Yang, Q., A survey on transfer learning. {\itshape IEEE Transactions on knowledge and data engineering}, 2009, \textbf{22}, 1345--1359.

\bibitem[\protect\citeauthoryear{Qian {\itshape{et~al.}}}{2021}]{qian2021spatiotemporal}
Qian, R., Meng, T., Gong, B., Yang, M.H., Wang, H., Belongie, S. and Cui, Y., Spatiotemporal contrastive video representation learning. In {\itshape Proceedings of the }{\itshape Proceedings of the IEEE/CVF Conference on Computer Vision and Pattern Recognition}, pp. 6964--6974, 2021.

\bibitem[\protect\citeauthoryear{Qiu {\itshape{et~al.}}}{2020}]{qiu2020pre}
Qiu, X., Sun, T., Xu, Y., Shao, Y., Dai, N. and Huang, X., Pre-trained models for natural language processing: A survey. {\itshape Science China Technological Sciences}, 2020, \textbf{63}, 1872--1897.

\bibitem[\protect\citeauthoryear{Radford {\itshape{et~al.}}}{2019}]{radford2019language}
Radford, A., Wu, J., Child, R., Luan, D., Amodei, D., Sutskever, I. {\itshape et~al.}, Language models are unsupervised multitask learners. {\itshape OpenAI blog}, 2019, \textbf{1}, 9.

\bibitem[\protect\citeauthoryear{Rapach {\itshape{et~al.}}}{2010}]{rapach2010out}
Rapach, D.E., Strauss, J.K. and Zhou, G., Out-of-sample equity premium prediction: Combination forecasts and links to the real economy. {\itshape The Review of Financial Studies}, 2010, \textbf{23}, 821--862.

\bibitem[\protect\citeauthoryear{Rho {\itshape{et~al.}}}{2024}]{rho2024language}
Rho, S., Smith, L., Li, T., Levine, S., Peng, X.B. and Ha, S., Language Guided Skill Discovery. {\itshape arXiv preprint arXiv:2406.06615}, 2024.

\bibitem[\protect\citeauthoryear{Rousseeuw and Driessen}{1999}]{rousseeuw1999fast}
Rousseeuw, P.J. and Driessen, K.V., A fast algorithm for the minimum covariance determinant estimator. {\itshape Technometrics}, 1999, \textbf{41}, 212--223.

\bibitem[\protect\citeauthoryear{Ruder and Plank}{2018}]{ruder2018strong}
Ruder, S. and Plank, B., Strong baselines for neural semi-supervised learning under domain shift. {\itshape arXiv preprint arXiv:1804.09530}, 2018.

\bibitem[\protect\citeauthoryear{Sharpe}{1964}]{sharpe1964capital}
Sharpe, W.F., Capital asset prices: A theory of market equilibrium under conditions of risk. {\itshape The journal of finance}, 1964, \textbf{19}, 425--442.

\bibitem[\protect\citeauthoryear{Song {\itshape{et~al.}}}{2020}]{song2020mpnet}
Song, K., Tan, X., Qin, T., Lu, J. and Liu, T.Y., Mpnet: Masked and permuted pre-training for language understanding. {\itshape Advances in Neural Information Processing Systems}, 2020, \textbf{33}, 16857--16867.

\bibitem[\protect\citeauthoryear{Touvron {\itshape{et~al.}}}{2023}]{touvron2023llama}
Touvron, H., Lavril, T., Izacard, G., Martinet, X., Lachaux, M.A., Lacroix, T., Rozi{\`e}re, B., Goyal, N., Hambro, E., Azhar, F. {\itshape et~al.}, Llama: Open and efficient foundation language models. {\itshape arXiv preprint arXiv:2302.13971}, 2023.

\bibitem[\protect\citeauthoryear{Uysal {\itshape{et~al.}}}{2023}]{uysal2023end}
Uysal, A.S., Li, X. and Mulvey, J.M., End-to-end risk budgeting portfolio optimization with neural networks. {\itshape Annals of Operations Research}, 2023, pp. 1--30.

\bibitem[\protect\citeauthoryear{Van~Aelst and Rousseeuw}{2009}]{van2009minimum}
Van~Aelst, S. and Rousseeuw, P., Minimum volume ellipsoid. {\itshape Wiley Interdisciplinary Reviews: Computational Statistics}, 2009, \textbf{1}, 71--82.

\bibitem[\protect\citeauthoryear{Vaswani {\itshape{et~al.}}}{2017}]{vaswani2017attention}
Vaswani, A., Shazeer, N., Parmar, N., Uszkoreit, J., Jones, L., Gomez, A.N., Kaiser, {\L}. and Polosukhin, I., Attention is all you need. {\itshape Advances in neural information processing systems}, 2017, \textbf{30}.

\bibitem[\protect\citeauthoryear{Wan {\itshape{et~al.}}}{2021}]{wan2021sentiment}
Wan, X., Yang, J., Marinov, S., Calliess, J.P., Zohren, S. and Dong, X., Sentiment correlation in financial news networks and associated market movements. {\itshape Scientific reports}, 2021, \textbf{11}, 3062.

\bibitem[\protect\citeauthoryear{Wei {\itshape{et~al.}}}{2022}]{wei2022masked}
Wei, C., Fan, H., Xie, S., Wu, C.Y., Yuille, A. and Feichtenhofer, C., Masked feature prediction for self-supervised visual pre-training. In {\itshape Proceedings of the }{\itshape Proceedings of the IEEE/CVF Conference on Computer Vision and Pattern Recognition}, pp. 14668--14678, 2022.

\bibitem[\protect\citeauthoryear{Welch and Goyal}{2008}]{welch2008comprehensive}
Welch, I. and Goyal, A., A comprehensive look at the empirical performance of equity premium prediction. {\itshape The Review of Financial Studies}, 2008, \textbf{21}, 1455--1508.

\bibitem[\protect\citeauthoryear{Wen {\itshape{et~al.}}}{2020}]{wen2020fast}
Wen, Q., Zhang, Z., Li, Y. and Sun, L., Fast RobustSTL: Efficient and robust seasonal-trend decomposition for time series with complex patterns. In {\itshape Proceedings of the }{\itshape Proceedings of the 26th ACM SIGKDD International Conference on Knowledge Discovery \& Data Mining}, pp. 2203--2213, 2020.

\bibitem[\protect\citeauthoryear{Woo {\itshape{et~al.}}}{2022}]{etsformer}
Woo, G., Liu, C., Sahoo, D., Kumar, A. and Hoi, S., Etsformer: Exponential smoothing transformers for time-series forecasting. {\itshape arXiv preprint arXiv:2202.01381}, 2022.

\bibitem[\protect\citeauthoryear{Wood {\itshape{et~al.}}}{2023}]{wood2023few}
Wood, K., Kessler, S., Roberts, S.J. and Zohren, S., Few-Shot Learning Patterns in Financial Time-Series for Trend-Following Strategies. {\itshape arXiv preprint arXiv:2310.10500}, 2023.

\bibitem[\protect\citeauthoryear{Wu {\itshape{et~al.}}}{2021{\natexlab{a}}}]{wu2021autoformer}
Wu, H., Xu, J., Wang, J. and Long, M., Autoformer: Decomposition transformers with auto-correlation for long-term series forecasting. {\itshape Advances in neural information processing systems}, 2021{\natexlab{a}}, \textbf{34}, 22419--22430.

\bibitem[\protect\citeauthoryear{Wu {\itshape{et~al.}}}{2021{\natexlab{b}}}]{autoformer}
Wu, H., Xu, J., Wang, J. and Long, M., Autoformer: Decomposition transformers with auto-correlation for long-term series forecasting. {\itshape Advances in Neural Information Processing Systems}, 2021{\natexlab{b}}, \textbf{34}, 22419--22430.

\bibitem[\protect\citeauthoryear{Xiao {\itshape{et~al.}}}{2021}]{xiao2021region}
Xiao, T., Reed, C.J., Wang, X., Keutzer, K. and Darrell, T., Region similarity representation learning. In {\itshape Proceedings of the }{\itshape Proceedings of the IEEE/CVF International Conference on Computer Vision}, pp. 10539--10548, 2021.

\bibitem[\protect\citeauthoryear{Yang {\itshape{et~al.}}}{2019}]{yang2019xlnet}
Yang, Z., Dai, Z., Yang, Y., Carbonell, J., Salakhutdinov, R.R. and Le, Q.V., Xlnet: Generalized autoregressive pretraining for language understanding. {\itshape Advances in neural information processing systems}, 2019, \textbf{32}.

\bibitem[\protect\citeauthoryear{Yue {\itshape{et~al.}}}{2022}]{yue2022ts2vec}
Yue, Z., Wang, Y., Duan, J., Yang, T., Huang, C., Tong, Y. and Xu, B., Ts2vec: Towards universal representation of time series. In {\itshape Proceedings of the }{\itshape Proceedings of the AAAI Conference on Artificial Intelligence}, Vol. ~36, pp. 8980--8987, 2022.

\bibitem[\protect\citeauthoryear{Yun {\itshape{et~al.}}}{2019}]{yun2019cutmix}
Yun, S., Han, D., Oh, S.J., Chun, S., Choe, J. and Yoo, Y., Cutmix: Regularization strategy to train strong classifiers with localizable features. In {\itshape Proceedings of the }{\itshape Proceedings of the IEEE/CVF international conference on computer vision}, pp. 6023--6032, 2019.

\bibitem[\protect\citeauthoryear{Yun {\itshape{et~al.}}}{2022}]{yun2022patch}
Yun, S., Lee, H., Kim, J. and Shin, J., Patch-level representation learning for self-supervised vision transformers. In {\itshape Proceedings of the }{\itshape Proceedings of the IEEE/CVF conference on computer vision and pattern recognition}, pp. 8354--8363, 2022.

\bibitem[\protect\citeauthoryear{Zhai {\itshape{et~al.}}}{2019}]{zhai2019s4l}
Zhai, X., Oliver, A., Kolesnikov, A. and Beyer, L., S4l: Self-supervised semi-supervised learning. In {\itshape Proceedings of the }{\itshape Proceedings of the IEEE/CVF international conference on computer vision}, pp. 1476--1485, 2019.

\bibitem[\protect\citeauthoryear{Zhang {\itshape{et~al.}}}{2021}]{zhang2021universal}
Zhang, C., Zhang, Z., Cucuringu, M. and Zohren, S., A universal end-to-end approach to portfolio optimization via deep learning. {\itshape arXiv preprint arXiv:2111.09170}, 2021.

\bibitem[\protect\citeauthoryear{Zhang {\itshape{et~al.}}}{2017}]{zhang2017mixup}
Zhang, H., Cisse, M., Dauphin, Y.N. and Lopez-Paz, D., mixup: Beyond empirical risk minimization. {\itshape arXiv preprint arXiv:1710.09412}, 2017.

\bibitem[\protect\citeauthoryear{Zhang {\itshape{et~al.}}}{2020}]{zhang2020deep}
Zhang, Z., Zohren, S. and Roberts, S., Deep learning for portfolio optimization. {\itshape arXiv preprint arXiv:2005.13665}, 2020.

\bibitem[\protect\citeauthoryear{Zhou {\itshape{et~al.}}}{2022}]{zhou2022fedformer}
Zhou, T., Ma, Z., Wen, Q., Wang, X., Sun, L. and Jin, R., Fedformer: Frequency enhanced decomposed transformer for long-term series forecasting. In {\itshape Proceedings of the }{\itshape International Conference on Machine Learning}, pp. 27268--27286, 2022.

\end{thebibliography}
